\documentclass[a4paper,fleqn]{cas-dc}

\usepackage[numbers,sort&compress]{natbib}

\usepackage{caption}
\usepackage{pifont} 
\usepackage{threeparttable}
\usepackage{arydshln}

\usepackage{hyperref}
\hypersetup{colorlinks=true, citecolor=blue, linkcolor=blue, urlcolor=blue}

\usepackage{wrapfig} 
\usepackage{flushend}
\usepackage{appendix}
\usepackage{bbding}
\usepackage{algorithm}
\usepackage{listings}

\definecolor{darkgreen}{RGB}{0,100,0}

\lstdefinestyle{mystyle}{
    language=Python,
    basicstyle=\small\ttfamily\bfseries,
    keywordstyle=\color{blue},
    commentstyle=\color{darkgreen},
    showstringspaces=false,
    breaklines=true,
    tabsize=4
}

\def\tsc#1{\csdef{#1}{\textsc{\lowercase{#1}}\xspace}}
\tsc{WGM}
\tsc{QE}
\tsc{EP}
\tsc{PMS}
\tsc{BEC}
\tsc{DE}

\begin{document}
\let\WriteBookmarks\relax
\def\floatpagepagefraction{1}
\def\textpagefraction{.001}

\shorttitle{TC-MGC: Text-Conditioned Multi-Grained Contrastive Learning for Text-Video Retrieval}

\shortauthors{Xiaolun Jing et al.}

\title [mode = title]{TC-MGC: Text-Conditioned Multi-Grained Contrastive Learning for Text-Video Retrieval}                      
	
\author[1,2]{Xiaolun Jing}[orcid=0000-0002-1159-062X]

\ead{jingxiaolun@sjtu.edu.cn}

\credit{Conceptualization of this study, Methodology, Experiment}

\address[1]{Ningbo Artificial Intelligence Institute, Shanghai Jiao Tong University, China}
\address[2]{Department of Automation, Shanghai Jiao Tong University, China}

\author[1,2]{Genke Yang}[orcid=0000-0003-3492-0211]
\cormark[1]
\cortext[1]{Corresponding author}


\ead{gkyang@sjtu.edu.cn}

\author[1,2]{Jian Chu}

\ead{chujian@sjtu.edu.cn}

\begin{abstract}
Motivated by the success of coarse-grained or fine-grained contrast in text-video retrieval, there emerge multi-grained contrastive learning methods which focus on the integration of contrasts with different granularity. However, due to the wider semantic range of videos, the text-agnostic video representations might encode misleading information not described in texts, thus impeding the model from capturing precise cross-modal semantic correspondence. To this end, we propose a Text-Conditioned Multi-Grained Contrast framework, dubbed TC-MGC. Specifically, our model employs a language-video attention block to generate aggregated frame and video representations conditioned on the word's and text's attention weights over frames. To filter unnecessary similarity interactions and decrease trainable parameters in the Interactive Similarity Aggregation (ISA) module, we design a Similarity Reorganization (SR) module to identify attentive similarities and reorganize cross-modal similarity vectors and matrices. Next, we argue that the imbalance problem among multi-grained similarities may result in over- and under-representation issues. We thereby introduce an auxiliary Similarity Decorrelation Regularization (SDR) loss to facilitate cooperative relationship utilization by similarity variance minimization on matching text-video pairs. Finally, we present a Linear Softmax Aggregation (LSA) module to explicitly encourage the interactions between multiple similarities and promote the usage of multi-grained information. Empirically, TC-MGC achieves competitive results on multiple text-video retrieval benchmarks, outperforming X-CLIP model by +2.8\% (+1.3\%), +2.2\% (+1.0\%), +1.5\% (+0.9\%) relative (absolute) improvements in text-to-video retrieval R@1 on MSR-VTT, DiDeMo and VATEX, respectively. Our code is publicly available at \href{https://github.com/JingXiaolun/TC-MGC}{https://github.com/JingXiaolun/TC-MGC}.
\end{abstract}


\maketitle

\section{Introduction}
\label{Introduction}
Text-video retrieval (TVR) aims to retrieve relevant videos based on text queries (text-to-video retrieval, T2V) and search semantically matching texts based on video queries (video-to-text retrieval, V2T). With the explosive growth of videos on the Internet during the past decade, TVR has become increasingly essential and attracted widespread attention from the academia community. People also desire a better text-video retrieval system as quickly finding target videos has been a part of daily lives.

The recent breakthroughs in large-scale contrastive image-language pre-training benefit TVR significantly. One of the representative models, CLIP \cite{radford2021learning} employs a dual-stream network architecture to learn generalizable multi-modal knowledge from large-scale contrastive learning among 400M image-text pairs. As a pioneering work, CLIP4Clip \cite{luo2022clip4clip} successfully extends the image-text knowledge of CLIP \cite{radford2021learning} into the TVR task to perform coarse-grained contrast, resulting in significant performance improvements on text-video retrieval benchmarks. However, CLIP4Clip \cite{luo2022clip4clip} simply conducts global alignment between sentence and video representations, thus lacking the ability to capture fine-grained semantic details. To this, some previous works \cite{yao2021filip, zou2022tokenflow} propose fine-grained contrastive
frameworks to explore the contrast between each word in the sentence and each frame in the video. For example, Fine-grained Interactive Language-Image Pre-training (FILIP) \cite{yao2021filip} is presented to leverage a cross-modal late interaction mechanism for finer-level semantic alignment. TokenFlow \cite{zou2022tokenflow} designs a new model-agnostic formulation to achieve fine-grained cross-modal alignment. Despite the success of these methods, they are limited to single-grained contrast and insufficient to simultaneously consider fine-grained information (fine-grained contrast) and contextual information (coarse-grained contrast). Therefore, several works \cite{ma2022x, wang2023unified} follow the paradigm of multi-grained contrastive learning to address this weakness. X-CLIP \cite{ma2022x} introduces cross-grained contrasts to lower the negative effects of unnecessary frames and unimportant words. UCoFiA \cite{wang2023unified} accomplishes the unification of multi-grained alignment from patch-word, frame-sentence, and video-sentence contrasts.
Although these approaches have shown satisfying results, cross-modal semantic contrast still remains challenging due to the semantic range discrepancy between the text and video.

\begin{figure*}[!t]
	\centerline{\includegraphics[width=0.9\textwidth]{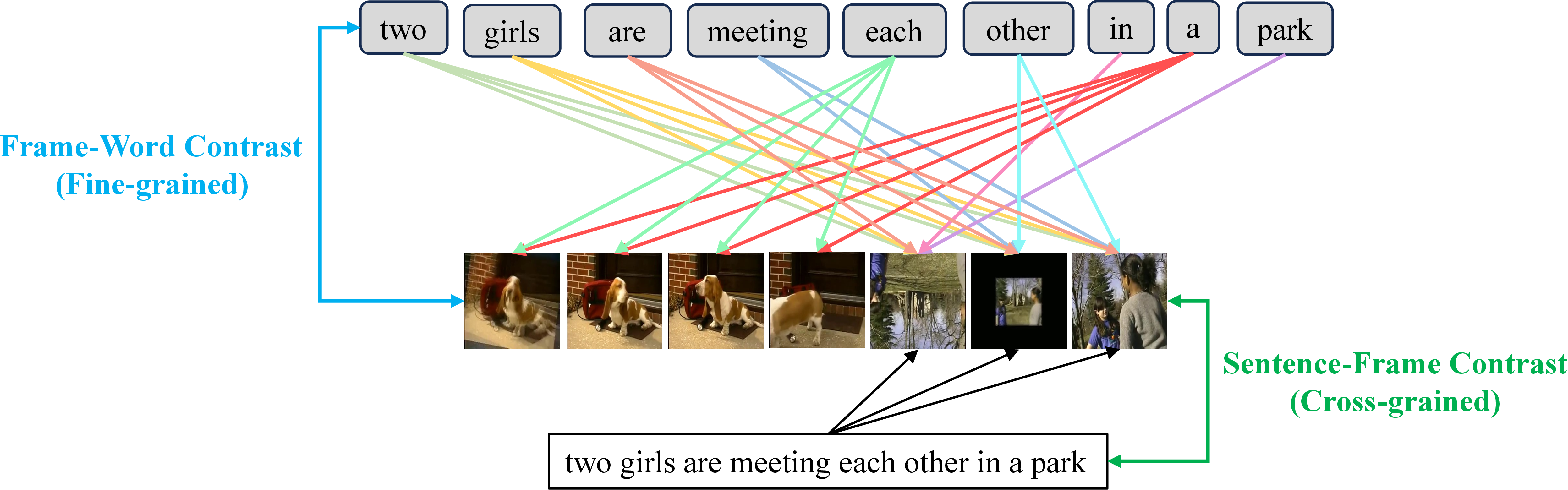}}
	\caption{Illustration of the multi-grained contrasts between frame and sentence (word) representations, including sentence-frame (cross-grained) and frame-word (fine-grained) contrasts. The connections indicate that the texts are semantic-relevant to sub-regions of videos.}
	\label{fig:contrast between sent-frame and word-frame}
\end{figure*}

As shown in Fig. \ref{fig:contrast between sent-frame and word-frame}, a video is composed of multiple frames depicting various scenes, and a sentence consists of several words expressing different semantics. Both global sentences and individual words are partially semantic-relevant to video frames. Concretely, the sentence is semantic-aligned with the last three frames but irrelevant to others. Some words (\textit{e.g.,} girls and park) only keep high relevance to sub-regions of video frames. However, most current methods \cite{gorti2022x, zhang2024text} mainly focus on the video representation refinement via cross-attention or softmax-based interactions between sentence and frames without considering the frame representations refinement, which is also critical for fine-grained semantic alignment.

To this end, we present TC-MGC, a Text-Conditioned Multi-Grained Contrast model for text-video retrieval. The main idea behind our method is to refine video and frame representations through the cross-attention mechanism between multi-grained textual and visual representations. Specifically, our approach begins with textual and visual representations extraction through modality-specific encoders. Next, we use a language-video attention block to generate semantic-relevant video and frame representations in a text-guided manner. Finally, we perform multi-grained contrasts to obtain similarity scores, vectors, and matrices for cross-modal semantic alignment.

Inspired by the significant improvements brought by the Interactive Similarity Aggregation module (ISA) in UCoFiA \cite{wang2023unified}, we employ ISA and its bidirectional variant (Bi-ISA) modules that use feature interactions in the 
cross-modal similarity vectors/matrices aggregation. However, not all feature interactions are necessary. Specifically, content words in the sentence like nouns and verbs express concrete semantics while function words, such as articles and prepositions, contain less useful information. The interactions involving these unimportant features may be noisy and hinder the model from capturing precise cross-modal correspondence. To address this issue, we propose a similarity reorganization (SR) module to identify attentive similarities. For the video-word contrast, our SR module preserves the attentive similarities and removes the inattentive similarities. For the sentence-frame contrast, considering the integrity of visual information, we fuse the inattentive similarities into one similarity. For the frame-word contrast, we design a bidirectional variant (Bi-SR) to preserve attentive similarities and remove inattentive similarities.

Next, once multi-grained scores are obtained, we can average them as a retrieval score. However, we find that there exists a highly imbalanced problem across different scores. Specifically, sometimes one specific similarity score might be much higher than the other similarities, meaning that this similarity is over-represented and will lower the utilization of other similarities. To overcome this limitation, we introduce an auxiliary Similarity Decorrelation Regularization (SDR) loss to incorporate variance minimization among matching text-video data into the training objective. Besides, considering the relationships modeling among different scores, we transfer the idea of ISA to  similarities aggregation, and present a new Linear Softmax Aggregation (LSA) module to facilitate multi-grained information interaction before aggregation. 

In short, our main contributions in this work can be summarized as follows: \par
$\bullet$ We propose a novel Text-Conditioned Multi-Grained Contrast (TC-MGC) framework to explore multi-grained contrasts between textual and semantic-relevant visual representations. \par
$\bullet$ We present three effective designs: the Similarity Reorganization (SR) module, the Similarity Decorrelation Regularization (SDR) loss and the Linear Softmax Aggregation (LSA) module for similarity vectors/matrices reorganization, over-representation/under-representation issues alleviation and multi-grained scores aggregation. \par
$\bullet$ We conduct extensive experiments on three benchmark datasets of MSR-VTT \cite{xu2016msr}, DiDeMo \cite{anne2017localizing} and VATEX \cite{wang2019vatex} to demonstrate the merits of our approach.

The rest of this paper is organized as follows. Section \ref{Related Work} reviews the related work. Section \ref{Methodology} describes each component of our proposed approach. Section \ref{Experiments} presents the experimental results and analysis, followed by the limitations and discussions in section \ref{Limitations and Discussions}. Section \ref{Conclusion} concludes this study. 

\section{Related Work}
\label{Related Work}
\subsection{Contrastive Learning}
Most contrastive learning methods can be divided into two categories: multi-view contrastive learning and multi-modal contrastive learning. The former leverages multiple views of the same data to facilitate feature representation learning, while the latter integrates diverse information to improve semantic understanding across different modalities. In terms of multi-view contrastive learning, recent years have witnessed continuous advances \cite{yang2021partially, liu2021contrastive, liu2023contrastive}. The authors of \cite{yang2021partially} present MvCLN to handle the partially view-aligned problem by endowing contrastive learning with the noise-robust contrastive loss. TupleInfoNCE \cite{liu2021contrastive} designs a novel contrastive learning objective to contrast multi-modal anchor tuples with challenging negative samples. CMK \cite{liu2023contrastive} implicitly embeds the views into a joint semantic space to ensure all views resemble each other and promote diverse views learning. As for multi-modal contrastive learning, the large-scale contrastive image-text pre-training \cite{radford2021learning, jia2021scaling} has injected new impetus into the prosperity of various downstream cross-modal tasks \cite{luo2022clip4clip,mokady2021clipcap,ye2023video}. 
For example, the authors of CLIP4Clip \cite{luo2022clip4clip} introduce CLIP \cite{radford2021learning} into text-video retrieval and achieve impressive results. ClipCap \cite{
mokady2021clipcap} uses CLIP encoding as a prefix to the caption and fine-tunes a language model to generate the image captions. CCVQA \cite{ye2023video} proposes a CLIP-guided visual-text attention mechanism to boost cross-modal learning for video question answering. Similar to CLIP4Clip \cite{luo2022clip4clip}, our work is built upon CLIP \cite{radford2021learning} for the text-video retrieval task.

\subsection{Text-Video Retrieval}
Text-video retrieval is a fundamental but challenging task in multi-modal understanding. Previously, prototypical approaches \cite{miech2018learning, gabeur2020multi, croitoru2021teachtext, liu2019use, dzabraev2021mdmmt} focus on designing task-specific or modality-specific fusion mechanisms for cross-modal semantic alignment. Later, the end-to-end paradigm of text-video pre-training from raw text/video has gained large popularity. HowTo100M \cite{miech2019howto100m}, MIL-NCE \cite{miech2020end}, ActBERT \cite{zhu2020actbert} and VideoBERT \cite{sun2019videobert} are all such works. With the prominent success of large-scale pre-training model CLIP \cite{radford2021learning}, several works attempt to transfer the image-text knowledge of CLIP \cite{radford2021learning} to the video domain. We refer to methods using CLIP \cite{radford2021learning} for feature extraction as CLIP-based methods. CLIP4Clip \cite{luo2022clip4clip} is the first work to apply CLIP \cite{radford2021learning} knowledge to text-video retrieval and investigate three similarity calculators for coarse-grained contrast between text and video features. CenterCLIP \cite{zhao2022centerclip} designs a multi-segment token clustering algorithm to select the most representative tokens. TS2-Net \cite{liu2022ts2} introduces a token selection module to find the most informative tokens in both temporal and spatial dimensions. DRL \cite{wang2022disentangled} designs a Weighted Token-wise Interaction mechanism to exploit the pair-wise correlations. X-Pool \cite{gorti2022x} attempts to generate an aggregated video representation according to text-guided attention. X-CLIP \cite{ma2022x} presents a multi-grained contrastive learning framework to filter out unnecessary fine-grained features. UCoFiA \cite{wang2023unified} accomplishes the effective unification of multi-grained alignments by jointly considering the similarity of different granularity. However, above CLIP-based methods except X-Pool \cite{gorti2022x} focus on either single-grained or multi-grained contrast between text and text-agnostic video representations. Furthermore, Unlike X-Pool \cite{gorti2022x} limited to single-grained contrast between text and text-conditioned video representations, we propose a text-conditioned multi-grained contrastive learning method for text-video retrieval, by considering all the contrasts between textual (sentence and word) and text-conditioned visual (video and frame) representations with different granularity.

\subsection{Token Reorganization}
The token reorganization has a long history in single-modality models acceleration. In the text-only domain, PoWER-BERT \cite{goyal2020power} first exploits redundancy pertaining to the word-vectors from intermediate encoder outputs for speeding up the inference process. Following the line of PoWER-BERT \cite{goyal2020power}, some works are also dedicated to token pruning without altering the network architecture. For example, SpAtten \cite{wang2021spatten} presents a cascade token pruning mechanism to remove unimportant tokens for attention computation reduction. LTP \cite{kim2022learned} designs a Learned Token Pruning scheme to adaptively prune away unimportant tokens as an input sequence passes through transformer layers. Additionally, similar token pruning approaches are also presented in the vision-only domain. DynamicViT \cite{rao2021dynamicvit} and A-ViT \cite{yin2022vit}
are all such works. Meanwhile, another line of merging inattentive tokens has been explored in some recent works. EViT \cite{liang2022not} proposes a token reorganization method to achieve attentive tokens identification and inattentive tokens fusion. SPViT \cite{kong2022spvit} designs a dynamic attention-based multi-head token selector for informative tokens selection and uninformative tokens combination. Despite the above token reorganization methods proving to be effective, the applications are limited to single-modality models. In this work, we extend token pruning and merging mechanisms into cross-modal similarity vectors/matrices reorganization.

\section{Methodology}
\label{Methodology}

\begin{figure*}[!t]
	\centerline{\includegraphics[width=0.9\textwidth]{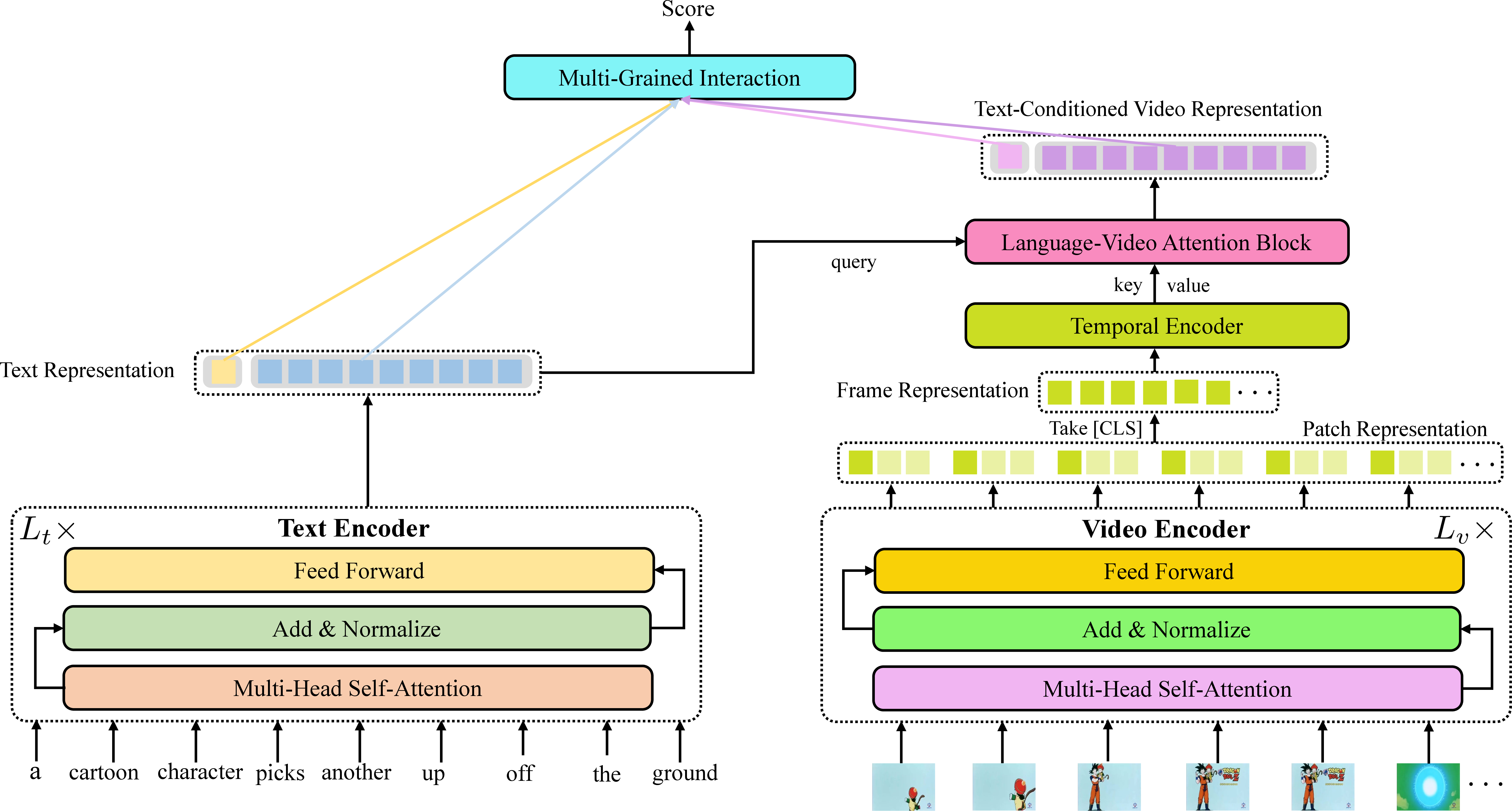}}
	\caption{The pipeline of TC-MGC. Given pair-wise text-video data, CLIP encoders simultaneously extract textual and visual representations, of which the extracted frame features are fed into the temporal encoder block for sequential modeling. Through language-video attention block, video representations with different granularity are regenerated in a text-guided manner. Finally, multi-grained interaction is implemented on the textual representations and text-conditioned visual representations to obtain the similarity score.}
	\label{fig: main_structure}
\end{figure*}

In this section, we incrementally present each component of our proposed TC-MGC, whose pipeline is depicted in Fig. \ref{fig: main_structure}. Specifically, we first introduce the multi-grained textual and visual representations extraction in Section \ref{sec: feature representation}. We then explain the mechanism of language-video attention block in Section \ref{sec: cross-modal language-video attention}, which aims to generate semantic-relevant video-level and frame-level visual representations. We also elaborate on the details of multi-grained contrastive learning in Section \ref{sec: multi-grained contrastive learning}, followed by the similarity reorganization (SR), interactive similarity aggregation (ISA) modules in Section \ref{sec: similarity reorganization} and \ref{sec: interactive similarity attention}. Finally, we describe the linear softmax aggregation (LSA) module and objective function in Section \ref{sec: linear softmax attention} and \ref{sec: objective function}, respectively.

\subsection{Feature Representation}
\label{sec: feature representation}
Given a set of $N$ sentences $\mathcal{T}=\{t_{i}\}_{i=1}^{N}$ and corresponding videos $\mathcal{V}=\{v_{i}\}_{i=1}^{N}$, our goal is to use CLIP-Pretrained text encoder $g$ and video encoder $h$ for textual and visual representation extraction, respectively. For a text $t_{i} \in \mathcal{T}$, the text encoder $g(t_{i})$ with $L_{t}(12)$ transformer layers  produces sentence-level textual feature $t_{i}^{'} \in \mathbb{R}^{d}$ and word-level textual feature $\hat{t}_{i} \in \mathbb{R}^{m \times d}$, where $m$ is the length of $t_{i}$ and $d$ is the size of feature dimension. For a video $v_{i} \in \mathcal{V}$, we first sample video frames with the sampling rate of 1 frame per second (FPS), followed by dividing them into disjoint patches prepended with a [CLS] token. Then, the video encoder $h(v_{i})$ with $L_{v}$(12) transformer layers is used to extract the patch representation. After that, combining [CLS] representation from each frame together and taking them as frame-level feature $\hat{v}_{i} \in \mathbb{R}^{n \times d}$, where n is the number of frames. 

Since $\hat{v}_{i} \in \mathbb{R}^{n \times d}$ are extracted from separate frames without considering the temporal information in videos, we employ a temporal encoder with position embedding $P$ to model the temporal relationship. Specifically, it is a 3-layer transformer encoder, which contains multi-head self-attention of 8 attention heads and feed-forward networks. The dimension of query, key, and value features is set as 512. The above operation is formulated as:
\begin{equation}
    \bar{v}_{i}=\textup{Transformer-Enc}(\hat{v}_{i}+P),
\end{equation}
where $\bar{v}_{i} \in \mathbb{R}^{n \times d}$ represents the final frame-level feature for the video $v_{i}$.

\begin{figure*}[!t]
	\centerline{\includegraphics[width=0.9\textwidth]{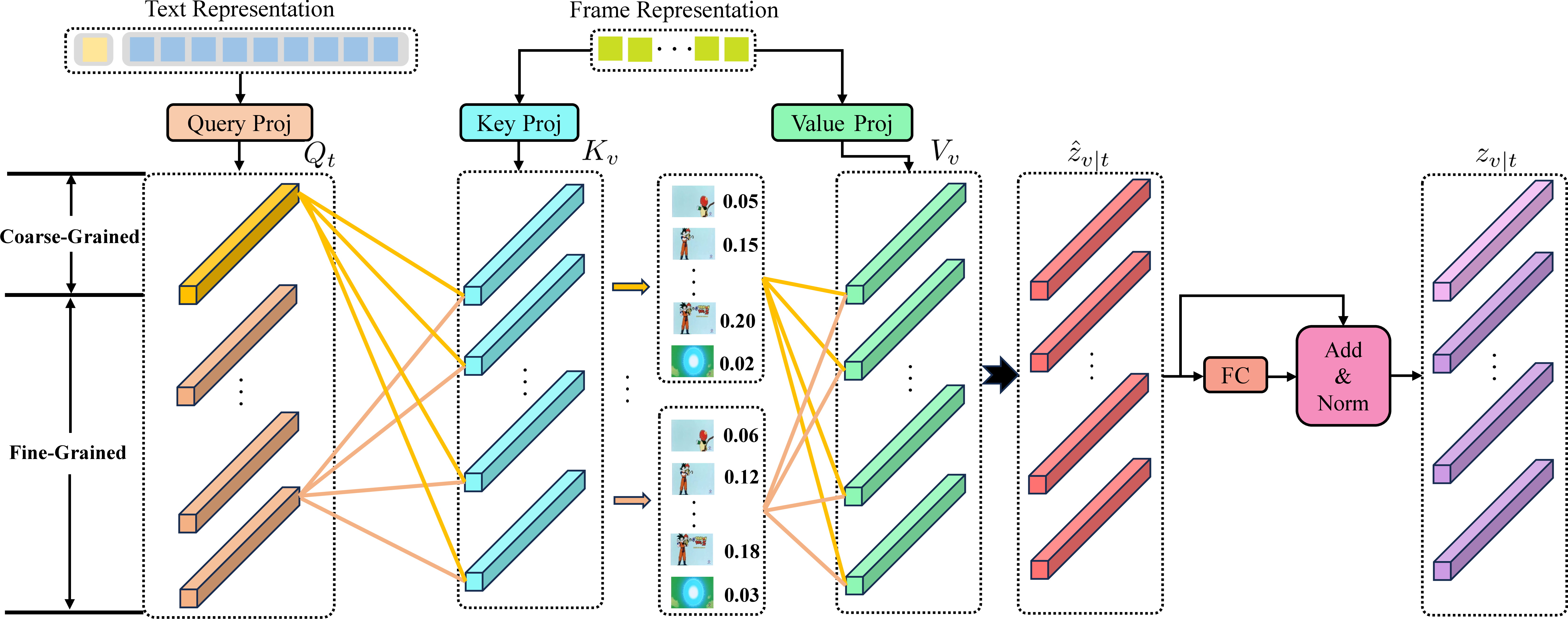}}
	\caption{The diagram of language-video attention block. For the textual representations, query projection is employed to obtain $Q_{t}$, including query-projected coarse-grained sentence embedding and fine-grained word embeddings. We similarly use key and value projections to obtain $K_{v}$ and $V_{v}$ from frame representations. After relevance weights calculation between textual and frame embeddings through scaled dot product, we aggregate frame embeddings with computed attention scores to obtain semantic-relevant video representations $\hat{z}_{v|t}$, which are passed through a fully connected layer and residual connection to obtain sentence-conditioned video representation and word-conditioned frame representations.}
	\label{fig: cross_attn_block}
\end{figure*}

\subsection{Language-Video Attention}
\label{sec: cross-modal language-video attention}
Previous multi-grained approaches \cite{ma2022x, wang2022disentangled} focus on sophisticated interaction mechanisms between specific textual and visual entities. However, as explained in Section \ref{Introduction}, the inherent cross-modal semantic discrepancy between multi-grained textual and visual representations may harm the retrieval performance. Therefore, we leverage the language-video attention block in Fig. \ref{fig: cross_attn_block} to explicitly generate sentence-conditioned video representation and word-conditioned frame representations, respectively. 

\noindent\textbf{Sentence-Conditioned Video Representation. }
Formally, we first use layer normalization (LN), followed by projection matrices to project a sentence embedding $t_{i}^{'} \in \mathbb{R}^{d}$ into a single query $Q_{t_{i}}^{c} \in \mathbb{R}^{1 \times d_{p}}$, as well as frame embeddings $\bar{v}_{i} \in \mathbb{R}^{n \times d}$ into key $K_{v_{i}} \in \mathbb{R}^{n \times d_{p}}$ and value $V_{v_{i}} \in \mathbb{R}^{n \times d_{p}}$ matrices, where $d_{p}$ denotes the size of projection dimension. The projections are formulated as:
\begin{align}
	Q_{t_{i}}^{c}&=\textup{LN}(t_{i}^{'})W_{Q}, \\
	K_{v_{i}}&=\textup{LN}(\bar{v}_{i})W_{K}, \label{eq: key_projection} \\
	V_{v_{i}}&=\textup{LN}(\bar{v}_{i})W_{V}, \label{eq: value_projection}
\end{align}
where $W_{Q} \in \mathbb{R}^{d \times d_{p}}$, $W_{K} \in \mathbb{R}^{d \times d_{p}}$ and $W_{V} \in \mathbb{R}^{d \times d_{p}}$ are projection matrices.

Then the scaled dot product is adapted to obtain relevancy weights from query-projected sentence embedding to key-projected frame embeddings, which is utilized to aggregate value-projected frame embeddings into video embedding:
\begin{equation}
	\textup{Attention}(Q_{t_{i}}^{c}, K_{v_{i}}, V_{v_{i}})=\textup{softmax}(\frac{Q_{t_{i}}^{c}K_{v_{i}}^{T}}{\sqrt{d_{p}}})V_{v_{i}}.
\end{equation}

Through cross-modal interactions between query-projected sentence embedding and key-projected frame embeddings, the sentence embedding assigns larger weights to frames embeddings with high relevance, as well as smaller weights to semantic-irrelevant frame embeddings. As such, the obtained weights only aggregate sub-regions of frames depending on the sentence content, which can greatly reduce the negative effect of superfluous information.

In order to align the sentence and video embeddings in a joint space, we project the aggregated video representation into $\mathbb{R}^{n \times d}$ by employing a weight matrices $W_{O} \in \mathbb{R}^{d_{p} \times d}$:
\begin{equation}
	\hat{z}_{v_{i}|t_{i}}^{c}=\textup{LN}(\textup{Attention}(Q_{t_{i}}^{c}, K_{v_{i}}, V_{v_{i}})W_{O}),
\end{equation}
where $\hat{z}_{v_{i}|t_{i}}^{c} \in \mathbb{R}^{d}$ is the aggregated video embedding conditioned on the sentence embedding.
Similar to residual connection in original transformer, we introduce a fully connected layer together with residual connection to enhance network capacity for more complex reasoning. The formulation is shown as follows:
\begin{equation}
	z_{v_{i}|t_{i}}^{c}=\textup{LN}(\textup{FC}(\hat{z}_{v_{i}|t_{i}}^{c})+\hat{z}_{v_{i}|t_{i}}^{c}),
\end{equation}
where FC refers to the fully connected layer, and $z_{v_{i}|t_{i}}^{c} \in \mathbb{R}^{d}$ is the final sentence-conditioned video embedding.

\begin{figure*}[!t]
	\centerline{\includegraphics[width=0.9\textwidth]{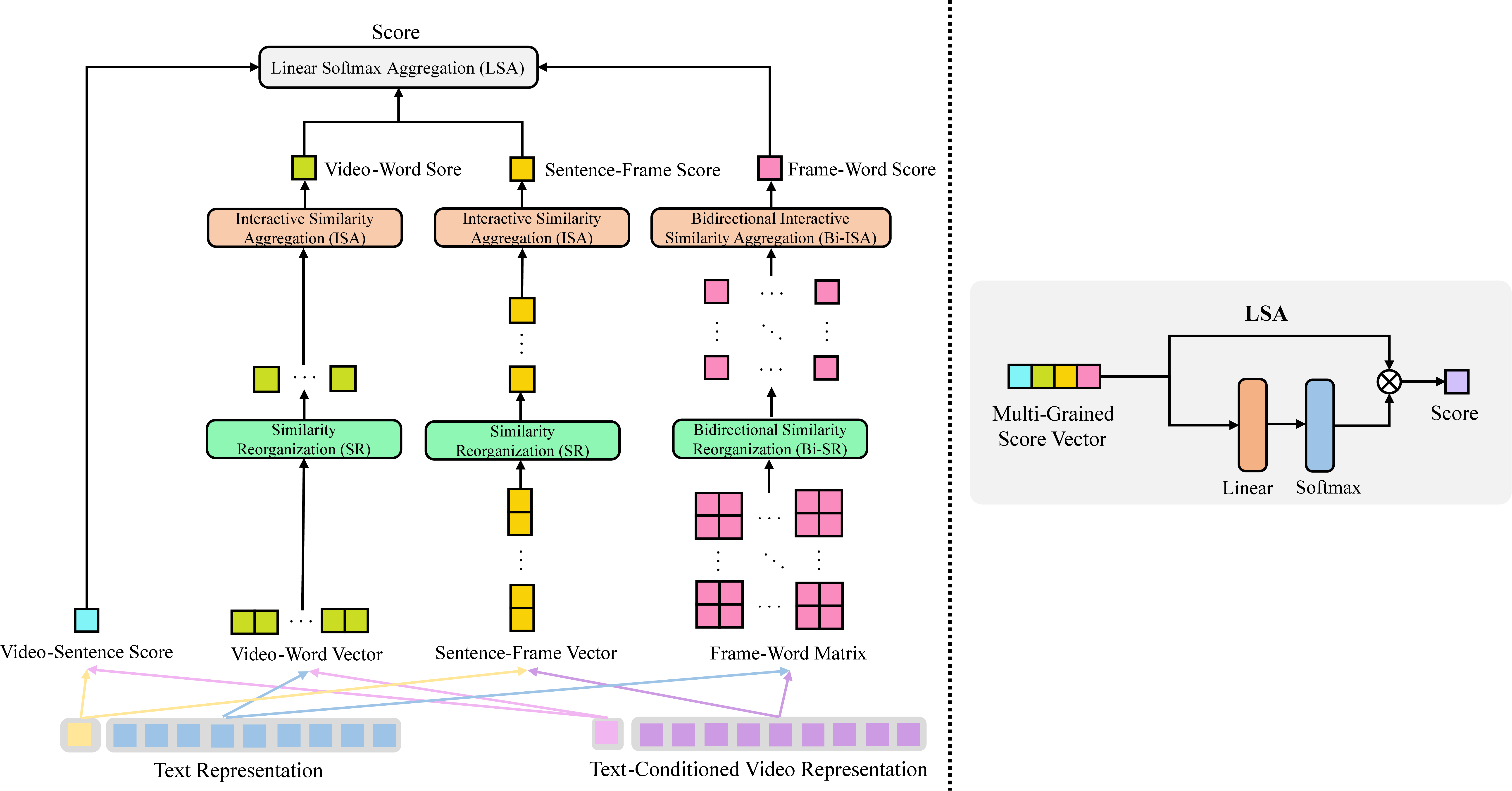}}
	\caption{Left: the illustration of multi-grained interaction mechanism. We first use matrix multiplication to obtain video-sentence similarity score, video-word and sentence-frame similarity vectors, frame-word similarity matrices respectively, followed by SR and Bi-SR modules to achieve similarity vectors and matrices reorganization. Next, we perform ISA and Bi-ISA modules on the reorganized similarity vectors and matrices to generate instance-level scores. Finally, we employ LSA module to achieve multi-grained scores aggregation. Right: the overview of LSA, which leverages the cascade of linear and softmax layers to calculate the weights of different instance-level scores. }
	\label{fig: multi-grained_contrast}
\end{figure*}

\noindent\textbf{Word-Conditioned Frame Representations. }
Similarly, due to the key and value projections presentation in Eq. \ref{eq: key_projection}-\ref{eq: value_projection}, we only formulate the query projection as:
\begin{align}
	Q_{t_{i}}^{f}&=\textup{LN}(\hat{t}_{i})W_{Q},
\end{align}
where $\hat{t}_{i} \in \mathbb{R}^{m \times d}$ is word-level features, and $Q_{t_{i}}^{f} \in \mathbb{R}^{m \times d}$ is the query-projected word embeddings. The aggregated frame embeddings are computed through:
\begin{equation}
	\textup{Attention}(Q_{t_{i}}^{f}, K_{v_{i}}, V_{v_{i}})=\textup{softmax}(\frac{Q_{t_{i}}^{f}K_{v_{i}}^{T}}{\sqrt{d_{p}}})V_{v_{i}}, 
\end{equation}
\begin{align}
	\hat{z}_{v_{i}|t_{i}}^{f}&=\textup{LN}(\textup{Attention}(Q_{t_{i}}^{f}, K_{v_{i}}, V_{v_{i}})W_{O}), \label{12} \\
	z_{v_{i}|t_{i}}^{f}&=\textup{LN}(\textup{FC}(\hat{z}_{v_{i}|t_{i}}^{f})+\hat{z}_{v_{i}|t_{i}}^{f}), \label{13}
\end{align}
where $\hat{z}_{v_{i}|t_{i}}^{f} \in \mathbb{R}^{m \times d}$ denotes the aggregated frame embeddings conditioned on the word embeddings, and $z_{v_{i}|t_{i}}^{f} \in \mathbb{R}^{m \times d}$ is the final word-conditioned frame representations.

\subsection{Multi-Grained Contrastive Learning}
\label{sec: multi-grained contrastive learning}
When the multi-grained textual and text-conditioned visual representations are obtained, we conduct multi-grained interactions in the left part of Fig. \ref{fig: multi-grained_contrast} to obtain video-sentence score, video-word and sentence-frame vectors, frame-word matrix, respectively. 

\noindent\textbf{Video-Sentence Contrast. }
For the given sentence representation $t_{i}^{'} \in \mathbb{R}^{d}$ and sentence-conditioned video representation $z_{v_{i}|t_{i}}^{c} \in \mathbb{R}^{d}$, we use the global dot product to compute the similarity score as follows:
\begin{equation}
	s_{v-s}=(z_{v_{i}|t_{i}}^{c})^{T}(t_{i}^{'}),
\end{equation}
where $s_{v-s} \in \mathbb{R}^{1}$ is the similarity score between sentence-conditioned video and sentence representations.

\noindent\textbf{Video-Word Contrast. }
Given the word representations $\hat{t}_{i} \in \mathbb{R}^{m \times d}$ and sentence-conditioned video representation $z_{v_{i}|t_{i}}^{c} \in \mathbb{R}^{d}$, we use matrix multiplication to calculate the similarity vector, which can be formulated as:
\begin{equation}
	\hat{s}_{v-w}=(z_{v_{i}|t_{i}}^{c})^{T}(\hat{t}_{i})^{T},
\end{equation}
where $\hat{s}_{v-w} \in \mathbb{R}^{1 \times m}$ refers to the similarity vector between sentence-conditioned video and word representations. \par

\noindent\textbf{Sentence-Frame Contrast. }
Similar to video-word contrast, the matrix multiplication is conducted on the sentence representation $t_{i}^{'} \in \mathbb{R}^{d}$ and word-conditioned frame representations $z_{v_{i}|t_{i}}^{f} \in \mathbb{R}^{m \times d}$ to obtain the similarity vector:
\begin{equation}
	\hat{s}_{s-f}=z_{v_{i}|t_{i}}^{f}t_{i}^{'},
\end{equation}
where $\hat{s}_{s-f} \in \mathbb{R}^{m \times 1}$ denotes the similarity vector between sentence and word-conditioned frame representations.

\noindent\textbf{Frame-Word Contrast. }
The fine-grained similarity matrix between word and word-conditioned frame representations can be computed using the matrix multiplication:
\begin{equation}
	\hat{s}_{f-w}=z_{v_{i}|t_{i}}^{f}(\hat{t}_{i})^{T},
\end{equation}
where $\hat{s}_{f-w} \in \mathbb{R}^{m \times m}$ is the similarity matrix, of which the individual element denotes the similarity score between each word-conditioned frame representation and word representation. 

\begin{figure*}[!t]
	\centerline{\includegraphics[width=0.9\textwidth]{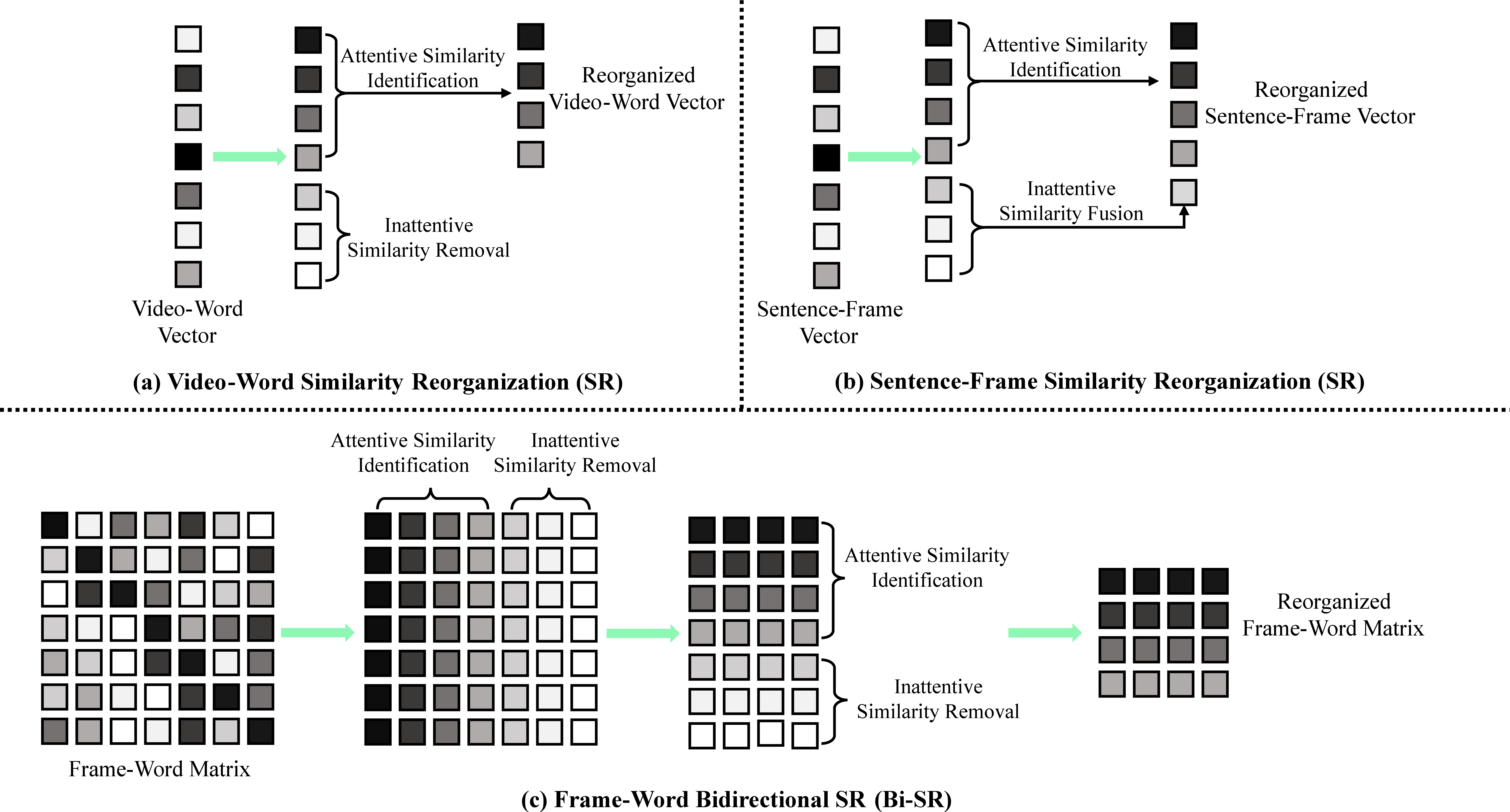}}
	\caption{Similarity Reorganization modules (SR). (a) We identify and rearrange the attentive similarities as the reorganized video-word vector. (b) We preserve the attentive similarities and fuse the inattentive similarities into one similarity, which are concatenated to generate the reorganized sentence-frame vector. (c) We extend SR module to bidirectional SR (Bi-SR) to obtain the reorganized frame-word matrix. }
	\label{fig: similarity reorganization}
\end{figure*}

\subsection{Similarity Reorganization (SR)}
\label{sec: similarity reorganization}
As mentioned in Section \ref{Introduction}, ISA module typically aggregates cross-modal similarity vectors/matrices by considering all feature interactions in the similarity weights calculation. However, we find the usage of all interactions may be unnecessary. For instance, during the video-word similarity vector aggregation, content words with specific semantics, such as nouns and verbs, are more likely than function words (\textit{e.g.,} articles and prepositions) to be aligned with visual content. As a result, the feature interactions involving function words may introduce noise and impede the precise cross-modal correspondence capture. The  sentence-frame similarity vector and frame-word similarity matrix aggregations have similar problems. To this end, we propose a simple yet effective similarity reorganization (SR) module. As shown in Fig. \ref{fig: similarity reorganization}, 
the idea of SR is to jointly consider attentive similarities ‌preservation and inattentive others removal/fusion while reorganizing the similarity vectors/matrices. The reorganized similarity vectors/matrices are used for ISA aggregation. In this way, the noisy feature interactions can be greatly eliminated, thus boosting the performance with a considerable margin.

\noindent\textbf{Video-Word Similarity Vector. } 
Given the similarity vector $\hat{s}_{v-w} \in \mathbb{R}^{1 \times m}$, we define the similarity keep rate as $r$, and select the words with $k$ largest ($k=m \times r$) similarities as the reorganized similarity vector $\bar{s}_{v-w} \in \mathbb{R}^{1 \times k}$.

\noindent\textbf{Sentence-Frame Similarity Vector. } 
Given the similarity vector $\hat{s}_{s-f} \in \mathbb{R}^{m \times 1}$, we perform top-$k$ largest elements identification to obtain the attentive similarity vector and the remaining $m-k$ similarities are fused into one similarity through the softmax-based weighted combination, followed by  the concatenation operation to generate our reorganized similarity vector $\bar{s}_{s-f} \in \mathbb{R}^{(k+1) \times 1}$.

\noindent\textbf{Frame-Word Similarity Matrix. } 
Given the similarity matrix $\hat{s}_{f-w} \in \mathbb{R}^{m \times m}$, we first perform the SR module on the word direction to generate first-reorganized similarity matrix $\check{s}_{f-w} \in \mathbb{R}^{m \times k}$, which can be formulated as:
\begin{equation}
	\check{s}_{f-w(i,*)}=[\hat{s}_{f-w(i,1)}, \cdots, \hat{s}_{f-w(i,j)}, \cdots, \hat{s}_{f-w(i,k)}], 
\end{equation}
where $i \in [1, m]$, $\hat{s}_{f-w(i,j)} \in \mathbb{R}^{1}$ represents the $j$-th largest element in the $i$-th row of the frame-word similarity matrix.

Next, we can obtain the final reorganized similarity matrix $\bar{s}_{f-w} \in \mathbb{R}^{k \times k}$ by further leveraging the SR module on the frame dimension for attentive similarities identification and inattentive similarities removal:
\begin{equation}
   	\bar{s}_{f-w(*,i)}=[\check{s}_{f-w(1,i)}, \cdots, \check{s}_{f-w(j,i)}, \cdots, \check{s}_{f-w(k,i)}]^{T}, 
\end{equation}
where $i \in [1, k]$, $\check{s}_{f-w(j,i)} \in \mathbb{R}^{1}$ represents the $j$-th largest element in the $i$-th column of the first-reorganized frame-word similarity matrix. 

\subsection{Interactive Similarity Aggregation (ISA)}
\label{sec: interactive similarity attention}
\begin{figure}[!t]
    \centerline{\includegraphics[width=0.5\textwidth]{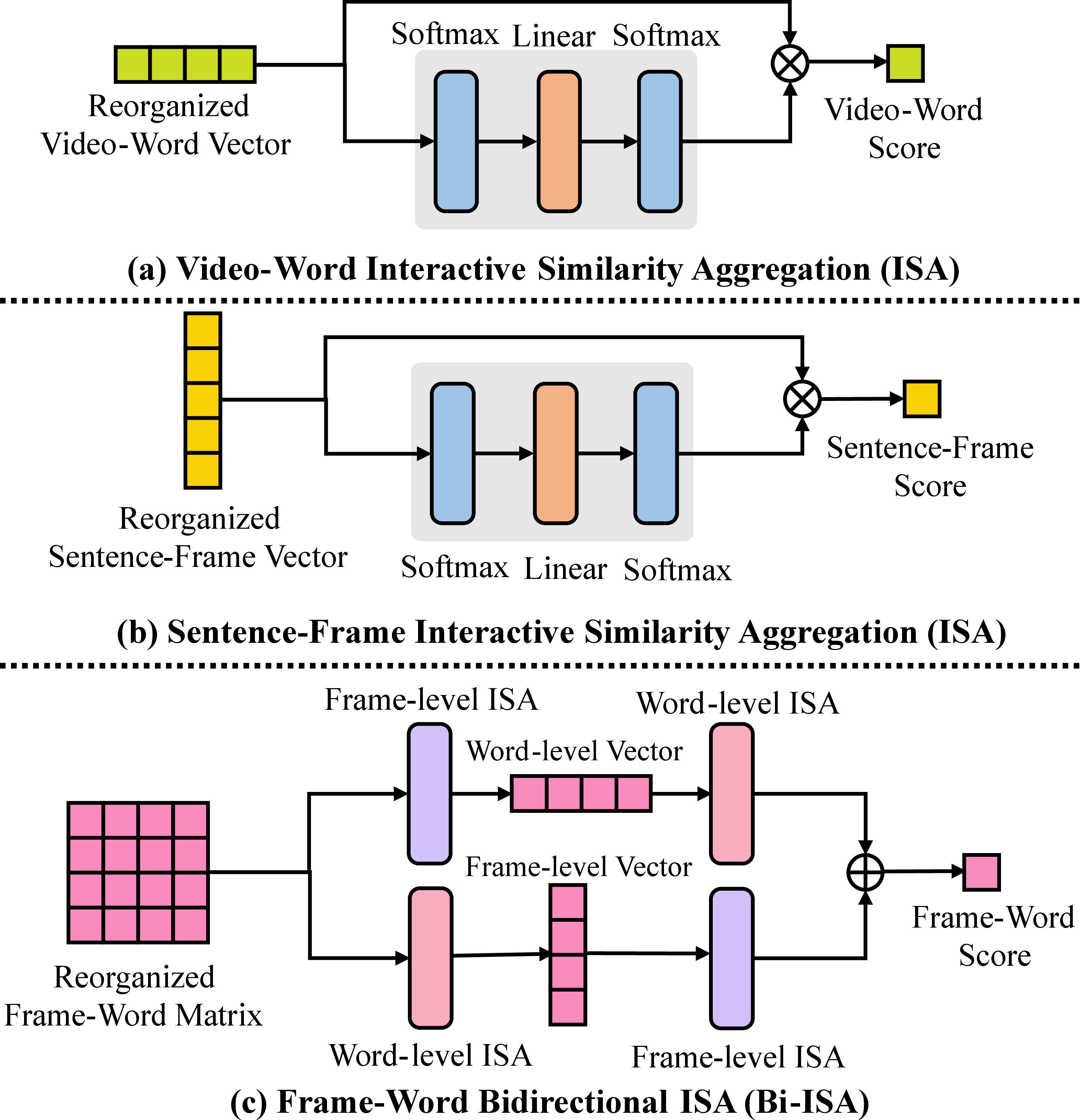}}
    \caption{Interactive Similarity Aggregation module (ISA). (a) We employ the ISA module to aggregate the reorganized video-word vector into the video-word score. (b) We employ the ISA module to obtain the sentence-frame score from the reorganized sentence-frame vector. (c) We leverage the bidirectional ISA (Bi-ISA) module to aggregate the reorganized frame-word matrix into the frame-word score. }
    \label{fig: interactive similarity attention}
\end{figure}

Unlike previous methods \cite{ma2022x, liu2022ts2} using a softmax linear in the similarity vectors/matrices aggregation for cross-modal relevance capture, we directly employ ISA and Bi-ISA modules from \cite{wang2023unified} shown in Fig. \ref{fig: interactive similarity attention} to generate instance-level scores from the reorganized similarity vectors/matrices. The core of ISA module is to take the interaction between different features into consideration via a linear layer while computing the weights of different similarities.

\noindent\textbf{Video-Word Similarity Vector. }
For the reorganized video-word similarity vector $\bar{s}_{v-w} \in \mathbb{R}^{1 \times k}$, we first use a softmax layer to compute the cross-modal relevance, followed by a linear layer to encourage similarity interactions. Then another softmax layer is utilized to obtain the final weight of similarity vector. The whole process can be formulated as:
\begin{align}
    \textup{w}_{v-w}&=\textup{softmax}(f_{l_{1}}(\textup{softmax}(\bar{s}_{v-w}))), \\
    s_{v-w}&=\sum_{i=1}^{k} \textup{w}_{v-w} \bar{s}_{v-w},
\end{align} 
where $f_{l_{1}}(\cdot)$ represents a linear layer initialized with the identity matrix ($\mathbb{R}^{k \times k}$), $\textup{w}_{v-w} \in \mathbb{R}^{1 \times k}$ is the weight of the similarity vector, $s_{v-w} \in \mathbb{R}^{1}$ is the instance-level similarity between video and word representations.

\noindent\textbf{Sentence-Frame Similarity Vector. } 
Given the reorganized sentence-frame similarity vector $\bar{s}_{s-f} \in \mathbb{R}^{(k+1) \times 1}$, the instance-level similarity score between sentence and frame representations can be computed as:
\begin{align}
    \textup{w}_{s-f}&=\textup{softmax}(f_{l_{2}}(\textup{softmax}(\bar{s}_{s-f}))), \\
    s_{s-f}&=\sum_{i=1}^{m} \textup{w}_{s-f} \bar{s}_{s-f},
\end{align} 
where $f_{l_{2}}(\cdot)$ denotes a linear layer initialized with the identity matrix ($\mathbb{R}^{(k+1) \times (k+1)}$), $\textup{w}_{s-f} \in \mathbb{R}^{(k+1) \times 1}$ is the weight of the similarity vector, $s_{s-f} \in \mathbb{R}^{1}$ is the sentence-frame similarity score. 

\noindent\textbf{Frame-Word Similarity Matrix. } 
Since the reorganized frame-word similarity matrix $\bar{s}_{f-w} \in \mathbb{R}^{k \times k}$ contains similarity scores between $k$ frames and $k$ words, we employ a Bi-ISA module to perform ISA operation on frame and word directions respectively. For the frame direction, we first adopt a frame-level ISA module to obtain a word-level similarity vector, which can be represented as:
\begin{align}
	\textup{w}_{f_{1}}&=\textup{softmax}(f_{l_{3}}(\textup{softmax}(\bar{s}_{f-w}))), \\
	\tilde{s}_{w}&=\sum_{i=1}^{k} \textup{w}_{f_{1}(i,*)} \bar{s}_{f-w(i,*)},
\end{align}
where $f_{l_{3}}(\cdot)$ represents a linear layer initialized with the identity matrix ($\mathbb{R}^{k \times k}$), $\textup{w}_{f_{1}} \in \mathbb{R}^{k \times k}$ is the weight of frame-level similarity vector, $\tilde{s}_{w} \in \mathbb{R}^{1 \times k}$ is the word-level similarity vector.

Next, the word-level ISA is applied on the word-level similarity vector to obtain the frame-then-word similarity score:
\begin{align}
	\textup{w}_{w_{1}}&=\textup{softmax}(f_{l_{4}}(\textup{softmax}(\tilde{s}_{w}))), \\
	\tilde{s}_{f-w}&=\sum_{i=1}^{k} \textup{w}_{w_{1}(1,i)} \tilde{s}_{w(1,i)},
\end{align}
where $f_{l_{4}}(\cdot)$ denotes a linear layer initialized with the identity matrix ($\mathbb{R}^{k \times k}$), $\textup{w}_{w_{1}} \in \mathbb{R}^{1 \times k}$ is the weight of word-level similarity score, $\tilde{s}_{f-w} \in \mathbb{R}^{1}$ is the frame-then-word similarity score.

Similarly, we can compute the word-then-frame similarity score $\tilde{s}_{w-f} \in \mathbb{R}^{1}$ on the 
word direction:
\begin{align}
	\textup{w}_{w_{2}}&=\textup{softmax}(f_{l_{5}}(\textup{softmax}(\bar{s}_{f-w}))), \\
	\tilde{s}_{f}&=\sum_{i=1}^{k} \textup{w}_{w_{2}(*,i)} \bar{s}_{f-w(*,i)}, \\
	\textup{w}_{f_{2}}&=\textup{softmax}(f_{l_{6}}(\textup{softmax}(\tilde{s}_{f}))), \\
	\tilde{s}_{w-f}&=\sum_{i=1}^{k} \textup{w}_{f_{2}(i,1)} \tilde{s}_{f(i,1)},
\end{align}
where $f_{l_{5}}(\cdot)$ and $f_{l_{6}}(\cdot)$ represent two linear layers initialized with the identity matrices ($\mathbb{R}^{k \times k}$), $\textup{w}_{w_{2}} \in \mathbb{R}^{k \times k}$ is the weight of word-level similarity vector, $\tilde{s}_{f} \in \mathbb{R}^{k \times 1}$ is the frame-level similarity vector, $\textup{w}_{f_{2}} \in \mathbb{R}^{k \times 1}$ is the weight of frame-level similarity score, $\tilde{s}_{w-f} \in \mathbb{R}^{1}$ is the word-then-frame similarity score.

Finally, the instance-level similarity score $s_{f-w} \in \mathbb{R}^{1}$ is represented as the average value of frame-then-word and word-then-frame similarity scores:
\begin{equation}
	s_{f-w}=\frac{\tilde{s}_{f-w}+\tilde{s}_{w-f}}{2}.
\end{equation}

\subsection{Linear Softmax Aggregation (LSA)}
\label{sec: linear softmax attention}
As for the multi-grained similarities aggregation, existing methods adopt average-based or softmax-based weighted combinations of the similarity vector to generate instance-level similarity score. However, both weighted combinations fail to capture the information between different similarities. The computed weights only focus on the multi-grained relevance and ignore the interaction between different similarities. To address this, we transfer the core of ISA to multi-grained similarities aggregation and present a Linear Softmax Aggregation (LSA) module shown in the right part of Fig. \ref{fig: multi-grained_contrast}. 

The idea of LSA module is to jointly consider the multi-grained relevance and the interaction between different similarities while computing the weights of different granularity. The LSA module applies a linear layer to encourage interactions between different similarities, followed by a softmax layer to obtain the final weights of different similarities. Specifically, given the obtained multi-grained similarity vector $\tilde{s}(t_{i}, v_{i})=\{s_{v-s}, s_{v-w}, s_{s-f}, s_{f-w}\}$ ($\mathbb{R}^{1 \times 4}$), the LSA module can be formulated as:
\begin{align}
	\textup{w}_{s}&=\textup{softmax}(f_{l_{7}}(\tilde{s}(t_{i}, v_{i}))), \\
	s(t_{i}, v_{i})&= \sum_{k=1}^{4} (\textup{w}_{s})_{k} \tilde{s}(t_{i}, v_{i})_{k},
\end{align}
where $f_{l_{7}}(\cdot)$ denotes a linear layer initialized with the identity matrices ($\mathbb{R}^{4 \times 4}$), $\textup{w}_{s} \in \mathbb{R}^{1 \times 4}$ is the weight of different similarity scores, and $s(t_{i}, v_{i}) \in \mathbb{R}^{1}$ is the final similarity score between text and video.

\subsection{Objective Function}
\label{sec: objective function}
Given a batch of $B$ text-video pairs, the model will generate a $B \times B$ similarity matrix. By considering $B$ matching text-video pairs as positives and other $B^{2}-B$ pair-wise text-video combinations in a batch as negatives, we adopt the symmetric InfoNCE loss to optimize the model's parameters:
\begin{align}
	\mathcal{L}_{t2v}&=-\frac{1}{B} \sum_{i=1}^{B} \textup{log} \frac{e^{s(t_{i}, v_{i}) \cdot \lambda}}{\sum_{j=1}^{B}e^{s(t_{i}, v_{j})\cdot \lambda}}, \\
	\mathcal{L}_{v2t}&=-\frac{1}{B} \sum_{i=1}^{B} \textup{log} \frac{e^{s(t_{i}, v_{i}) \cdot \lambda}}{\sum_{j=1}^{B}e^{s(t_{j}, v_{i}) \cdot \lambda}}, \\
	\mathcal{L}_{\text{InfoNCE}}&=\mathcal{L}_{t2v}+\mathcal{L}_{v2t},
\end{align}
where the loss $\mathcal{L}_{\text{InfoNCE}}$ is the sum of text-to-video loss $\mathcal{L}_{t2v}$ and video-to-text loss $\mathcal{L}_{v2t}$, and $\lambda$ is a scaling temperature parameter of softmax.

Moreover, inspired by the channel decorrelation regularization approach in DRL \cite{wang2022disentangled}, we conduct similarity decorrelation regularization and employ the variance minimization computation var($\cdot$) among matching text-video pairs for over- and under-representation issues alleviation:
\begin{equation}
	\mathcal{L}_{SDR}=\frac{1}{B} \sum_{i=1}^{B} \text{var}(\tilde{s}(t_{i}, v_{i})).
\end{equation} 

Hence, the total training loss $\mathcal{L}_{all}$ is defined as:
\begin{equation}
	\mathcal{L}_{all}=\mathcal{L}_{\text{InfoNCE}}+\alpha \mathcal{L}_{SDR}, \label{eq: total_loss}
\end{equation}
where $\alpha$ is the weighting parameter.

\section{Experiments}
\label{Experiments}

\subsection{Experimental Settings}
\textbf{Datasets.} We evaluate TC-MGC on three popular text-video retrieval benchmarks: 
\textbf{(a) MSR-VTT} \cite{xu2016msr} contains 10,000 videos with a length range from 10 to 32 seconds, and each video is paired with 20 human-labeled captions. We adopt two training data splits, ``Training-7K'' and ``Training-9K'', which follow the data splits from HowTo100M \cite{miech2019howto100m} and MMT \cite{gabeur2020multi}, respectively. The test data in both splits is ``test 1k-A'', which contains 1,000 text-video pairs following JSFusion \cite{yu2018joint}. Unless otherwise specified, we use the ``Training-9K'' split by default. 
\textbf{(b) DiDeMo} \cite{anne2017localizing} consists of 10,000 videos and 40,000 captions. All textual descriptions for a given video are concatenated into one single query during video-paragraph retrieval. 
\textbf{(c) VATEX} \cite{wang2019vatex} contains 34,991 videos with multilingual annotations per video. Following HGR's \cite{chen2020fine} data splits, there are 25,991 videos used for training, 1,500 videos for validation, and 1,500 videos for testing. 

\textbf{Evaluation Protocols.} We adopt standard retrieval metrics: recall at rank K (R@K, higher is better), median rank (MdR, lower is better), and mean rank (MnR, lower is better) to evaluate the retrieval performance. R@K is defined as the percentage of correct samples among the top-K retrieved points to the query sample. R@1, R@5, and R@10 are reported. MdR computes the median rank of groundtruth in the retrieved ranking list and MnR measures the mean rank of groundtruth in the retrieved ranking list. Meanwhile, we take the sum of all R@K results in T2V and V2T tasks as RSum. Additionally, to show the overall retrieval performance, we also sum together the two RSum as SumR, which is the main concern in our experiments. Note that for RSum and SumR, the higher score means the better(indicated as $\uparrow$).

\begin{table*}
	\caption{Retrieval results on MSR-VTT. Speed is the inference time per video during evaluation on a NVIDIA GeForce RTX 3090 GPU. $\dag$ denotes that results are obtained by our re-training. Bold denotes the best performance. ``–”: result is unavailable. ``NeurComp'' refers to Neurocomputing.}
	\label{table: MSR-VTT-7K Results}
	\centering
	\normalsize
	\begin{threeparttable}
			\resizebox{\textwidth}{!}{
				\begin{tabular}{l|l|c|cccccc|cccccc|c}
					\hline
					\multirow{2}*{Methods} & \multirow{2}*{Venue} & \multirow{2}*{\makecell[c]{Speed\\(ms)}} & \multicolumn{6}{c|}{Text $\rightarrow$ Video} &  \multicolumn{6}{c|}{Video $\rightarrow$ Text} & \multirow{2}*{SumR$\uparrow$}\\ 
					\cline{4-9} \cline{10-15} 
					&&& R@1$\uparrow$  & R@5$\uparrow$  & R@10$\uparrow$ & MdR$\downarrow$ &  MnR$\downarrow$ & RSum$\uparrow$
					& R@1$\uparrow$  & R@5$\uparrow$  & R@10$\uparrow$ & MdR$\downarrow$ &  MnR$\downarrow$ & RSum$\uparrow$\\
					\hline
					HowTo100M \cite{miech2019howto100m} & ICCV'19 & - & 14.9 & 40.2 & 52.8 & 9.0 & - & 107.9 & 16.8 & 41.7 & 55.1 & 8.0 & - & 113.6 & 221.5 \\
					ActBERT \cite{zhu2020actbert} & CVPR'20 & - & 8.6 & 23.4 & 33.1 & 36.0 & - & 65.1 & - & - & - & - & - & - & - \\
					HERO \cite{li2020hero} & ArXiv'20 & - & 16.8 & 43.4 & 57.7 & - & - & 117.9 & - & - & - & - & - & - & - \\
					NoiseE \cite{amrani2021noise} & ArXiv'20 & - & 17.4 & 41.6 & 53.6 & 8.0 & - & 112.6 & - & - & - & - & - & - & - \\
					ClipBERT \cite{lei2021less} & CVPR'21 & - & 22.0 & 46.8 & 59.9 & 6.0 & - & 128.7 & - & - & - & - & - & - & - \\
					\hline
					\textbf{\textit{CLIP-ViT-B/32}} \\
					\hdashline[1pt/5pt]
					$\textup{CLIP4Clip-meanP}^{\dag}$ \cite{luo2022clip4clip} & NeurComp'22 & 14.9 & 42.9 & 67.9 & 79.2 & \textbf{2.0} & 16.6 & 190.0 & 42.2 & 69.7 & 80.1 & \textbf{2.0} & 12.7 & 192.0 & 382.0 \\
					$\textup{CLIP4Clip-seqLSTM}^{\dag}$ \cite{luo2022clip4clip} & NeurComp'22 & 54.7 & 42.3 & 67.3 & 79.0 & \textbf{2.0} & 16.4 & 188.6 & 42.5 & 69.5 & 79.6 & \textbf{2.0} & 12.6 & 191.6 & 380.2 \\
					$\textup{CLIP4Clip-seqTransf}^{\dag}$ \cite{luo2022clip4clip} & NeurComp'22 & 215.0 & 41.9 & 69.1 & 79.0 & \textbf{2.0} & 17.1 & 190.0 & 40.5 & 68.7 & 77.5 & \textbf{2.0} & 12.7 & 186.7 & 376.7 \\
					$\textup{CLIP4Clip-tightTransf}^{\dag}$ \cite{luo2022clip4clip} & NeurComp'22 & 587.3 & 38.3 & 68.7 & 79.1 & \textbf{2.0} & 16.4 & 186.1 & 37.3 & 67.7 & 78.5 & \textbf{2.0} & 14.3 & 183.5 & 369.6 \\
					$\textup{X-Pool}^{\dag}$ \cite{gorti2022x} & CVPR'22 & 330.2 & 42.4 & 69.6 & 80.6 & \textbf{2.0} & 15.0 & 192.6 & 43.7 & 70.7 & \textbf{82.8} & \textbf{2.0} & 9.6 & 197.2 & 389.8 \\
					$\textup{CenterCLIP}$ \cite{zhao2022centerclip} & SIGIR’22 & 42.5 & 43.7 & 71.3 & 80.2 & \textbf{2.0} & 16.2 & 195.2 & 43.2 & 71.0 & 80.4 & \textbf{2.0} & 12.3 & 194.6 & 389.8 \\
					$\textup{DRL}^{\dag}$ \cite{wang2022disentangled} & ArXiv'22 & 90.3 & 44.7 & 70.8 & 79.2 & \textbf{2.0} & 15.1 & 194.7 & 43.7 & 69.0 & 80.2 & \textbf{2.0} & 10.2 & 192.9 & 387.6 \\
					$\textup{UCoFiA}^{\dag}$ \cite{wang2023unified} & ICCV'23 & 136.1 & 44.0 & 70.8 & 80.0 & \textbf{2.0} & 16.3 & 194.8 & 42.4 & 70.2 &  79.5 & \textbf{2.0} & 11.7 & 192.1 & 386.9 \\
                    $\textup{X-CLIP}^{\dag}$ \cite{ma2022x} (base) & ArXiv'22 & 75.6 & 44.9 & 69.6 & 80.1 & \textbf{2.0} & 16.5 & 194.6 & 43.3 & 69.3 & 78.2 & \textbf{2.0} & 11.5 & 190.8 & 385.4 \\
					\rowcolor{gray!25}
					TC-MGC (ours) & & 672.7 & \textbf{45.5} & \textbf{71.6} & \textbf{81.4} & \textbf{2.0} & \textbf{14.2} & \textbf{198.5} & \textbf{44.2} & \textbf{72.2} & 81.4 & \textbf{2.0} & \textbf{9.1} & \textbf{197.8} & \textbf{396.3} \\
					\hline
					\textbf{\textit{CLIP-ViT-B/16}} \\
					\hdashline[1pt/5pt]
					$\textup{CLIP4Clip-meanP}^{\dag}$ \cite{luo2022clip4clip} & NeurComp'22 & 16.2 & 45.9 & 72.0 & 81.6 & \textbf{2.0} & 14.5 & 199.5 & 45.7 & 72.2 & 81.3 & \textbf{2.0} & 11.7 & 199.2 & 398.7 \\
					$\textup{CLIP4Clip-seqLSTM}^{\dag}$ \cite{luo2022clip4clip} & NeurComp'22 & 54.4 & 45.5 & 72.6 & 81.9 & \textbf{2.0} & 14.5 & 200.0 & 45.9 & 72.4 & 81.6 & \textbf{2.0} & 11.5 & 199.9 & 399.9 \\
					$\textup{CLIP4Clip-seqTransf}^{\dag}$ \cite{luo2022clip4clip} & NeurComp'22 & 203.3 & 46.2 & 71.3 & 81.0 & \textbf{2.0} & 15.0 & 198.5 & 43.9 & 71.0 & 80.5 & \textbf{2.0} & 12.0 & 195.4 & 393.9 \\
					$\textup{CLIP4Clip-tightTransf}^{\dag}$ \cite{luo2022clip4clip} & NeurComp'22 & 589.2 & 40.6 & 71.6 & 80.4 & \textbf{2.0} & 13.9 & 192.6 & 40.8 & 69.9 & 79.9 & \textbf{2.0} & 10.9 & 190.6 & 393.2 \\
					$\textup{X-Pool}^{\dag}$ \cite{gorti2022x} & CVPR'22 & 345.3 & 44.4 & 71.5 & 81.6 & \textbf{2.0} & \textbf{12.8} & 197.5 & 45.5 & 74.2 & 84.1 & \textbf{2.0} & \textbf{8.6} & 203.8 & 401.3 \\
					CenterCLIP \cite{zhao2022centerclip} & SIGIR’22 & 84.2 & 47.5 & 74.4 & 82.5 & \textbf{2.0} & 13.7 & 204.4 & 46.9 & 73.4 & 83.2 & \textbf{2.0} & 9.3 & 203.5 & 407.9 \\
					$\textup{DRL}^{\dag}$ \cite{wang2022disentangled} & ArXiv'22 & 43.6 & 47.0 & 73.5 & 82.1 & \textbf{2.0} & 13.1 & 202.6 & 46.1 & 74.8 & 82.8 & \textbf{2.0} & 9.2 & 203.7 & 406.3 \\
					$\textup{UCoFiA}^{\dag}$ \cite{wang2023unified} & ICCV'23 & 142.0 & \textbf{48.8} & 74.1 & 81.3 & \textbf{2.0} & 13.1 & 204.2 & 46.9 & 74.2 & 82.9 & \textbf{2.0} & 9.1 & 204.0 & 408.2 \\
                    $\textup{X-CLIP}^{\dag}$ \cite{ma2022x} (base) & ArXiv'22 & 98.0 & 48.2 & \textbf{75.1} & 82.8 & \textbf{2.0} & 14.0 & 206.1 & \textbf{47.1} & 74.8 & 83.1 & \textbf{2.0} & 9.5 & 205.0 & 411.1 \\
					\rowcolor{gray!25}
					TC-MGC (ours) & & 719.1 & 48.7 & 74.6 & \textbf{83.2} & \textbf{2.0} & 13.0 & \textbf{206.5} & 45.7 & \textbf{75.1} & \textbf{84.3} & \textbf{2.0} & 8.9 & \textbf{205.1} & \textbf{411.6} \\
					\hline
				\end{tabular}
			}
			\begin{tablenotes}    
				\footnotesize              
				\item \qquad \qquad \qquad \qquad \qquad \qquad \qquad \qquad \qquad \qquad \qquad \qquad (a) Training on Training-7K       
			\end{tablenotes}           
		\end{threeparttable}  
	\end{table*}
	
	\begin{table*}
		\label{table: MSR-VTT-9K Results}
		\centering
		\normalsize
		\begin{threeparttable}
				\resizebox{\textwidth}{!}{
					\begin{tabular}{l|l|c|cccccc|cccccc|c}
						\hline
						\multirow{2}*{Methods} & \multirow{2}*{Venue} & \multirow{2}*{\makecell[c]{Speed\\(ms)}} & \multicolumn{6}{c|}{Text $\rightarrow$ Video} &  \multicolumn{6}{c|}{Video $\rightarrow$ Text} & \multirow{2}*{SumR$\uparrow$}\\ 
						\cline{4-9} \cline{10-15} 
						&&& R@1$\uparrow$  & R@5$\uparrow$  & R@10$\uparrow$ & MdR$\downarrow$ &  MnR$\downarrow$ & RSum$\uparrow$
						& R@1$\uparrow$  & R@5$\uparrow$  & R@10$\uparrow$ & MdR$\downarrow$ &  MnR$\downarrow$ & RSum$\uparrow$\\
						\hline
						CE \cite{liu2019use} & ArXiv'19 & - & 20.9 & 48.8 & 62.4 & 6.0 & 28.2 & 132.1 & 20.6 & 50.3 & 64.0 & 5.3 & - & 134.9 & 267.0 \\
						MMT \cite{gabeur2020multi} & ECCV’20 & - & 26.6 & 57.1 & 69.6 & 4.0 & 24.0 & 153.3 & 27.0 & 57.5 & 69.7 & 3.7 & - & 154.2 & 307.5 \\
						MDMMT \cite{dzabraev2021mdmmt} & CVPR’21 & - & 38.9 & 69.0 & 79.7 & 2.0 & 16.5 & 187.6 & - & - & - & - & - & - & - \\
						Frozen \cite{bain2021frozen} & ICCV’21 & - & 31.0 & 59.5 & 70.5 & 3.0 & - & 161.0 & - & - & - & - & - & - & - \\
						HiT \cite{liu2021hit} & ICCV’21 & - & 30.7 & 60.9 & 73.2 & 2.6 & - & 164.8 & 32.1 & 62.7 & 74.1 & 3.0 & - & 168.9 & 333.7 \\
						TMVM  \cite{lin2022text} & NeurIPS’22 & - & 36.2 & 64.2 & 75.7 & 3.0 & - & 176.1 & 34.8 & 63.8 & 73.7 & 3.0 & - & 172.3 & 348.4 \\
						\hline
						\textbf{\textit{CLIP-ViT-B/32}} \\
						\hdashline[1pt/5pt]
						CLIP4Clip-meanP \cite{luo2022clip4clip} & NeurComp'22 & 14.4 & 43.1 & 70.4 & 80.8 & \textbf{2.0} & 16.2 & 194.3 & 43.1 & 70.5 & 81.2 & \textbf{2.0} & 12.4 & 194.8 & 389.1 \\
						CLIP4Clip-seqLSTM \cite{luo2022clip4clip} & NeurComp'22 & 52.5 & 42.5 & 70.8 & 80.7 & \textbf{2.0} & 16.7 & 194.0 & 42.8 & 71.0 & 80.4 & \textbf{2.0} & 12.3 & 194.2 & 388.2 \\
						CLIP4Clip-seqTransf \cite{luo2022clip4clip} & NeurComp'22 & 199.6 & 44.5 & 71.4 & 81.6 & \textbf{2.0} & 15.3 & 197.5 & 42.7 & 70.9 & 80.6 & \textbf{2.0} & 11.6 & 194.2 & 391.7 \\
						CLIP4Clip-tightTransf \cite{luo2022clip4clip} & NeurComp'22 & 582.7 & 40.2 & 71.5 & 80.5 & \textbf{2.0} & 13.4 & 192.2 & 40.6 & 69.5 & 79.5 & \textbf{2.0} & 13.6 & 189.6 & 381.8 \\
						CenterCLIP \cite{zhao2022centerclip} & SIGIR’22 & 41.4 & 44.2 & 71.6 & 82.1 & \textbf{2.0} & 15.1 & 197.9 & 42.8 & 71.7 & 82.2 & \textbf{2.0} & 10.9 & 196.7 & 394.6 \\
						X-Pool \cite{gorti2022x} & CVPR'22 & 346.0 & 46.9 & 72.8 & 82.2 & \textbf{2.0} & 14.3 & 201.9 &  44.4 & 73.3 & \textbf{84.0} & \textbf{2.0} & 9.0 & 201.7 & 403.6 \\
						$\textup{DRL}^{\dag}$ \cite{wang2022disentangled} & ArXiv'22 & 95.4 & \textbf{47.5} & 73.8 & 83.6 & \textbf{2.0} & 13.3 & 204.9 & 46.3 & 72.7 & 82.5 & \textbf{2.0} & 9.5 & 201.5 & 406.4 \\
						$\textup{UCoFiA}^{\dag}$ \cite{wang2023unified} & ICCV'23 & 134.3 & 46.8 & 72.6 & 82.4 & \textbf{2.0} & 13.9 & 201.8 & 45.3 & 72.7 & 82.2 & \textbf{2.0} & 9.8 & 200.2 & 402.0 \\
                        X-CLIP \cite{ma2022x} (base) & ArXiv'22 & 75.7 & 46.1 & 73.0 & 83.1 & \textbf{2.0} & 13.2 & 202.2 & \textbf{46.8} & 73.3 & \textbf{84.0} & \textbf{2.0} & 9.1 & \textbf{204.1} & 406.3 \\
						\rowcolor{gray!25}
						TC-MGC (ours) & & 673.1 & 47.4 & \textbf{74.8} & \textbf{84.2} & \textbf{2.0} & \textbf{12.8} & \textbf{206.4} & 45.9 & \textbf{74.5} & 83.3 & \textbf{2.0} & \textbf{8.6} & 203.7 & \textbf{410.1} \\
						\hline
						\textbf{\textit{CLIP-ViT-B/16}} \\
						\hdashline[1pt/5pt]
						$\textup{CLIP4Clip-meanP}^{\dag}$ \cite{luo2022clip4clip} & NeurComp'22 & 16.3 & 45.3 & 73.3 & 83.0 & \textbf{2.0} & 13.0 & 201.6 & 44.8 & 73.2 & 82.2 & \textbf{2.0} & 9.6 & 200.2 & 401.8 \\
						$\textup{CLIP4Clip-seqLSTM}^{\dag}$ \cite{luo2022clip4clip} & NeurComp'22 & 54.7 & 44.3 & 72.0 & 82.2 & \textbf{2.0} & 13.7 & 198.5 & 44.3 & 73.4 & 82.4 & \textbf{2.0} & 10.3 & 200.1 & 398.6 \\
						$\textup{CLIP4Clip-seqTransf}^{\dag}$ \cite{luo2022clip4clip} & NeurComp'22 & 204.5 & 46.4 & 72.1 & 82.0 & \textbf{2.0} & 14.7 & 200.5 & 45.4 & 73.4 & 82.4 & \textbf{2.0} & 10.7 & 201.2 & 401.7 \\
						$\textup{CLIP4Clip-tightTransf}^{\dag}$ \cite{luo2022clip4clip} & NeurComp'22 & 585.7 & 42.9 & 71.7 & 81.5 & \textbf{2.0} & 13.3 & 196.1 & 41.9 & 71.0 & 80.7 & \textbf{2.0} & 10.1 & 193.6 & 389.7 \\
						CenterCLIP \cite{zhao2022centerclip} & SIGIR’22 & 82.0 & 48.4 & 73.8 & 82.0 & \textbf{2.0} & 13.8 & 204.2 & 47.7 & 75.0 & 83.3 & \textbf{2.0} & 10.2 & 206.0 & 410.2 \\		
						$\textup{X-Pool}^{\dag}$ \cite{gorti2022x} & CVPR'22 & 353.7 & \textbf{49.7} & 74.7 & 84.2 & \textbf{2.0} & 12.3 & 208.6 & 48.1 & 76.0 & \textbf{85.5} & \textbf{2.0} & 8.1 & \textbf{209.6} & 418.2 \\
						$\textup{DRL}^{\dag}$ \cite{wang2022disentangled} & ArXiv'22 & 62.1 & 49.4 & \textbf{76.4} & 84.2 & \textbf{2.0} & 13.2 & 210.0 & 47.0 & \textbf{77.1} & 84.4 & \textbf{2.0} & 9.2 & 208.5 & 418.5 \\
						$\textup{UCoFiA}^{\dag}$ \cite{wang2023unified} & ICCV'23 & 131.7 & \textbf{49.7} & 75.7 & 84.2 & \textbf{2.0} & 12.6 & 209.6 & 48.1 & 76.3 & 84.4 & \textbf{2.0} & 8.8 & 208.8 & 418.4 \\
                        $\textup{X-CLIP}^{\dag}$ \cite{ma2022x} (base)& ArXiv'22 & 95.7 & 49.4 & 75.7 & 84.4 & \textbf{2.0} & \textbf{12.2} & 209.5 & \textbf{48.6} & 75.2 & 84.6 & \textbf{2.0} & \textbf{8.0} & 208.4 & 417.9 \\
						\rowcolor{gray!25}
						TC-MGC (ours) & & 712.8 & 49.0 & 75.7 & \textbf{85.4} & \textbf{2.0} & 13.2 & \textbf{210.1} & 46.4 & \textbf{77.1} & 85.3 & \textbf{2.0} & 8.8 & 208.8 & \textbf{418.9} \\
						\hline
					\end{tabular}
				}
				\begin{tablenotes}    
					\footnotesize              
					\item \qquad \qquad \qquad \qquad \qquad \qquad \qquad \qquad \qquad \qquad \qquad \qquad (b) Training on Training-9K       
				\end{tablenotes}           
			\end{threeparttable}  
		\end{table*}
	
\textbf{Implementation Details.} Our baseline method is X-CLIP \cite{ma2022x}. All experiments are performed on 4 NVIDIA GeForce RTX 3090 24GB GPUs using PyTorch. Following previous works \cite{luo2022clip4clip, wang2021actionclip}, the text encoder and video encoder are initialized from the public CLIP checkpoints (ViT-B/32). The linear embedding initialized with the identity matrices is utilized in the ISA and LSA modules to enhance the expression ability. We set the initial learning rate as 1e-7 for CLIP encoders, and 1e-4 for other modules.
We use Adam \cite{kingma2014adam} with a cosine learning rate schedule \cite{loshchilov2016sgdr} to optimize our model. For MSR-VTT and VATEX, we set the training epoch, batch size, max token length, and the max frame length to 5, 128, 32, and 12. We configure the training epoch, max token length and max frame length as 20, 64 and 64 in DiDeMo. Due to GPU memory limitations, we also reduce the batch size of DiDeMo to 64. During training, we set the similarity keep rate $r=0.1$, and the SDR loss weight $\alpha=0.5$. All videos are compressed to 3 FPS with the resolution of 224 in height or width for training process acceleration. 
	
\subsection{Retrieval Results}
We compare TC-MGC with recent works on three text-video retrieval datasets. The details of compared CLIP-based methods are listed as follows: \par
$\bullet$ CLIP4Clip \cite{luo2022clip4clip} transfers the knowledge of CLIP to text-video retrieval and designs three similarity calculators. \par
$\bullet$ X-Pool \cite{gorti2022x} designs a cross-modal attention mechanism to generate video representation in a text-guided manner. \par
$\bullet$ CenterCLIP \cite{zhao2022centerclip} introduces a token clustering module to find the most representative tokens. \par
$\bullet$ TS2-Net \cite{liu2022ts2} designs a token selection module to select informative tokens in both temporal and spatial dimensions. \par
$\bullet$ X-CLIP \cite{ma2022x} utilizes multi-grained contrastive learning to reduce the negative effect of unimportant information. \par
$\bullet$ DRL \cite{wang2022disentangled} applies a weighted token-wise interaction mechanism to exploit the pair-wise correlations. \par
$\bullet$ UCoFiA \cite{wang2023unified} presents an interactive similarity aggregation module to mitigate the effect of irrelevant clues in cross-modal similarity aggregation. \par

Note that all CLIP-based methods employ CLIP-ViT-B/32 or CLIP-ViT-B/16 as backbone without considering post-processing operations.

\begin{table*}
	\caption{Retrieval results on DiDeMo. Speed is the inference time per video during evaluation on a NVIDIA GeForce RTX 3090 GPU. $\dag$ denotes re-training. Bold denotes the best performance. ``–”: result is unavailable. ``NeurComp'' refers to Neurocomputing.}
	\label{table: DiDeMo Results}
	\centering
		\resizebox{\textwidth}{!}{
			\begin{tabular}{l|l|c|cccccc|cccccc|c}
				\hline
				\multirow{2}*{Methods} & \multirow{2}*{Venue} & \multirow{2}*{\makecell[c]{Speed\\(ms)}} & \multicolumn{6}{c|}{Text $\rightarrow$ Video} &  \multicolumn{6}{c|}{Video $\rightarrow$ Text} & \multirow{2}*{SumR$\uparrow$} \\ 
				\cline{4-9} \cline{10-15} 
				&&& R@1$\uparrow$  & R@5$\uparrow$  & R@10$\uparrow$ & MdR$\downarrow$ &  MnR$\downarrow$ & RSum$\uparrow$
				& R@1$\uparrow$  & R@5$\uparrow$  & R@10$\uparrow$ & MdR$\downarrow$ &  MnR$\downarrow$ & RSum$\uparrow$\\
				\hline
				S2VT \cite{venugopalan2014translating} & ArXiv'14 & - & 11.9 & 33.6 & - & 13.0 & - & - & 13.2 & 33.6 & - & 15.0 & - & - & - \\
				FSE \cite{zhang2018cross} & ECCV'18 & - & 13.9 & 36.0 & - & 11.0 & - & - &  13.1 & 33.9 & - & 12.0 & - & - & -  \\
				CE \cite{liu2019use} & ArXiv'19 & - & 16.1 & 41.1 & - & 8.3 & 43.7 & - & 15.6 & 40.9 & - & 8.2 & 42.4 & - & -  \\
				Frozen \cite{bain2021frozen} & ICCV'21 & - & 34.6 & 65.0 & 74.7 & 3.0 & - & 174.3 & - & - & - & - & - & - & -  \\
				TMVM  \cite{lin2022text} & NeurIPS’22 & - & 36.5 & 64.9 & 75.4 & 3.0 & - & 176.8 & - & - & - & - & - & - & -  \\
				\hline
				\textbf{\textit{CLIP-ViT-B/32}} \\
				\hdashline[1pt/5pt]
				$\textup{CLIP4Clip-meanP}^{\dag}$ \cite{luo2022clip4clip} & NeurComp'22 & 18.8 & 40.8 & 67.8 & 77.4 & \textbf{2.0} & 20.8 & 186.0 & 40.7 & 67.3 & 77.5 & \textbf{2.0} & 15.4 & 185.5 & 371.5 \\
				$\textup{CLIP4Clip-seqLSTM}^{\dag}$ \cite{luo2022clip4clip} & NeurComp'22 & 65.9 & 40.8 & 67.8 & 77.1 & \textbf{2.0} & 21.5 & 185.7 & 40.1 & 66.8 & 77.4 & \textbf{2.0} & 15.7 & 184.3 & 370.0 \\
				$\textup{CLIP4Clip-seqTransf}^{\dag}$ \cite{luo2022clip4clip} & NeurComp'22 & 198.2 & 40.6 & 66.3 & 76.1 & \textbf{2.0} & 19.8 & 183.0 & 40.5 & 65.6 & 75.9 & \textbf{2.0} & 15.5 & 182.0 & 365.0 \\
				$\textup{TS2-Net}^{\dag}$ \cite{liu2022ts2} & ECCV'22 & 723.6 & 42.5 & 69.3 & 77.8 & \textbf{2.0} & 18.8 & 189.6 & 42.2 & 68.2 & 78.6 & \textbf{2.0} & 13.6 & 189.0 & 378.6 \\
				$\textup{UCoFiA}^{\dag}$ \cite{wang2023unified} & ICCV'23 & 337.2 & 45.2 & 72.4 & 81.3 & \textbf{2.0} & \textbf{13.9} & 198.9 & 44.8 & 72.0 & 81.5 & \textbf{2.0} & 10.7 & 198.3 & 397.2 \\
                $\textup{X-CLIP}^{\dag}$ \cite{ma2022x} (base) & ArXiv'22 & 238.1 & 44.7 & \textbf{72.7} & 80.6 & \textbf{2.0} & 15.9 & 198.0 & 44.5 & 72.3 & 82.0 & \textbf{2.0} & 12.6 & 198.8 & 396.8 \\
				\rowcolor{gray!25}
				TC-MGC (ours) & & 1944.7 & \textbf{45.7} & \textbf{72.7} & \textbf{81.9} & \textbf{2.0} & 15.0 & \textbf{200.3} & \textbf{45.4} & \textbf{73.3} & \textbf{82.9} & \textbf{2.0} & \textbf{10.5} & \textbf{201.6} & \textbf{401.9} \\
				\hline
				\textbf{\textit{CLIP-ViT-B/16}} \\
				\hdashline[1pt/5pt]
				$\textup{CLIP4Clip-meanP}^{\dag}$ \cite{luo2022clip4clip} & NeurComp'22 & 22.6 & 42.8 & 71.6 & 81.7 & \textbf{2.0} & 16.6 & 196.1 & 43.1 & 71.8 & 80.8 & \textbf{2.0} & 12.2 & 195.7 & 391.8 \\
				$\textup{CLIP4Clip-seqLSTM}^{\dag}$ \cite{luo2022clip4clip} & NeurComp'22 & 83.6 & 43.5 & 70.8 & 80.3 & \textbf{2.0} & 17.5 & 194.6 & 41.7 & 70.6 & 81.6 & \textbf{2.0} & 12.6 & 193.9 & 388.5 \\
				$\textup{CLIP4Clip-seqTransf}^{\dag}$ \cite{luo2022clip4clip} & NeurComp'22 & 216.3 & 42.8 & 72.6 & 80.8 & \textbf{2.0} & 15.9 & 196.2 & 42.4 & 71.9 & 81.0 & \textbf{2.0} & 11.9 & 195.3 & 391.5 \\
				$\textup{TS2-Net}^{\dag}$ \cite{liu2022ts2} & ECCV'22 & 738.8 & 44.7 & 73.6 & 82.8 & \textbf{2.0} & 16.2 & 201.1 & 44.8 & 72.2 & 82.0 & \textbf{2.0} & 11.8 & 199.0 & 400.1 \\
				$\textup{UCoFiA}^{\dag}$ \cite{wang2023unified} & ICCV'23 & 338.0 & \textbf{48.9} & 75.4 & \textbf{84.9} & \textbf{2.0} & \textbf{11.6} & \textbf{209.2} & 45.6 & 74.2 & 83.0 & \textbf{2.0} & \textbf{8.8} & 202.8 & 412.0 \\
                $\textup{X-CLIP}^{\dag}$ \cite{ma2022x} (base) & ArXiv'22 & 255.6 & 46.1 & 74.7 & 82.7 & \textbf{2.0} & 18.6 & 203.5 & 45.2 & \textbf{75.6} & 82.5 & \textbf{2.0} & 12.4 & 203.3 & 406.8 \\
				\rowcolor{gray!25}
				TC-MGC (ours) & & 1625.4 & 48.6 & \textbf{76.2} & 83.9 & \textbf{2.0} & 13.4 & 208.7 & \textbf{47.7} & 74.8 & \textbf{83.8} & \textbf{2.0} & 9.4 & \textbf{206.3} & \textbf{415.0} \\
				\hline
			\end{tabular}
		}
	\end{table*}

\subsubsection{MSR-VTT Results} 
\noindent\textbf{Retrieval Performance.} As shown in Table \ref{table: MSR-VTT-7K Results}, our approach significantly outperforms previous methods on both data splits. For  ``Training-7K'' split, building on CLIP-ViT-B/32, our model achieves 396.3 for SumR that measures the overall performance, surpassing the baseline by +2.8\% (+10.9\%) relative (absolute) improvements. Our model also improves the mean rank (MnR) from 16.5 to 14.2 (T2V) and from 11.5 to 9.1 (V2T).
Since X-CLIP also conducts multi-grained interactions, we owe the performance improvements to our text-conditioned multi-grained video representations output by the language-video attention block. Even compared with the recent competitor UCoFiA, our model yields +9.4\% absolute improvement on SumR. For ``Training-9K'' split, TC-MGC outperforms all compared models, with the best result being SumR=410.1 for the overall performance. After being equipped with stronger CLIP-ViT-B/16, the SumR of TC-MGC can be further improved to 411.6 and 418.9 on two data splits, which achieve +0.5\% and +1.0\% absolute performance boosts from the baseline. These results clearly verify the effectiveness of our proposed method.

\noindent\textbf{Computational Overhead.} We compare our method with compared models in terms of inference time and add a column to the Table \ref{table: MSR-VTT-7K Results} about the comparison of speed. From the table, we observe that the increased inference time leads to significant performance improvement. Taking ``Training-9K'' split as an example, our TC-MGC takes approximately ninefold inference time to accomplish per video evaluation compared to the baseline  (\textit{i.e.,} 673.1ms \textit{v.s.,} 75.7ms). The usage of stronger CLIP-ViT-B/16 presents similar phenomenon. This may be because the language-video attention block, which involves intensive attention computations. However, we find that the additional computation cost improves the SumR metric from 406.3 to 410.1 (CLIP-ViT-B/32) and from 417.9 to 418.9 (CLIP-ViT-B/16), which indicates the increased inference time is acceptable.

\subsubsection{DiDeMo Results} 
\noindent\textbf{Retrieval Performance.} Table \ref{table: DiDeMo Results} showcases the results on the DiDeMo dataset. Similar to MSR-VTT, our TC-MGC also outperforms compared CLIP-based methods by a considerable margin on CLIP-ViT-B/32 and CLIP-ViT-B/16. In particular, compared to the baseline, TC-MGC with CLIP-ViT-B/32 improves the R@1 metric from 44.7 to 45.7 (T2V) and from 44.5 to 45.4 (V2T). Meanwhile, the T2V and V2T MnR metrics are promoted to 15.0 and 10.5, respectively.
When compared with the recent methods, \textit{i.e.,} TS2-Net and UCoFiA, our method achieves +23.3\% and +4.7\% absolute performance gain at SumR metric. These results verify our motivation that building a text-conditioned multi-grained alignment is useful for longer videos retrieval. By using stronger CLIP-ViT-B/16 as backbone, the R@1 metrics of our model can be further improved to 48.6 and 47.7, with +2.5\% absolute improvement over the baseline, showing the generalization and robustness of our approach.

\noindent\textbf{Computational Overhead.} To experimentally examine the computational cost between our method and the baseline, we adopt the speed metric to represent the inference time as shown in Table \ref{table: DiDeMo Results}. Overall, our method takes excessive time during the video-paragraph retrieval task. Specifically, building on CLIP-ViT-B/32, TC-MGC takes 1944.7ms to perform per video evaluation, which is approximately 8 times increased. Beyond that, TC-MGC with stronger CLIP-ViT-B/16 has a higher computational cost (\textit{i.e.,} 1625.4ms \textit{v.s.,} 255.6ms). The inference time growth can be attributed to intensive attention computation in the language-video attention block, especially longer videos retrieval. However, it is also worth noting that the increased inference time brings significant performance gains over the baseline, which exhibits the additional computation cost is deserving.

\begin{table*}
	\caption{Retrieval results on VATEX. Speed is the inference time per video during evaluation on a NVIDIA GeForce RTX 3090 GPU. $\dag$ denotes re-training. Bold denotes the best performance. ``–”: result is unavailable. ``NeurComp'' refers to Neurocomputing.}
	\label{table: VATEX Results}
	\centering
		\resizebox{\textwidth}{!}{
			\begin{tabular}{l|l|c|cccccc|cccccc|c}
				\hline
				\multirow{2}*{Methods} & \multirow{2}*{Venue} & \multirow{2}*{\makecell[c]{Speed\\(ms)}} & \multicolumn{6}{c|}{Text $\rightarrow$ Video} &  \multicolumn{6}{c|}{Video $\rightarrow$ Text} & \multirow{2}*{SumR$\uparrow$} \\ 
				\cline{4-9} \cline{10-15} 
				&&&  R@1$\uparrow$  & R@5$\uparrow$  & R@10$\uparrow$ & MdR$\downarrow$ &  MnR$\downarrow$ & RSum$\uparrow$
				& R@1$\uparrow$  & R@5$\uparrow$  & R@10$\uparrow$ & MdR$\downarrow$ &  MnR$\downarrow$ & RSum$\uparrow$\\
				\hline
				HGR \cite{chen2020fine} & CVPR'20 & - & 35.1 & 73.5 & 83.5 & 2.0 & - & 192.1 & - & - & - & - & - & - & - \\ 
				SUPPORT \cite{patrick2020support} & ICLR'21 & - & 44.6 & 81.8 & 89.5 & \textbf{1.0} & - & 215.9 & 58.1 & 83.8 & 90.9 & \textbf{1.0} & - & 232.8 & 448.7 \\
				\hline
				\textbf{\textit{CLIP-ViT-B/32}} \\
				\hdashline[1pt/5pt]
				$\textup{CLIP4Clip-meanP}^{\dag}$ \cite{luo2022clip4clip} & NeurComp'22 & 29.5 & 57.3 & 88.6 & 94.3 & \textbf{1.0} & 4.0 & 240.2 & 73.9 & 96.1 & 98.7 & \textbf{1.0} & \textbf{1.8} & 268.7 & 508.9 \\
				$\textup{CLIP4Clip-seqLSTM}^{\dag}$ \cite{luo2022clip4clip} & NeurComp'22 & 104.0 & 57.3 & 88.7 & 94.2 & \textbf{1.0} & 4.0 & 240.2 & 73.8 & 96.3 & 99.2 & \textbf{1.0} & \textbf{1.8} & 269.3 & 509.5 \\
				$\textup{CLIP4Clip-seqTransf}^{\dag}$ \cite{luo2022clip4clip} & NeurComp'22 & 387.7 & 58.7 & 89.3 & 94.7 & \textbf{1.0} & 3.7 & 242.7 & 74.8 & 96.7 & 98.6 & \textbf{1.0} & \textbf{1.8} & 270.1 & 512.8 \\
				$\textup{TS2-Net}^{\dag}$ \cite{liu2022ts2} & ECCV'22 & 441.9 & 59.5 & \textbf{89.8} & \textbf{95.0} & \textbf{1.0} & \textbf{3.6} & 244.3 & 74.6 & 96.3 & 98.9 & \textbf{1.0} & \textbf{1.8} & 269.8 & 514.1 \\
				$\textup{UCoFiA}^{\dag}$ \cite{wang2023unified} & ICCV'23 & 1815.8 & 59.3 & 88.8 & 94.3 & \textbf{1.0} & 4.1 & 242.4 & 73.0 & 96.1 & 98.9 & \textbf{1.0} & 1.9 & 268.0 & 510.4 \\
                $\textup{X-CLIP}^{\dag}$ \cite{ma2022x} (base) & ArXiv'22 & 488.4 & 59.1 & 88.9 & 94.2 & \textbf{1.0} & 3.9 & 242.2 & 74.8 & \textbf{97.3} & \textbf{99.0} & \textbf{1.0} & \textbf{1.8} & \textbf{271.1} & 513.3 \\
				\rowcolor{gray!25}
				TC-MGC (ours) & & 3203.5 & \textbf{60.0} & 89.5 & 94.9 & \textbf{1.0} & 3.8 & \textbf{244.4} & \textbf{75.3} & 97.0 & 98.6 & \textbf{1.0} & 1.9 & 270.9 & \textbf{515.3} \\
				\hline
				\textbf{\textit{CLIP-ViT-B/16}} \\
				\hdashline[1pt/5pt]
				$\textup{CLIP4Clip-meanP}^{\dag}$ \cite{luo2022clip4clip} & NeurComp'22 & 29.6 & 62.0 & 91.1 & 95.8 & \textbf{1.0} & 3.3 & 248.9 & 78.3 & 97.9 & 99.1 & \textbf{1.0} & \textbf{1.5} & 275.3 & 524.2 \\
				$\textup{CLIP4Clip-seqLSTM}^{\dag}$ \cite{luo2022clip4clip} & NeurComp'22 & 101.2 & 62.4 & 91.1 & 95.9 & \textbf{1.0} & 3.2 & 249.4 & 77.9 & 97.8 & 99.2 & \textbf{1.0} & \textbf{1.5} & 274.9 & 524.3 \\
				$\textup{CLIP4Clip-seqTransf}^{\dag}$ \cite{luo2022clip4clip} & NeurComp'22 & 396.3 & 63.3 & \textbf{92.0} & 96.2 & \textbf{1.0} & \textbf{3.0} & 251.5 & \textbf{80.1} & \textbf{98.2} & 99.1 & \textbf{1.0} & \textbf{1.5} & \textbf{277.4} & \textbf{528.9} \\
				$\textup{TS2-Net}^{\dag}$ \cite{liu2022ts2} & ECCV'22 & 461.3 & 62.3 & 91.5 & 96.2 & \textbf{1.0} & 3.1 & 250.0 & 77.4 & 97.5 & 99.3 & \textbf{1.0} & 1.6 & 274.2 & 524.2 \\
				$\textup{UCoFiA}^{\dag}$ \cite{wang2023unified} & ICCV'23 & 1840.8 & \textbf{64.3} & 91.6 & 95.9 & \textbf{1.0} & 3.3 & \textbf{251.8} & 79.8 & 97.6 & 99.0 & \textbf{1.0} & \textbf{1.5} & 276.4 & 528.2 \\
                $\textup{X-CLIP}^{\dag}$ \cite{ma2022x} (base) & ArXiv'22 & 480.5 & 62.8 & 91.4 & 96.1 & \textbf{1.0} & 3.1 & 250.3 & 78.5 & 97.8 & 99.2 & \textbf{1.0} & 1.6 & 275.5 & 525.8 \\
				\rowcolor{gray!25}
				TC-MGC (ours) & & 5974.2 & 63.4 & 91.9 & \textbf{96.3} & \textbf{1.0} & \textbf{3.0} & 251.6 & 77.7 & 97.9 & \textbf{99.4} & \textbf{1.0} & \textbf{1.5} & 275.0 & 526.6 \\
				\hline
			\end{tabular}
		}
	\end{table*}

\subsubsection{VATEX Results} 
\noindent\textbf{Retrieval Performance.} The comparison results on the VATEX dataset are presented in Table \ref{table: VATEX Results}, which provides retrieval performance and computational cost (Speed) on CLIP-ViT-B/32 and CLIP-ViT-B/16. From the table, building on CLIP-ViT-B/32, our method notably outperforms compared CLIP-based methods in terms of recall metrics. In particular, compared to the baseline, TC-MGC improves the R@1 metric from 59.1 to 60.0 (T2V) and from 74.8 to 75.3 (V2T). In addition, TC-MGC achieves +2.0\% absolute improvement than the baseline on SumR . Even compared with TS2-Net and UCoFiA, our method still surpasses them by +1.2\% and +4.9\% in SumR. These results show the importance of text-conditioned multi-grained alignment. After being equipped with stronger CLIP-ViT-B/16, our TC-MGC also maintains admirable performance compared to the baseline, with better overall result SumR = 526.6. The above results show that our method can consistently  promote the performance with a large margin under both backbone settings.
	
\noindent\textbf{Computational Overhead.} By analyzing the speed and SumR metrics in Table \ref{table: VATEX Results}, we find that our method has significant performance gains and acceptable inference time growth compared to the baseline. Specifically, our TC-MGC with CLIP-ViT-B/32 outperforms the baseline by +2.0\% on SumR, with being nearly 7 times increased in terms of inference time. Additionally, with the employment of stronger CLIP-ViT-B/16, our method also achieves performance boost over the baseline via a higher computational cost (\textit{i.e.,} 5974.2ms \textit{v.s.,} 480.5ms). Since the language-video attention block in our TC-MGC contains intensive attention computations, the speed during per video evaluation suffers a great decrease while presenting better performance, indicating the additional computational cost is valuable.
	
\subsection{Ablation Study}
We evaluate the effectiveness of each component in our TC-MGC  by comprehensive ablation experiments on MSR-VTT with the 1k-A test split.
	
\begin{table}[htbp]
		\caption{Ablation study of text-conditioned contrast mechanism. The softmax-based weighted combination is applied on the cross-modal similarity vectors/matrices. For simplicity, video-sent and sent-frame stand for video-sentence and sentence-frame, respectively. TCC is short for text-conditioned contrast.}
		\label{table: w/wo text-conditioned contrastive learning}
		\centering
		\resizebox{0.5\textwidth}{!}{
			\Huge
			\renewcommand{\arraystretch}{1.1}
			\begin{tabular}{c|c|cccc|cccc|c}
				\hline
				\multirow{2}*{Module} & \multirow{2}*{TCC} &  \multicolumn{4}{c|}{Text $\rightarrow$ Video} &  \multicolumn{4}{c|}{Video $\rightarrow$ Text} & \multirow{2}*{SumR$\uparrow$} \\
				\cline{3-6} \cline{7-10} 
				&& R@1$\uparrow$  & R@5$\uparrow$  & R@10$\uparrow$ & RSum$\uparrow$
				& R@1$\uparrow$  & R@5$\uparrow$  & R@10$\uparrow$ & RSum$\uparrow$ \\
				\hline
				\multirow{2}*{Video-Sent} & & 43.0 & 70.7 & 81.6 & 195.3 & 43.0 & 70.2 & 81.2 & 194.4 & 389.7 \\
				~ & \Checkmark & \textbf{46.6} & \textbf{73.1} & \textbf{83.1} & \textbf{202.8} & \textbf{45.1} & \textbf{72.5} & \textbf{81.4} & \textbf{199.0} & \textbf{401.8} \\
				\hline
				\multirow{2}*{Video-Word} & & 42.8 & 69.9 & 80.1 & 192.8 & 43.2 & 70.1 & 80.5 & 193.8 & 386.6 \\
				~ & \Checkmark & \textbf{46.5} & \textbf{73.5} & \textbf{83.7} & \textbf{203.7} & \textbf{45.8} & \textbf{73.5} & \textbf{82.0} &\textbf{201.3} & \textbf{405.0} \\
				\hline
				\multirow{2}*{Sent-Frame} & & 42.7 & 69.6 & 81.3 & 193.6 & 43.1 & 70.7 & \textbf{82.1} & 195.9 & 389.5 \\
				~ & \Checkmark & \textbf{47.2} & \textbf{72.9} & \textbf{83.8} & \textbf{203.9} & \textbf{44.3} & \textbf{73.2} & 81.7 & \textbf{199.2} & \textbf{403.1} \\
				\hline
				\multirow{2}*{Frame-Word} & & 42.7 & 69.5 & 81.3 & 193.5 & 42.8 & 70.8 & \textbf{81.7} & 195.3 & 388.8 \\
				~ & \Checkmark & \textbf{46.7} & \textbf{74.6} & \textbf{83.0} &\textbf{204.3} & \textbf{44.2} & \textbf{73.0} & 80.9 & \textbf{198.1} & \textbf{402.4} \\
				\hline
			\end{tabular}
		}
\end{table}

\subsubsection{Text-Conditioned Contrast Mechanism} To justify the effectiveness of text-conditioned contrast mechanism, we conduct an ablation study to compare four single-grained contrasts with and without the text-conditioned visual information. As shown in Table \ref{table: w/wo text-conditioned contrastive learning}, the usage of sentence-conditioned video representation in video-sentence contrast achieves 46.6 T2V R@1, outperforming the baseline results by +8.4\% (+3.6\%) relative (absolute) improvements. Similarly, we observe that compared to the frame-word baseline results, a better SumR of 402.4 with +13.6\% absolute improvement is obtained by the word-conditioned frame representations, further demonstrating the superiority of text-conditioned contrast mechanism.

\begin{table*}[htbp]
    \caption{T2V results comparison between two variants of similarity reorganization for the video-word similarity vector. The baseline results are obtained by the ISA implementation on the computed similarity vector. The values of params indicate learnable parameter statistics in the linear layer of ISA. The number in blue is the gap of the corresponding value w.r.t. the baseline. }
    \label{table: similarity reorganization on video-word similarity vector}
    \centering
    \resizebox{\textwidth}{!}{
        \renewcommand{\arraystretch}{1.02}
        \begin{tabular}{ccccccccccc}
            \hline
            Keep Rate & R@1$\uparrow$  & R@5$\uparrow$  & R@10$\uparrow$ & RSum$\uparrow$ & Params(K) & R@1$\uparrow$  & R@5$\uparrow$  & R@10$\uparrow$ & RSum$\uparrow$ & Params(K)\\
            \hline
            baseline & 45.8 & 75.1 & 83.6 & 204.5 & 1.024 & 45.8 & 75.1 & 83.6 & 204.5 & 1.024 \\
            \hline
            & \multicolumn{5}{c}{inattentive similarities removal} & \multicolumn{5}{c}{inattentive similarities fusion} \\
            \cmidrule(lr){2-6} \cmidrule(r){7-11}
            0.1 & 46.4\textcolor{blue}{(+0.6)} & 74.3\textcolor{blue}{(-0.8)} &  83.9\textcolor{blue}{(+0.3)}  & 204.6\textcolor{blue}{(+0.1)} & \textbf{0.009}\textcolor{blue}{(-99.1\%)} & \textbf{47.2}\textcolor{blue}{(+1.4)} & 73.9\textcolor{blue}{(-1.2)} & 84.3\textcolor{blue}{(+0.7)} &  \textbf{205.4}\textcolor{blue}{(+0.9)} & \textbf{0.016}\textcolor{blue}{(-98.4\%)}  \\
            0.2 & \textbf{47.7}\textcolor{blue}{(+1.9)} & 74.0\textcolor{blue}{(-1.1)} & 84.2\textcolor{blue}{(+0.6)} & \textbf{205.9}\textcolor{blue}{(+1.4)} & 0.036\textcolor{blue}{(-96.5\%)} & 46.4\textcolor{blue}{(+0.6)} & 74.4\textcolor{blue}{(-0.7)} & 84.3\textcolor{blue}{(+0.7)} &  205.1\textcolor{blue}{(+0.6)} & 0.049\textcolor{blue}{(-95.2\%)} \\
            0.3 & 46.4\textcolor{blue}{(+0.6)} & 74.3\textcolor{blue}{(-0.8)} & 83.3\textcolor{blue}{(-0.3)} & 204.0\textcolor{blue}{(-0.5)} & 0.081\textcolor{blue}{(-92.1\%)} & 46.4\textcolor{blue}{(+0.6)} & 74.5\textcolor{blue}{(-0.6)} & 84.1\textcolor{blue}{(+0.5)} &  205.0\textcolor{blue}{(+0.5)} & 0.100\textcolor{blue}{(-90.2\%)}\\  
            0.4 & 46.6\textcolor{blue}{(+0.8)} & \textbf{75.3}\textcolor{blue}{(+0.2)} & 83.5\textcolor{blue}{(-0.1)} & 205.4\textcolor{blue}{(+0.9)} & 0.144\textcolor{blue}{(-85.9\%)} & 46.5\textcolor{blue}{(+0.7)} & 74.5\textcolor{blue}{(-0.6)} & 84.0\textcolor{blue}{(+0.4)} &  205.0\textcolor{blue}{(+0.5)} & 0.169\textcolor{blue}{(-83.5\%)} \\
            0.5 & 46.6\textcolor{blue}{(+0.8)} & 75.2\textcolor{blue}{(+0.1)} & 83.6\textcolor{blue}{(-0.0)} & 205.4\textcolor{blue}{(+0.9)} & 0.256\textcolor{blue}{(-75.0\%)} & 46.2\textcolor{blue}{(+0.4)} & 73.7\textcolor{blue}{(-1.4)} & 83.8\textcolor{blue}{(+0.2)} &  203.7\textcolor{blue}{(-0.8)} & 0.289\textcolor{blue}{(-71.8\%)}\\
            0.6 & 46.3\textcolor{blue}{(+0.5)} & 74.4\textcolor{blue}{(-0.7)} & 83.9\textcolor{blue}{(+0.3)} & 204.6\textcolor{blue}{(+0.1)} & 0.361\textcolor{blue}{(-64.7\%)} & 46.3\textcolor{blue}{(+0.5)} & 74.6\textcolor{blue}{(-0.5)} & \textbf{84.4}\textcolor{blue}{(+0.8)} &  205.3\textcolor{blue}{(+0.8)} & 0.400\textcolor{blue}{(-60.9\%)} \\
            0.7 & 47.0\textcolor{blue}{(+1.2)} & 74.3\textcolor{blue}{(-0.8)} & 83.9\textcolor{blue}{(+0.3)} & 205.2\textcolor{blue}{(+0.7)} & 0.484\textcolor{blue}{(-52.7\%)} & 46.3\textcolor{blue}{(+0.5)} & \textbf{75.1}\textcolor{blue}{(-0.0)} & 83.0\textcolor{blue}{(-0.6)} &  204.4\textcolor{blue}{(-0.1)} & 0.529\textcolor{blue}{(-48.3\%)} \\
            0.8 & 46.1\textcolor{blue}{(+0.3)} & 74.6\textcolor{blue}{(-0.5)} & \textbf{84.5}\textcolor{blue}{(+0.9)} & 205.2\textcolor{blue}{(+0.7)} & 0.625\textcolor{blue}{(-39.0\%)} & 46.9\textcolor{blue}{(+1.1)} & 74.4\textcolor{blue}{(-0.7)} & 83.6\textcolor{blue}{(-0.0)} &  204.9\textcolor{blue}{(+0.4)} & 0.676\textcolor{blue}{(-34.0\%)} \\
            0.9 & 46.1\textcolor{blue}{(+0.3)} & 74.2\textcolor{blue}{(-0.9)} & 84.0\textcolor{blue}{(+0.4)} & 204.3\textcolor{blue}{(-0.2)} & 0.784\textcolor{blue}{(-23.4\%)} & 46.8\textcolor{blue}{(+1.0)} & 74.2\textcolor{blue}{(-0.9)} & 83.8\textcolor{blue}{(+0.2)} &  204.8\textcolor{blue}{(+0.3)} & 0.841\textcolor{blue}{(-17.9\%)} \\
            \hline
        \end{tabular}
    }
\end{table*}

\begin{table*}[hbtp]
    \caption{T2V results comparison between two variants of similarity reorganization for the sentence-frame similarity vector. The baseline results are obtained by the ISA implementation on the computed similarity vector. The values of params indicate learnable parameter statistics in the linear layer of ISA. The number in blue is the gap of the corresponding value w.r.t. the baseline. }
    \label{table: similarity reorganization on sentence-frame similarity vector}
    \centering
    \resizebox{\textwidth}{!}{
        \renewcommand{\arraystretch}{1.02}
        \begin{tabular}{ccccccccccc}
            \hline
            Keep Rate & R@1$\uparrow$  & R@5$\uparrow$  & R@10$\uparrow$ & RSum$\uparrow$ & Params(K) & R@1$\uparrow$  & R@5$\uparrow$  & R@10$\uparrow$ & RSum$\uparrow$ & Params(K) \\
            \hline
            baseline & 47.1 & 74.8 & 83.7 & 205.6 & 1.024 & 47.1 & 74.8 & 83.7 & 205.6 & 1.024 \\
            \hline 
            & \multicolumn{5}{c}{inattentive similarities removal} & \multicolumn{5}{c}{inattentive similarities fusion} \\
            \cmidrule(lr){2-6} \cmidrule(r){7-11}
            0.1 & 46.7\textcolor{blue}{(-0.4)} & 75.0\textcolor{blue}{(+0.2)} & \textbf{84.4}\textcolor{blue}{(+0.7)} & \textbf{206.1}\textcolor{blue}{(+0.5)} & \textbf{0.009}\textcolor{blue}{(-99.1\%)} & 47.7\textcolor{blue}{(+0.6)} & 75.0\textcolor{blue}{(+0.2)} & 83.9\textcolor{blue}{(+0.2)} &  \textbf{206.6}\textcolor{blue}{(+1.0)} & \textbf{0.016}\textcolor{blue}{(-98.4\%)}\\
            0.2 & 47.4\textcolor{blue}{(+0.3)} & 74.0\textcolor{blue}{(-0.8)} & 83.4\textcolor{blue}{(-0.3)} & 204.8\textcolor{blue}{(-0.8)} & 0.036\textcolor{blue}{(-96.5\%)} & 47.8\textcolor{blue}{(+0.7)} & 74.5\textcolor{blue}{(-0.3)} & 83.2\textcolor{blue}{(-0.5)} &  205.5\textcolor{blue}{(-0.1)} & 0.049\textcolor{blue}{(-95.2\%)} \\
            0.3 & \textbf{48.1}\textcolor{blue}{(+1.0)} & 73.8\textcolor{blue}{(-1.0)} & 82.9\textcolor{blue}{(-0.8)} & 204.8\textcolor{blue}{(-0.8)} & 0.081\textcolor{blue}{(-92.1\%)} & 47.3\textcolor{blue}{(+0.2)} & 73.4\textcolor{blue}{(-1.4)} & 82.7\textcolor{blue}{(-1.0)} &  203.4\textcolor{blue}{(-2.2)} & 0.100\textcolor{blue}{(-90.2\%)} \\  
            0.4 & 46.8\textcolor{blue}{(-0.3)} & 74.9\textcolor{blue}{(+0.1)} & 83.5\textcolor{blue}{(-0.2)} & 205.2\textcolor{blue}{(-0.4)} & 0.144\textcolor{blue}{(-85.9\%)} & 47.5\textcolor{blue}{(+0.4)} & 74.9\textcolor{blue}{(+0.1)} & 82.7\textcolor{blue}{(-1.0)} &  205.1\textcolor{blue}{(-0.5)} & 0.169\textcolor{blue}{(-83.5\%)} \\
            0.5 & 46.9\textcolor{blue}{(-0.2)} & 74.3\textcolor{blue}{(-0.5)} & 82.9\textcolor{blue}{(-0.8)} & 204.1\textcolor{blue}{(-1.5)} & 0.256\textcolor{blue}{(-75.0\%)} & 47.2\textcolor{blue}{(+0.1)} & 74.2\textcolor{blue}{(-0.6)} & 83.1\textcolor{blue}{(-0.6)} &  204.5\textcolor{blue}{(-1.1)} & 0.289\textcolor{blue}{(-71.8\%)} \\
            0.6 & \textbf{48.1}\textcolor{blue}{(+1.0)} & 74.3\textcolor{blue}{(-0.5)} & 83.1\textcolor{blue}{(-0.6)} & 205.5\textcolor{blue}{(-0.1)} & 0.361\textcolor{blue}{(-64.7\%)} & 47.7\textcolor{blue}{(+0.6)} & 73.9\textcolor{blue}{(-0.9)} & \textbf{84.1}\textcolor{blue}{(+0.4)} &  205.7\textcolor{blue}{(+0.1)} & 0.400\textcolor{blue}{(-60.9\%)} \\
            0.7 & 47.0\textcolor{blue}{(-0.1)} & 73.7\textcolor{blue}{(-1.1)} & 84.1\textcolor{blue}{(+0.4)} & 204.8\textcolor{blue}{(-0.8)} & 0.484\textcolor{blue}{(-52.7\%)} & \textbf{48.0}\textcolor{blue}{(+0.9)} & 74.3 \textcolor{blue}{(-0.5)}& 84.0\textcolor{blue}{(+0.3)} & 
            206.3\textcolor{blue}{(+0.7)} & 0.529\textcolor{blue}{(-48.3\%)} \\
            0.8 & 46.3\textcolor{blue}{(-0.8)} & \textbf{75.1}\textcolor{blue}{(+0.3)} & 83.6\textcolor{blue}{(-0.1)} & 205.0\textcolor{blue}{(-0.6)} & 0.625\textcolor{blue}{(-39.0\%)} & 47.3\textcolor{blue}{(+0.2)} & 75.0\textcolor{blue}{(+0.2)} & 83.6\textcolor{blue}{(-0.1)} &  205.9\textcolor{blue}{(+0.3)} & 0.676\textcolor{blue}{(-34.0\%)} \\
            0.9 & 47.2\textcolor{blue}{(+0.1)} & 74.2\textcolor{blue}{(-0.6)} & 83.8\textcolor{blue}{(+0.1)} & 205.2\textcolor{blue}{(-0.4)} & 0.784\textcolor{blue}{(-23.4\%)} & 47.2\textcolor{blue}{(+0.1)} & \textbf{75.3}\textcolor{blue}{(+0.5)} & 82.5\textcolor{blue}{(-1.2)} &  205.0\textcolor{blue}{(-0.6)} & 0.841\textcolor{blue}{(-17.9\%)} \\
            \hline
        \end{tabular}
    }
\end{table*}

\subsubsection{Similarity Reorganization} To fully validate the effectiveness of our proposed SR module, we conduct an ablative study to compare three single-grained contrasts with and without the similarity reorganization. Meanwhile, we experimentally compare two variants of SR, namely vanilla inattentive similarities removal and inattentive similarities fusion, at keep rate range setting $r \in [0.1, 0.9]$.

\noindent\textbf{Video-Word Similarity Vector.} From Table \ref{table: similarity reorganization on video-word similarity vector}, we observe that both SR variants can achieve better retrieval performance with fewer parameters in the linear layer of ISA. For example, when $r=0.2$, we find that vanilla inattentive similarities removal obtains 47.7 T2V R@1, surpassing the baseline results by +1.9\% absolute improvement. The inattentive similarities fusion at this keep rate also promotes the T2V R@1 from 45.8 to 46.4 with 95.2\% trainable parameters decrease. We suppose that the attentive similarities enable the model to better capture cross-modal semantic correspondence. Additionally, the best T2V 47.7 R@1 and 205.9 RSum are obtained by vanilla inattentive similarities removal. The main reason may be that the fusion of semantically distant words inevitably brings noise, making it hard for the model to capture precise cross-modal correspondence. As a result, we use vanilla inattentive similarities removal to reorganize video-word similarity vector.

\begin{table*}[htbp]
    \caption{T2V results comparison between two variants of unidirectional similarity reorganization for the frame-word similarity matrix. The baseline results are obtained by the Bi-ISA implementation on the computed similarity matrix. The values of params indicate learnable parameter statistics in the linear layers of Bi-ISA. The number in blue is the gap of the corresponding value w.r.t. the baseline. }
    \label{table: unidirectional similarity reorganization on frame-word similarity vector}
    \centering
    \begin{threeparttable}
        \resizebox{\textwidth}{!}{
            \renewcommand{\arraystretch}{1.02}
            \begin{tabular}{ccccccccccc}
                \hline
                Keep Rate & R@1$\uparrow$  & R@5$\uparrow$  & R@10$\uparrow$ & RSum$\uparrow$ & Params(K) & R@1$\uparrow$  & R@5$\uparrow$  & R@10$\uparrow$ & RSum$\uparrow$ & Params(K) \\
                \hline
                baseline & 47.8 & 75.3 & 83.8 & 206.9 & 4.096 & 47.8 & 75.3 & 83.8 & 206.9 & 4.096 \\
                \hline
                & \multicolumn{5}{c}{inattentive similarities removal} & \multicolumn{5}{c}{inattentive similarities fusion} \\
                \cmidrule(lr){2-6} \cmidrule(r){7-11}
                0.1 & 46.6\textcolor{blue}{(-1.2)} & 74.3\textcolor{blue}{(-1.0)} &  83.3\textcolor{blue}{(-0.5)} & 204.2\textcolor{blue}{(-2.7)} & \textbf{2.066}\textcolor{blue}{(-49.6\%)} & 47.7\textcolor{blue}{(-0.1)} & 73.4\textcolor{blue}{(-1.9)} & 83.6\textcolor{blue}{(-0.2)} &  204.7\textcolor{blue}{(-2.2)} & \textbf{2.080}\textcolor{blue}{(-49.2\%)} \\
                0.2 & 46.9\textcolor{blue}{(-0.9)} & 72.9\textcolor{blue}{(-2.4)} & 83.1\textcolor{blue}{(-0.7)} & 202.9\textcolor{blue}{(-4.0)} & 2.120\textcolor{blue}{(-48.2\%)} & 47.8\textcolor{blue}{(-0.0)} & \textbf{74.6}\textcolor{blue}{(-0.7)} & 83.1\textcolor{blue}{(-0.7)} &  \textbf{205.5}\textcolor{blue}{(-1.4)} & 2.146\textcolor{blue}{(-47.6\%)}\\
                0.3 & 46.2\textcolor{blue}{(-1.6)} & 74.5\textcolor{blue}{(-0.8)} & 83.4\textcolor{blue}{(-0.4)} & 204.1\textcolor{blue}{(-2.8)} & 2.210\textcolor{blue}{(-46.0\%)} & 47.3\textcolor{blue}{(-0.5)} & 74.3\textcolor{blue}{(-1.0)} & 83.2\textcolor{blue}{(-0.6)} &  204.8\textcolor{blue}{(-2.1)} & 2.248\textcolor{blue}{(-45.1\%)} \\  
                0.4 & 46.0\textcolor{blue}{(-1.8)} & 74.5\textcolor{blue}{(-0.8)} & 84.0\textcolor{blue}{(+0.2)} & 204.5\textcolor{blue}{(-2.4)} & 2.336\textcolor{blue}{(-43.0\%)} & 47.3\textcolor{blue}{(-0.5)} & 74.2\textcolor{blue}{(-1.1)} & 83.2\textcolor{blue}{(-0.6)} &  204.7\textcolor{blue}{(-2.2)} & 2.386\textcolor{blue}{(-41.7\%)} \\
                0.5 & \textbf{48.3}\textcolor{blue}{(+0.5)} & 73.2\textcolor{blue}{(-2.1)} & 83.7\textcolor{blue}{(-0.1)} & 205.2\textcolor{blue}{(-1.7)} & 2.560\textcolor{blue}{(-37.5\%)} & 46.5\textcolor{blue}{(-1.3)} & 74.2\textcolor{blue}{(-1.1)} & 84.0\textcolor{blue}{(+0.2)} &  204.7\textcolor{blue}{(-2.2)} & 2.626\textcolor{blue}{(-35.9\%)} \\
                0.6 & 47.8\textcolor{blue}{(-0.0)} & 74.5\textcolor{blue}{(-0.8)} & 83.7\textcolor{blue}{(-0.1)} & \textbf{206.0}\textcolor{blue}{(-0.9)} & 2.770\textcolor{blue}{(-32.4\%)} & \textbf{48.0}\textcolor{blue}{(+0.2)} & 73.5\textcolor{blue}{(-1.8)} & 83.3\textcolor{blue}{(-0.5)} &  204.8\textcolor{blue}{(-2.1)} & 2.848\textcolor{blue}{(-30.5\%)}\\
                0.7 & 46.2\textcolor{blue}{(-1.6)} & 73.1\textcolor{blue}{(-2.2)} & 83.1\textcolor{blue}{(-0.7)} & 202.4\textcolor{blue}{(-4.5)} & 3.016\textcolor{blue}{(-26.4\%)} & 46.5\textcolor{blue}{(-1.3)} & 73.7\textcolor{blue}{(-1.6)} & 83.3\textcolor{blue}{(-0.5)} &  203.5\textcolor{blue}{(-3.4)} & 3.106\textcolor{blue}{(-24.2\%)} \\
                0.8 & 46.6\textcolor{blue}{(-1.2)} & 74.8\textcolor{blue}{(-0.5)} & 84.5\textcolor{blue}{(+0.7)} & 205.9\textcolor{blue}{(-1.0)} & 3.298\textcolor{blue}{(-19.5\%)} & 47.0\textcolor{blue}{(-0.8)} & 74.3\textcolor{blue}{(-1.0)} & 83.4\textcolor{blue}{(-0.4)} &  204.7\textcolor{blue}{(-2.2)} & 3.400\textcolor{blue}{(-17.0\%)} \\
                0.9 & 46.5\textcolor{blue}{(-1.3)} & \textbf{74.9}\textcolor{blue}{(-0.4)} & \textbf{84.6}\textcolor{blue}{(+0.8)} & \textbf{206.0}\textcolor{blue}{(-0.9)} & 3.616\textcolor{blue}{(-11.7\%)} & 46.2\textcolor{blue}{(-1.6)} & 74.0\textcolor{blue}{(-1.3)} & \textbf{84.2}\textcolor{blue}{(+0.4)} &  204.4\textcolor{blue}{(-2.5)} & 3.730\textcolor{blue}{(-8.9\%)} \\
                \hline		
            \end{tabular}
        }
        \begin{tablenotes}    
            \footnotesize              
            \item \qquad \qquad \qquad \qquad \qquad \qquad \qquad \qquad \qquad \qquad (a) Similarity reorganization on the word direction       
        \end{tablenotes}           
    \end{threeparttable}  
\end{table*}

\begin{table*}[htbp]
    \centering
    \begin{threeparttable}
        \resizebox{\textwidth}{!}{
            \renewcommand{\arraystretch}{1.02}
            \begin{tabular}{ccccccccccc}
                \hline
                Keep Rate & R@1$\uparrow$  & R@5$\uparrow$  & R@10$\uparrow$ & RSum$\uparrow$ & Params(K) & R@1$\uparrow$  & R@5$\uparrow$  & R@10$\uparrow$ & RSum$\uparrow$ & Params(K) \\
                \hline
                baseline & 47.8 & 75.3 & 83.8 & 206.9 & 4.096 & 47.8 & 75.3 & 83.8 & 206.9 & 4.096 \\
                \hline 
                & \multicolumn{5}{c}{inattentive similarities removal} & \multicolumn{5}{c}{inattentive similarities fusion} \\
                \cmidrule(lr){2-6} \cmidrule(r){7-11}
                0.1 & 46.4\textcolor{blue}{(-1.4)} & 74.2\textcolor{blue}{(-1.1)} & 82.8\textcolor{blue}{(-1.0)} & 203.4\textcolor{blue}{(-3.5)} & \textbf{2.066}\textcolor{blue}{(-49.6\%)} & 47.1\textcolor{blue}{(-0.7)} & 74.1\textcolor{blue}{(-1.2)} & 83.9\textcolor{blue}{(+0.1)} &  205.1\textcolor{blue}{(-1.8)} & \textbf{2.080}\textcolor{blue}{(-49.2\%)} \\
                0.2 & 46.5\textcolor{blue}{(-1.3)} & 74.0\textcolor{blue}{(-1.3)} & 83.4\textcolor{blue}{(-0.4)} & 203.9\textcolor{blue}{(-3.0)} & 2.120\textcolor{blue}{(-48.2\%)} & 47.4\textcolor{blue}{(-0.4)} & \textbf{75.3}\textcolor{blue}{(-0.0)} & 83.8\textcolor{blue}{(-0.0)} &  \textbf{206.5}\textcolor{blue}{(-0.4)} & 2.146\textcolor{blue}{(-47.6\%)} \\
                0.3 & 47.1\textcolor{blue}{(-0.7)} & 74.4\textcolor{blue}{(-0.9)} & 83.6\textcolor{blue}{(-0.2)} & 205.1\textcolor{blue}{(-1.8)} & 2.210\textcolor{blue}{(-46.0\%)} & 46.6\textcolor{blue}{(-1.2)} & 73.8\textcolor{blue}{(-1.5)} & \textbf{84.1}\textcolor{blue}{(+0.3)} &  204.5\textcolor{blue}{(-2.4)} & 2.248\textcolor{blue}{(-45.1\%)} \\  
                0.4 & 47.0\textcolor{blue}{(-0.8)} & 74.3\textcolor{blue}{(-1.0)} & \textbf{83.8}\textcolor{blue}{(-0.0)} & 205.1\textcolor{blue}{(-1.8)} & 2.336\textcolor{blue}{(-43.0\%)} & 47.9\textcolor{blue}{(+0.1)} & 74.5\textcolor{blue}{(-0.8)} & 83.4\textcolor{blue}{(-0.4)} &  205.8\textcolor{blue}{(-1.1)} & 2.386\textcolor{blue}{(-41.7\%)} \\
                0.5 & \textbf{47.9}\textcolor{blue}{(+0.1)} & 74.2\textcolor{blue}{(-1.1)} & 83.6\textcolor{blue}{(-0.2)} & \textbf{205.7}\textcolor{blue}{(-1.2)} & 2.560\textcolor{blue}{(-37.5\%)} & 45.9\textcolor{blue}{(-1.9)} & 73.9\textcolor{blue}{(-1.4)} & 83.1\textcolor{blue}{(-0.7)} &  202.9\textcolor{blue}{(-4.0)} & 2.626\textcolor{blue}{(-35.9\%)} \\
                0.6 & 46.9\textcolor{blue}{(-0.9)} & 73.8\textcolor{blue}{(-1.5)} & 83.4\textcolor{blue}{(-0.4)} & 204.1\textcolor{blue}{(-2.8)} & 2.770\textcolor{blue}{(-32.4\%)} & 47.5\textcolor{blue}{(-0.3)} & 74.6\textcolor{blue}{(-0.7)} & 83.4\textcolor{blue}{(-0.4)} &  205.5\textcolor{blue}{(-1.4)} & 2.848\textcolor{blue}{(-30.5\%)} \\
                0.7 & 47.3\textcolor{blue}{(-0.5)} & 73.2\textcolor{blue}{(-2.1)} & 83.0\textcolor{blue}{(-0.8)} & 203.5\textcolor{blue}{(-3.4)} & 3.016\textcolor{blue}{(-26.4\%)} & 46.2\textcolor{blue}{(-1.6)} & 73.7\textcolor{blue}{(-1.6)}& 83.4\textcolor{blue}{(-0.4)} &  203.3\textcolor{blue}{(-3.6)} & 3.106\textcolor{blue}{(-24.2\%)} \\
                0.8 & 46.6\textcolor{blue}{(-1.2)} & \textbf{74.7}\textcolor{blue}{(-0.6)} & 83.1\textcolor{blue}{(-0.7)} & 204.4\textcolor{blue}{(-2.5)} & 3.298\textcolor{blue}{(-19.5\%)} & \textbf{48.2}\textcolor{blue}{(+0.4)} & 75.0\textcolor{blue}{(-0.3)} & 82.8\textcolor{blue}{(-1.0)} &  206.0\textcolor{blue}{(-0.9)} & 3.400\textcolor{blue}{(-17.0\%)} \\
                0.9 & 47.0\textcolor{blue}{(-0.8)} & 74.1\textcolor{blue}{(-1.2)} & 83.2\textcolor{blue}{(-0.6)} & 204.3\textcolor{blue}{(-2.6)} & 3.616\textcolor{blue}{(-11.7\%)} & 46.5\textcolor{blue}{(-1.3)} & 73.8\textcolor{blue}{(-1.5)} & 83.4\textcolor{blue}{(-0.4)} &  203.7\textcolor{blue}{(-3.2)} & 3.730\textcolor{blue}{(-8.9\%)} \\
                \hline
            \end{tabular}
        }
        \begin{tablenotes}    
            \footnotesize              
            \item \qquad \qquad \qquad \qquad \qquad \qquad \qquad \qquad \qquad \qquad (b) Similarity reorganization on the frame direction      
        \end{tablenotes}           
    \end{threeparttable}  
\end{table*}

\noindent\textbf{Sentence-Frame Similarity Vector.} As shown in Table \ref{table: similarity reorganization on sentence-frame similarity vector}, it can be seen that when $r=0.1$, similarity reorganization with vanilla inattentive similarities removal achieves 206.1 T2V RSum with 99.1\% learnable parameters reduction in the linear layer of ISA. Surprisingly, the T2V RSum can be further promoted to 206.6 by inattentive similarities fusion. Although the improvement is small, there is no significant trainable parameter increase (0.009k vs 0.016k). Additionally, compared to the occasional performance decline in the inattentive similarities removal, the T2V R@1 with inattentive similarities fusion always presents positive fluctuation. We think that the inattentive similarities fusion is beneficial for temporal relationships capture and effective visual information preservation, thus generating better retrieval performance. In our experiments, we adopt inattentive similarities fusion in the sentence-frame similarity vector reorganization.

\noindent\textbf{Frame-Word Similarity Matrix.}  We conduct extensive experiments on unidirectional and bidirectional similarity reorganizations respectively. Specifically, the former is performed on the word and frame directions, and the latter contains four combinations of removal/fusion operations on the word and frame directions.


\begin{table*}[htbp]
	\caption{T2V results comparison between four variants of bidirectional similarity reorganization for the frame-word similarity matrix. The baseline results are obtained by the Bi-ISA implementation on the computed similarity matrix. The values of params indicate learnable parameter statistics in the linear layers of Bi-ISA. The number in blue is the gap of the corresponding value w.r.t. the baseline. }
	\label{table: bidirectional similarity reorganization on frame-word similarity vector}
	\centering
	\Huge
	\resizebox{\textwidth}{!}{
		\renewcommand{\arraystretch}{1.2}
		\begin{tabular}{c|c|c|c|c|ccccc}
			\hline
			\multirow{2}*{Keep Rate} & \multicolumn{4}{c|}{Similarity Reorganization} & \multirow{2}*{R@1$\uparrow$} & \multirow{2}*{R@5$\uparrow$}& \multirow{2}*{R@10$\uparrow$} & \multirow{2}*{RSum$\uparrow$} & \multirow{2}*{Params(K)} \\
			\cline{2-5} 
			& inattentive words removal & inattentive words fusion & inattentive frames removal & inattentive frames fusion \\
			\hline
			baseline & - & - & - & - & 47.8 & 75.3 & 83.8 & 206.9 & 4.096 \\
			\hline
			\multirow{4}*{0.1} & 
			~ \Checkmark & & \Checkmark & & 47.3\textcolor{blue}{(-0.5)} & \textbf{75.3}\textcolor{blue}{(-0.0)} &  83.6\textcolor{blue}{(-0.2)}  &   206.2\textcolor{blue}{(-0.7)} & \textbf{0.036}\textcolor{blue}{(-99.1\%)} \\
			~ & \Checkmark & & & \Checkmark & 47.4\textcolor{blue}{(-0.4)} & 74.3\textcolor{blue}{(-1.0)} &  83.9\textcolor{blue}{(+0.1)} & 205.6\textcolor{blue}{(-1.3)} & 0.050\textcolor{blue}{(-98.8\%)} \\
			~ & & \Checkmark & \Checkmark & & 46.5\textcolor{blue}{(-1.3)} & 74.5\textcolor{blue}{(-0.8)} &  83.3\textcolor{blue}{(-0.5)}  &  204.3\textcolor{blue}{(-2.6)} & 0.050\textcolor{blue}{(-98.8\%)} \\
			~ & & \Checkmark & & \Checkmark & 47.2\textcolor{blue}{(-0.6)} & 74.0\textcolor{blue}{(-1.3)} &  82.6\textcolor{blue}{(-1.2)}  &  203.8\textcolor{blue}{(-3.1)} & 0.064\textcolor{blue}{(-98.4\%)} \\
			\hline
			\multirow{4}*{0.2} & 
			~ \Checkmark & & \Checkmark & & 47.3\textcolor{blue}{(-0.5)} & 73.9\textcolor{blue}{(-1.4)} &  83.9\textcolor{blue}{(+0.1)}  &  205.1\textcolor{blue}{(-1.8)} & 0.144\textcolor{blue}{(-96.5\%)} \\
			~ & \Checkmark & & & \Checkmark & 46.1\textcolor{blue}{(-1.7)} & 74.4\textcolor{blue}{(-0.9)} &  84.1\textcolor{blue}{(+0.3)}  &  204.6\textcolor{blue}{(-2.3)} & 0.170\textcolor{blue}{(-95.8\%)} \\
			~ & & \Checkmark & \Checkmark & & 46.5\textcolor{blue}{(-1.3)} & 74.7\textcolor{blue}{(-0.6)} &  83.9\textcolor{blue}{(+0.1)}  &  205.1\textcolor{blue}{(-1.8)} & 0.170\textcolor{blue}{(-95.8\%)} \\
			~ & & \Checkmark & & \Checkmark & 45.8\textcolor{blue}{(-2.0)} & 74.6\textcolor{blue}{(-0.7)} &  83.8\textcolor{blue}{(-0.1)}  &  204.2\textcolor{blue}{(-2.7)} & 0.196\textcolor{blue}{(-95.2\%)} \\
			\hline
			\multirow{4}*{0.3} & 
			~ \Checkmark & & \Checkmark & & 46.7\textcolor{blue}{(-1.1)} & 74.5\textcolor{blue}{(-0.8)} &  84.4\textcolor{blue}{(+0.6)}  &  205.6\textcolor{blue}{(-1.3)} & 0.324\textcolor{blue}{(-92.1\%)} \\
			~ & \Checkmark & & & \Checkmark & 46.3\textcolor{blue}{(-1.5)} & 74.0\textcolor{blue}{(-1.3)} &  82.7\textcolor{blue}{(-1.1)}  &  203.0\textcolor{blue}{(-3.9)} & 0.362\textcolor{blue}{(-91.2\%)} \\
			~ & & \Checkmark & \Checkmark & & 47.0\textcolor{blue}{(-0.8)} & 74.4\textcolor{blue}{(-0.9)} &  82.6\textcolor{blue}{(-1.2)}  &  204.0\textcolor{blue}{(-2.9)} & 0.362\textcolor{blue}{(-91.2\%)} \\
			~ & & \Checkmark & & \Checkmark & 47.0\textcolor{blue}{(-0.8)} & 73.8\textcolor{blue}{(-1.5)} &  83.1\textcolor{blue}{(-0.7)}  &  203.9\textcolor{blue}{(-3.0)} & 0.400\textcolor{blue}{(-90.2\%)} \\
			\hline
			\multirow{4}*{0.4} & 
			~ \Checkmark & & \Checkmark & & 47.6\textcolor{blue}{(-0.2)} & 74.8\textcolor{blue}{(-0.5)} &  84.0\textcolor{blue}{(+0.2)}  &  206.4\textcolor{blue}{(-0.5)} & 0.576\textcolor{blue}{(-85.9\%)}\\
			~ & \Checkmark & & & \Checkmark & 46.9\textcolor{blue}{(-0.9)} & 75.1\textcolor{blue}{(-0.2)} &  83.7\textcolor{blue}{(-0.1)}  &  205.7\textcolor{blue}{(-1.2)} & 0.626\textcolor{blue}{(-84.7\%)} \\
			~ & & \Checkmark & \Checkmark & & 47.0\textcolor{blue}{(-0.8)} & 74.0\textcolor{blue}{(-1.3)} &  83.1\textcolor{blue}{(-0.7)}  &  204.1\textcolor{blue}{(-2.8)} & 0.626\textcolor{blue}{(-84.7\%)} \\
			~ & & \Checkmark & & \Checkmark & 46.9\textcolor{blue}{(-0.9)} & 73.1\textcolor{blue}{(-2.2)} &  82.8\textcolor{blue}{(-1.0)}  &  202.8\textcolor{blue}{(-4.1)} & 0.676\textcolor{blue}{(-83.5\%)} \\
			\hline
			\multirow{4}*{0.5} & 
			~ \Checkmark & & \Checkmark & & 46.3\textcolor{blue}{(-1.5)} & 75.0\textcolor{blue}{(-0.3)} &  83.9\textcolor{blue}{(+0.1)}  &  205.2\textcolor{blue}{(-1.7)} & 1.024\textcolor{blue}{(-75.0\%)} \\
			~ & \Checkmark & & & \Checkmark & 46.7\textcolor{blue}{(-1.1)} & 74.1\textcolor{blue}{(-1.2)} &  83.2\textcolor{blue}{(-0.6)}  &  204.0\textcolor{blue}{(-2.9)} & 1.090\textcolor{blue}{(-73.4\%)} \\
			~ & & \Checkmark & \Checkmark & & 46.8\textcolor{blue}{(-1.0)} & 74.0\textcolor{blue}{(-1.3)} &  83.3\textcolor{blue}{(-0.5)}  &  204.1\textcolor{blue}{(-2.8)} & 1.090\textcolor{blue}{(-73.4\%)} \\
			~ & & \Checkmark & & \Checkmark & \textbf{47.9}\textcolor{blue}{(+0.1)} & 74.1\textcolor{blue}{(-1.2)} &  84.4\textcolor{blue}{(+0.6)}  &   206.4\textcolor{blue}{(-0.5)} & 1.156\textcolor{blue}{(-71.8\%)} \\
			\hline
			\multirow{4}*{0.6} & 
			~ \Checkmark & & \Checkmark & & 47.1\textcolor{blue}{(-0.7)} & 73.5\textcolor{blue}{(-1.8)} &  83.3\textcolor{blue}{(-0.5)}  &  203.9\textcolor{blue}{(-3.0)} & 1.444\textcolor{blue}{(-64.7\%)} \\
			~ & \Checkmark & & & \Checkmark & 47.1\textcolor{blue}{(-0.7)} & 74.8\textcolor{blue}{(-0.5)} &  83.9\textcolor{blue}{(+0.1)}  &  205.8\textcolor{blue}{(-1.1)} & 1.522\textcolor{blue}{(-62.8\%)} \\
			~ & & \Checkmark & \Checkmark & & 46.7\textcolor{blue}{(-1.1)} & 73.9\textcolor{blue}{(-1.4)} &  83.3\textcolor{blue}{(-0.5)}  &  203.9\textcolor{blue}{(-3.0)} & 1.522\textcolor{blue}{(-62.8\%)} \\
			~ & & \Checkmark & & \Checkmark & 46.9\textcolor{blue}{(-0.9)} & 74.4\textcolor{blue}{(-0.9)} &  84.0\textcolor{blue}{(+0.2)}  &  205.3\textcolor{blue}{(-1.6)} & 1.600\textcolor{blue}{(-60.9\%)} \\
			\hline
			\multirow{4}*{0.7} & 
			~ \Checkmark & & \Checkmark & & 45.6\textcolor{blue}{(-2.2)} & 74.5\textcolor{blue}{(-0.8)} &  84.1\textcolor{blue}{(+0.3)}  &  204.2\textcolor{blue}{(-2.7)} & 1.936\textcolor{blue}{(-52.7\%)} \\
			~ & \Checkmark & & & \Checkmark & 47.8\textcolor{blue}{(-0.0)} & 74.1\textcolor{blue}{(-1.2)} &  82.1\textcolor{blue}{(-1.7)}  &  204.0\textcolor{blue}{(-2.9)} & 2.026\textcolor{blue}{(-50.5\%)} \\
			~ & & \Checkmark & \Checkmark & & 46.9\textcolor{blue}{(-0.9)} & 73.5\textcolor{blue}{(-1.8)} &  82.8\textcolor{blue}{(-1.0)}  &  203.2\textcolor{blue}{(-3.7)} & 2.026\textcolor{blue}{(-50.5\%)} \\
			~ & & \Checkmark & & \Checkmark & 47.5\textcolor{blue}{(-0.3)} & 74.5\textcolor{blue}{(-0.8)} &  \textbf{84.5}\textcolor{blue}{(+0.7)}  &  \textbf{206.5}\textcolor{blue}{(-0.4)} & 2.116\textcolor{blue}{(-48.3\%)}  \\
			\hline
			\multirow{4}*{0.8} & 
			~ \Checkmark & & \Checkmark & & 46.6\textcolor{blue}{(-1.2)} & 73.9\textcolor{blue}{(-1.4)} &  83.3\textcolor{blue}{(-0.5)}  &  203.8\textcolor{blue}{(-3.1)} & 2.500\textcolor{blue}{(-39.0\%)} \\
			~ & \Checkmark & & & \Checkmark & 46.4\textcolor{blue}{(-1.4)} & 73.8\textcolor{blue}{(-1.5)} &  83.0\textcolor{blue}{(-0.8)}  &  203.2\textcolor{blue}{(-3.7)} & 2.602\textcolor{blue}{(-36.5\%)} \\
			~ & & \Checkmark & \Checkmark & & 46.8\textcolor{blue}{(-1.0)} & 75.0\textcolor{blue}{(-0.3)} &  83.1\textcolor{blue}{(-0.7)}  &  204.9\textcolor{blue}{(-2.0)} & 2.602\textcolor{blue}{(-36.5\%)} \\
			~ & & \Checkmark & & \Checkmark & 45.6\textcolor{blue}{(-2.2)} & 73.5\textcolor{blue}{(-1.8)} &  83.9\textcolor{blue}{(+0.1)}  &  203.0\textcolor{blue}{(-3.9)} & 2.704\textcolor{blue}{(-34.0\%)} \\
			\hline
			\multirow{4}*{0.9} & 
			~ \Checkmark & & \Checkmark & & 46.9\textcolor{blue}{(-0.9)} & 74.2\textcolor{blue}{(-1.1)} &  83.7\textcolor{blue}{(-0.1)}  &  204.8\textcolor{blue}{(-2.1)} & 3.136\textcolor{blue}{(-23.4\%)} \\
			~ & \Checkmark & & & \Checkmark & 46.3\textcolor{blue}{(-1.5)} & 73.9\textcolor{blue}{(-1.4)} &  82.9\textcolor{blue}{(-0.9)}  &  203.1\textcolor{blue}{(-3.8)} & 3.250\textcolor{blue}{(-20.7\%)} \\
			~ & & \Checkmark & \Checkmark & & 47.3\textcolor{blue}{(-0.5)} & 73.6\textcolor{blue}{(-1.7)} &  82.9\textcolor{blue}{(-0.9)}  &  203.8\textcolor{blue}{(-3.1)} & 3.250\textcolor{blue}{(-20.7\%)} \\
			~ & & \Checkmark & & \Checkmark & 47.7\textcolor{blue}{(-0.1)} & 74.7\textcolor{blue}{(-0.6)} &  83.6\textcolor{blue}{(-0.2)}  &  206.0\textcolor{blue}{(-0.9)} & 3.364\textcolor{blue}{(-17.9\%)} \\
			\hline
		\end{tabular}
	}
\end{table*}

Table \ref{table: unidirectional similarity reorganization on frame-word similarity vector} (a) and (b) report the retrieval performance of two unidirectional similarity reorganization mechanisms. In the top table (a), we observe that although both SR variants can achieve the best T2V R@1 of 48.3 and 48.0 with slight improvements, the best overall T2V RSum of 206.0 and 205.5 are still lower than the baseline results. Similarly, we notice a slight performance decrease (\textit{i.e.,} 205.7/206.5 RSum vs 206.9 RSum) in the bottom table (b). The main reason may come from the similarity matrix's sensitivity to information completeness, and the unidirectional similarity reorganization inevitably brings information loss. Besides, we find that the removal operation is superior to the fusion operation on the word direction, while the fusion operation is a better alternative for the frame direction. This phenomenon is consistent with the fusion/removal results in video-word and sentence-frame similarity vectors reorganization.

Table \ref{table: bidirectional similarity reorganization on frame-word similarity vector} shows results for bidirectional similarity reorganization. By analyzing the table, it can be seen that the inconsistent similarity reorganization on word and frame directions may harm the retrieval performance. We suppose that the inconsistency may result in the mismatch problem of bidirectional information density, and increase the risk of under- and over-representations at textual or visual levels. Moreover, we observe that the bidirectional inattentive similarities removal performs better at smaller keep rates, while the inattentive similarities fusion adapts to larger keep rates better. The main reason can be explained from two perspectives. On the one hand, the fusion operation at small keep rates may concatenate semantically distant similarities together and bring excessive noisy interactions. On the other hand, large keep rates in the removal operation can not greatly decrease the similarity redundancy. Therefore, we leverage the bidirectional inattentive similarities removal to reorganize the frame-word similarity matrix.

\begin{table}[htbp]
	\caption{The correspondences of Exp1-Exp15 and different granularity. For simplicity, video-sent and sent-frame stand for video-sentence and sentence-frame, respectively.}
	\label{table: Exp1-Exp15 Correspondence}
	\centering
	\resizebox{0.5\textwidth}{!}{
		\begin{tabular}{l|c|c|c|c}
			\hline
			\multirow{2}*{ID} & \multicolumn{4}{c}{Contrastive Module} \\ 
			\cline{2-5} 
			& Video-Sent & Video-Word & Sent-Frame & Frame-Word \\
			\hline
			Exp1 & \Checkmark & &  \\
			Exp2 & & \Checkmark & & \\
			Exp3 & & & \Checkmark & \\
			Exp4 & & & & \Checkmark  \\
			\hline
			Exp5 & \Checkmark & \Checkmark & \\
			Exp6 & \Checkmark & & \Checkmark \\
			Exp7 & \Checkmark & & & \Checkmark \\
			Exp8 & & \Checkmark & \Checkmark \\
			Exp9 & & \Checkmark & & \Checkmark \\
			Exp10 & & & \Checkmark & \Checkmark \\
			\hline
			Exp11 & \Checkmark & \Checkmark & \Checkmark \\
			Exp12 & \Checkmark & \Checkmark & & \Checkmark \\
			Exp13 & \Checkmark & & \Checkmark & \Checkmark \\
			Exp14 & & \Checkmark & \Checkmark & \Checkmark \\
			\hline
			Exp15 & \Checkmark & \Checkmark & \Checkmark & \Checkmark \\ 
			\hline
		\end{tabular}
	}
\end{table}


\subsubsection{Similarity Decorrelation Regularization}
\begin{figure}[!t]
    \centerline{\includegraphics[width=0.5\textwidth]{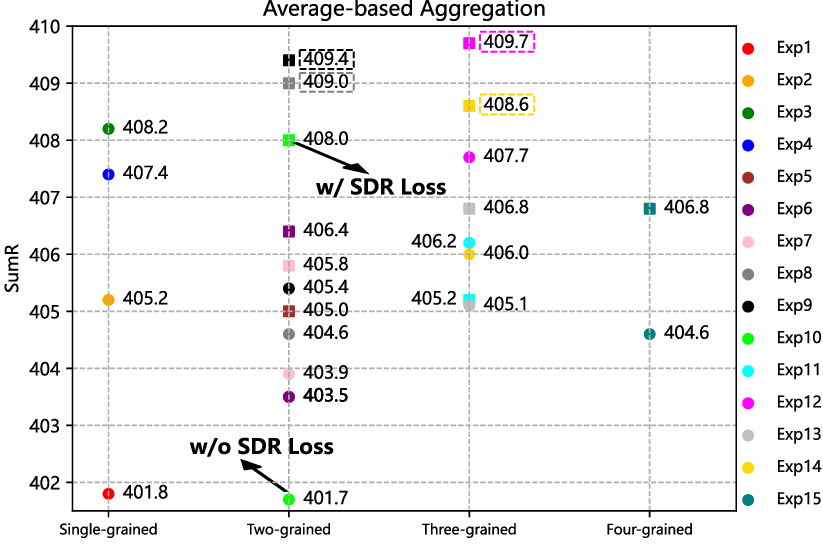}}
    \caption{Comparison of SumR with different granularity under averaged-based aggregation. The circle and square markers denote w/o and w/ SDR loss, respectively. The correspondences of Exp1-Exp15 and multi-grained contrasts are shown in Table \ref{table: Exp1-Exp15 Correspondence}. }
    \label{fig: visualization of w/ and wo/ sdr}
\end{figure}

To demonstrate the strength of the Similarity Decorrelation Regularization (SDR) loss, we also conduct an ablation study to compare different variants of TC-MGC with and without the SDR loss, and plot the comprehensive SumR comparison in Fig. \ref{fig: visualization of w/ and wo/ sdr}. Notably, the single-grained contrast results are obtained by the above SR and Bi-SR modules with keep rate $r=0.1$. For simplicity, we adopt the common average-based method to aggregate multi-grained similarity scores. From the retrieval results comparison, we observe that the auxiliary SDR loss typically brings significant performance gain. Specifically, TC-MGC with the sentence-frame and frame-word contrastive modules (\textit{i.e.}, Exp10) only achieves 401.7 SumR. However, when the SDR loss is utilized in the model optimization, the SumR can be promoted to 408.0 with +6.3\% absolute improvement. Similarly, the better SumR of 408.6, 406.8 are obtained in Exp14 and Exp15, outperforming the retrieval results without SDR loss by +2.6\% and +2.2\% absolute performance boost. Moreover, TC-MGC equipped with the SDR loss achieves better SumR (\textit{i.e.,} 409.7 SumR in Exp12) than the best single-grained SumR of 408.2 (\textit{i.e.,} Exp3). We suppose that the SDR loss can effectively alleviate over- and under-representation issues brought by the serious imbalance problem, thus generating better retrieval performance.

\begin{figure}[!t]
	\centerline{\includegraphics[width=0.5\textwidth]{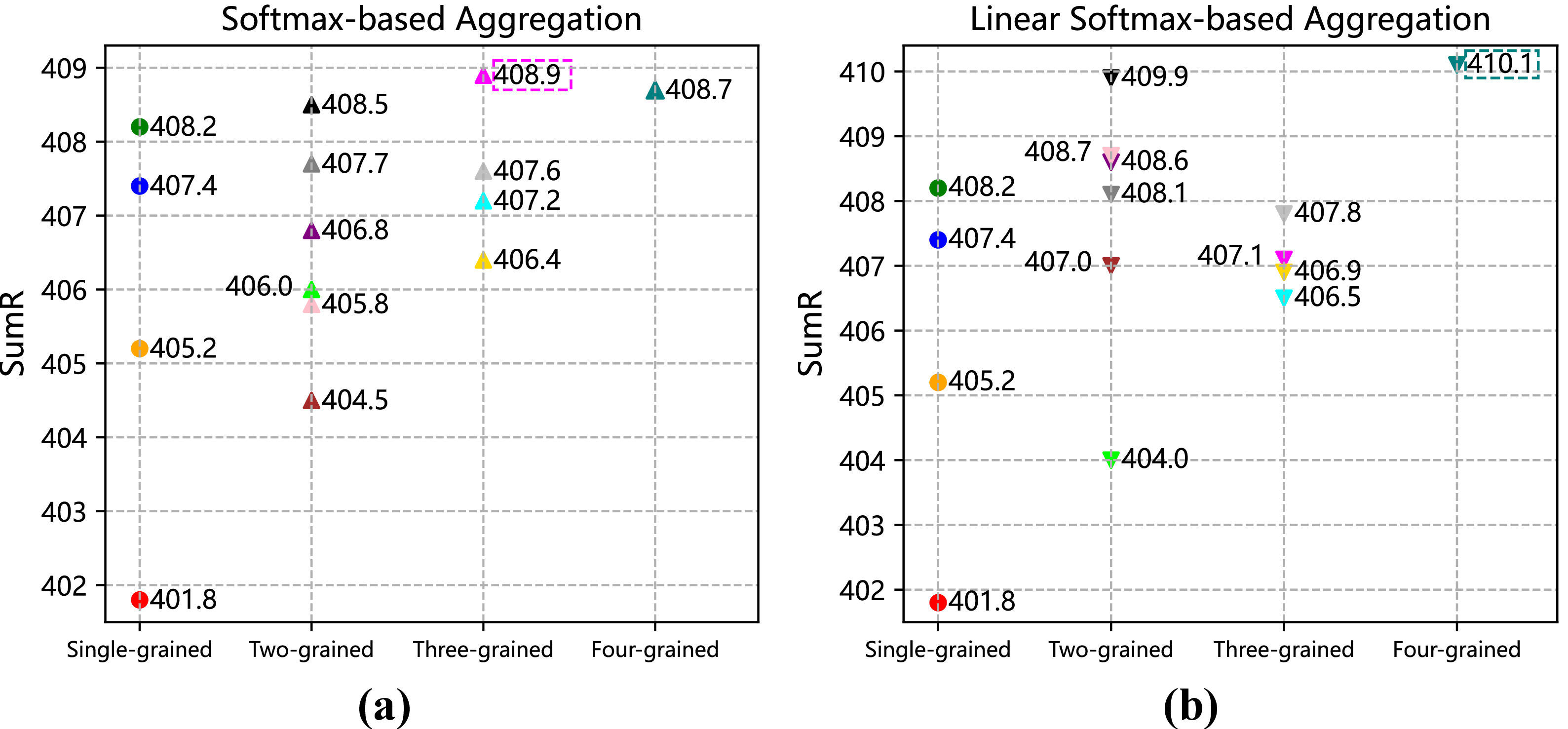}}
	\caption{(a) Comparison of SumR with different granularity under softmax-based aggregation. (b) Comparison of SumR with different granularity under linear softmax-based aggregation. The correspondences of marker color and experiment ID are given in the bottom part of Fig. \ref{fig: visualization of w/ and wo/ sdr}. The circle and triangle markers denote w/o and w/ SDR loss, respectively. }
	\label{fig: visualization of different aggregation}
\end{figure}

\subsubsection{Linear Softmax Aggregation} 
To verify the effectiveness of LSA module in multi-grained scores aggregation, we compare our model with the variant that uses softmax-based approach. Notably, the aforementioned SDR loss is applied in the multi-grained contrasts. The overall retrieval results SumR with different granularity are shown in Fig. \ref{fig: visualization of different aggregation}. With the number of contrastive modules increasing, the LSA module tends to achieve higher SumR. More specifically, compared to the lower SumR of 404.5 in Exp5, the employment of linear softmax-based aggregation can promote the SumR to 407.0 with +2.5\% absolute performance gain. Similarly, the better SumR of 408.7 is obtained in Exp7, outperforming the softmax-based result by +2.9\% absolute performance improvement. We think that the LSA module facilitates the information interactions among different scores, thus generating better retrieval performance.

Surprisingly, we observe that compared to 409.9, 410.1 SumR in Exp9 and Exp15, Exp12 and Exp14 achieve lower SumR of 407.1 and 406.9, respectively. The inferior results may result from the discrepancy of optimization difficulty. As shown in Fig. \ref{fig: qualitative results of sdr visualization}, video-word and sentence-frame similarities are in a larger magnitude of 1e-2, while the remaining similarities are only in a smaller magnitude of 1e-4. As a result, Exp9 and Exp15 generate larger initialized variances in the SDR loss optimization, thus promoting the difficulty of model optimization and bringing better performance. However, the smaller initialized variances in Exp12 and Exp14 greatly decrease the optimization difficulty, thus resulting in the significant performance loss.

\begin{table}[!t]
    \caption{Ablation study of temporal encoder. TE is short for temporal encoder.}
    \label{table: ablation study of temporal encoder}
    \centering
    \resizebox{0.5\textwidth}{!}{
        \Huge
        \renewcommand{\arraystretch}{1.1}
        \begin{tabular}{c|c|cccc|cccc|c}
            \hline
            \multirow{2}*{\makecell[c]{Base\\Model}} & \multirow{2}*{TE} & \multicolumn{4}{c|}{Text $\rightarrow$ Video} &  \multicolumn{4}{c|}{Video $\rightarrow$ Text} & \multirow{2}*{SumR$\uparrow$} \\ 
            \cline{3-6} \cline{7-10}
            && R@1$\uparrow$  & R@5$\uparrow$  & R@10$\uparrow$ & RSum$\uparrow$
            & R@1$\uparrow$  & R@5$\uparrow$  & R@10$\uparrow$ & RSum$\uparrow$\\
            \hline
            \multirow{2}*{ViT-B/32} & & 45.1 & 72.3 & 82.7 & 200.1 & 44.7 & 71.9 & 83.4 & 200.0 & 400.1 \\ 
            ~ & \Checkmark & \textbf{47.4} & \textbf{74.8} & \textbf{84.2} & \textbf{206.4} & \textbf{45.9} & \textbf{74.5} & \textbf{83.3} & \textbf{203.7} & \textbf{410.1} \\
            \hline
            \multirow{2}*{ViT-B/16} & & 47.5 & 75.1 & 84.2 & 206.8 & 45.5 & 75.0 & 85.2 & 205.7 & 412.5 \\ 
            ~ &  \Checkmark & \textbf{49.0} & \textbf{75.7} & \textbf{85.4} & \textbf{210.1} & \textbf{46.4} & \textbf{77.1} & \textbf{85.3} & \textbf{208.8} & \textbf{418.9} \\
            \hline
        \end{tabular}
    }
\end{table}

\subsubsection{Temporal Encoder} To examine the impact of the temporal encoder in TC-MGC, we compare the TC-MGC with and without the temporal encoder shown in Table \ref{table: ablation study of temporal encoder}. It can be found that based on either ViT-B/32 or ViT-B/16, TC-MGC with the temporal encoder consistently outperforms TC-MGC without the temporal encoder by considerable margins on SumR (+10.0\% and +6.4\% absolute improvements), which demonstrates that the temporal encoder is a key design to improving retrieval performance. We deem the reason is that the temporal encoder can effectively perceive the temporal relation across video frames, thus bringing significant performance gain.

\begin{table}[!t]
	\caption{Ablation study of residual connection to the FC layer. RC is short residual connection.}
	\label{table: ablation study of residual connection}
	\centering
	\resizebox{0.5\textwidth}{!}{
		\Huge
		\renewcommand{\arraystretch}{1.1}
		\begin{tabular}{c|c|cccc|cccc|c}
			\hline
			\multirow{2}*{\makecell[c]{Base\\Model}} & \multirow{2}*{RC} & \multicolumn{4}{c|}{Text $\rightarrow$ Video} &  \multicolumn{4}{c|}{Video $\rightarrow$ Text} & \multirow{2}*{SumR$\uparrow$} \\ 
			\cline{3-6} \cline{7-10}
			&& R@1$\uparrow$  & R@5$\uparrow$  & R@10$\uparrow$ & RSum$\uparrow$
			& R@1$\uparrow$  & R@5$\uparrow$  & R@10$\uparrow$ & RSum$\uparrow$\\
			\hline
			\multirow{2}*{ViT-B/32} & & 44.5 & 73.2 & 82.6 & 200.3 & 41.3 & 72.3 & 81.8 & 195.4 & 395.7 \\ 
			~ & \Checkmark & \textbf{47.4} & \textbf{74.8} & \textbf{84.2} & \textbf{206.4} & \textbf{45.9} & \textbf{74.5} & \textbf{83.3} & \textbf{203.7} & \textbf{410.1} \\
			\hline
			\multirow{2}*{ViT-B/16} & & 47.1 & 74.4 & 83.2 & 204.7 & 44.8 & 75.0 & 83.6 & 203.4 & 408.1 \\ 
			~ &  \Checkmark & \textbf{49.0} & \textbf{75.7} & \textbf{85.4} & \textbf{210.1} & \textbf{46.4} & \textbf{77.1} & \textbf{85.3} & \textbf{208.8} & \textbf{418.9} \\
			\hline
		\end{tabular}
	}
\end{table}

\begin{table}[!t]
	\caption{Retrieval results with different attention heads in language-video attention block.}
	\label{table: ablation study of different number of heads in cross-attn block}
	\centering
	\resizebox{0.5\textwidth}{!}{
		\Huge
		\renewcommand{\arraystretch}{1.1}
		\begin{tabular}{c|cccc|cccc|c}
			\hline
			\multirow{2}*{\makecell[c]{Attention\\Heads}} & \multicolumn{4}{c|}{Text $\rightarrow$ Video} &  \multicolumn{4}{c|}{Video $\rightarrow$ Text} & \multirow{2}*{SumR$\uparrow$}\\ 
			\cline{2-5} \cline{6-9}
			& R@1$\uparrow$  & R@5$\uparrow$  & R@10$\uparrow$ & RSum$\uparrow$
			& R@1$\uparrow$  & R@5$\uparrow$  & R@10$\uparrow$ & RSum$\uparrow$\\
			\hline
			1 & \textbf{47.4} & \textbf{74.8} & \textbf{84.2} & \textbf{206.4} & \textbf{45.9} & \textbf{74.5} & \textbf{83.3} & \textbf{203.7} & \textbf{410.1} \\
			2 & \textbf{47.4} & 73.9 & 83.3 & 204.6 & 45.1 & 73.6 & 82.8 & 201.5 & 406.1 \\
			4 & 45.3 & 74.0 & 82.4 & 201.7 & 45.1 & 73.4 & 82.3 & 200.8 & 402.5 \\
			8 & 45.1 & 73.6 & 82.4 & 201.1 & 44.1 & 73.0 & 82.8 & 199.9 & 401.0 \\
			\hline
		\end{tabular}
	}
\end{table}

\begin{table}[!t]
	\caption{Retrieval results with different temperature parameters $\lambda$ in softmax.}
	\label{table: ablation study of different temperature parameters in softmax}
	\centering
	\resizebox{0.5\textwidth}{!}{
		\Huge
		\renewcommand{\arraystretch}{1.1}
		\begin{tabular}{c|cccc|cccc|c}
			\hline
			\multirow{2}*{$\lambda$} & \multicolumn{4}{c|}{Text $\rightarrow$ Video} &  \multicolumn{4}{c|}{Video $\rightarrow$ Text} & \multirow{2}*{SumR$\uparrow$}\\ 
			\cline{2-5} \cline{6-9}
			& R@1$\uparrow$  & R@5$\uparrow$  & R@10$\uparrow$ & RSum$\uparrow$
			& R@1$\uparrow$  & R@5$\uparrow$  & R@10$\uparrow$ & RSum$\uparrow$ \\
			\hline
			1 & 46.3 & 75.1 & 84.2 & 205.6 & 45.7 & 73.7 & 82.8 & 202.2 & 407.8 \\
			10 & 46.5 & \textbf{75.4} & 83.9 & 205.8 & 45.6 & 74.3 & \textbf{83.6} & 203.5 & 409.3 \\
			100 & \textbf{47.4} & 74.8 & 84.2 & \textbf{206.4} & \textbf{45.9} & \textbf{74.5} & 83.3 & \textbf{203.7} & \textbf{410.1} \\
			1000 & 46.1 & 74.1 & \textbf{84.3} & 204.5 & \textbf{45.9} & 73.8 & 83.3 & 203.0 & 407.5 \\
			\hline
		\end{tabular}
 	}
\end{table}

\subsubsection{Language-Video Attention} To explore the impact of residual connection and attention heads in the language-video attention block, we first design an ablative study to quantitatively compare the TC-MGC with and without the residual connection, followed by the ablation experiments at various attention heads. 

\textbf{Residual Connection:} As shown in Table \ref{table: ablation study of residual connection}, TC-MGC without the residual connection suffers significant performance decrease based on either ViT-B/32 or ViT-B/16. For example, when the residual connection is not applied to the FC layer, TC-MGC with ViT-B/32 achieves lower T2V and V2T R@1 of 44.5 and 41.3, leading to 2.9\% and 4.6\% absolute performance decline. This may be because the residual connection to the FC layer provides additional capacity for more complex reasoning in the aggregation function. As a result, TC-MGC without the residual connection can not implement complex reasoning between textual and visual information, resulting in the under-representation of generated text-conditioned visual representations in cross-modal semantic alignment. 

\textbf{Attention Heads:} From Table \ref{table: ablation study of different number of heads in cross-attn block}, we observe that the single attention head achieves the best retrieval performance (\textit{i.e.,} 410.1 SumR), and the experimental results present monotonically decrease with more attention heads. The reason may be attributed to two aspects. On the one hand, 
single-head attention can focus on learning the precise cross-modal correspondence without being interfered by different heads in the multi-head attention. On the other hand, single-head attention can better capture global correlations, which may be more essential than local differences for text and video semantics alignment. 

\begin{table}[!t]
    \caption{Efficiency analysis of training time, parameters, FLOPs and memory usage. Here, CLIP4Clip refers to CLIP4Clip-seqTransf and X-CLIP serves as the baseline method. For a fair comparison, all models employ CLIP-ViT-B/32 with 64 mini-batch sizes. Experiments are performed with a 24GB NVIDIA GeForce RTX 3090 GPU.  }
    \label{table: ablation study of efficiecny analysis}
    \centering
    \resizebox{0.5\textwidth}{!}{
        \normalsize
        \renewcommand{\arraystretch}{1.1}
        \begin{tabular}{l|ccccc}
            \hline
            Methods & \makecell[c]{Training\\Time} & Parameters & FLOPs & \makecell[c]{Memory\\Usage} & SumR \\
            \hline
            CLIP4Clip \cite{luo2022clip4clip} & 3h 18min & 92.62M & \textbf{36.27G} & 2869.89MB & 391.7 \\
            X-Pool \cite{gorti2022x} & 3h 15min & \textbf{85.54M} & 37.34G & \textbf{2837.16MB} & 403.6 \\
            DRL \cite{wang2022disentangled} & 3h 10min & 93.15M & 36.28G & 2940.17MB & 406.4 \\
            UCoFiA \cite{wang2023unified} & \textbf{2h 21min} & 92.89M & 36.43G & 3037.43MB & 402.0 \\
            X-CLIP \cite{ma2022x} (base) & 3h 32min & 92.62M & \textbf{36.27G} & 2942.48MB & 406.3 \\    
            TC-MGC (ours) & 2h 37min & 92.89M & 37.39G & 2947.50MB & \textbf{410.1} \\
            \hline
        \end{tabular}
    }
\end{table}

\subsubsection{Temperature Parameter Analysis} To explore the effect of different $\lambda$, we also perform a group of ablation experiments by setting different temperature parameters $\lambda$ in softmax. As shown in Table \ref{table: ablation study of different temperature parameters in softmax}, the retrieval performance presents significant improvement with $\lambda$ increasing, and reaches a saturation point at $\lambda=100$, achieving 47.4 T2V R@1 and 206.4 T2V RSum. We suppose that when $\lambda$ is too small, the loss function tends to impose slight punishment on noisy features and hinder model from learning powerful discrimination capability.  We also observe that when $\lambda=1000$, the performance drops seriously with 2.6\% absolute decrease, which may be ascribed to the underestimation of certain critical features. As a result, we adopt $\lambda=100$ in our model to obtain the best retrieval performance.

\subsubsection{Efficiency Analysis} We provide detailed efficiency comparisons of our TC-MGC and recent methods including training time, parameters, FLOPs, and memory usage as shown in Table \ref{table: ablation study of efficiecny analysis}. Note that all efficiency metrics except training time are refined from integer to two decimal places for a more rigorous comparison. To further validate the effectiveness of our method, we add a column of SumR comparison to the Table \ref{table: ablation study of efficiecny analysis}. By analyzing the table, we gain the following key observations:  
i) In terms of training time, TC-MGC is slower than UCoFiA but still outperforms other methods, securing the second position. This may be because the SR module, which preserves attentive similarities in the cross-modal similarity vectors/matrices, can ease the training computational overhead and promote the computational efficiency. ii) Compared to X-CLIP, TC-MGC obtains better SumR of 410.1 while bringing acceptable parameters and FLOPs growth (+0.3\% and +3.1\% relative improvements) that are attributed to the introduced language-video attention block. iii) Moreover, despite a slight increase in memory usage than the baseline method, our overall metric SumR is significantly better. The memory usage growth can be categorized into two aspects. On the one hand, the computed attention weight matrix in the language-video attention block requires storage. On the other hand, the SR module needs to save index information for attentive similarities selection. 

\subsection{Qualitative Analysis}
To further understand and analyze our proposed method, we visualize several text-to-video retrieval results and the word and frame features.

\begin{figure*}[!t]
    \centerline{\includegraphics[width=0.83\textwidth]{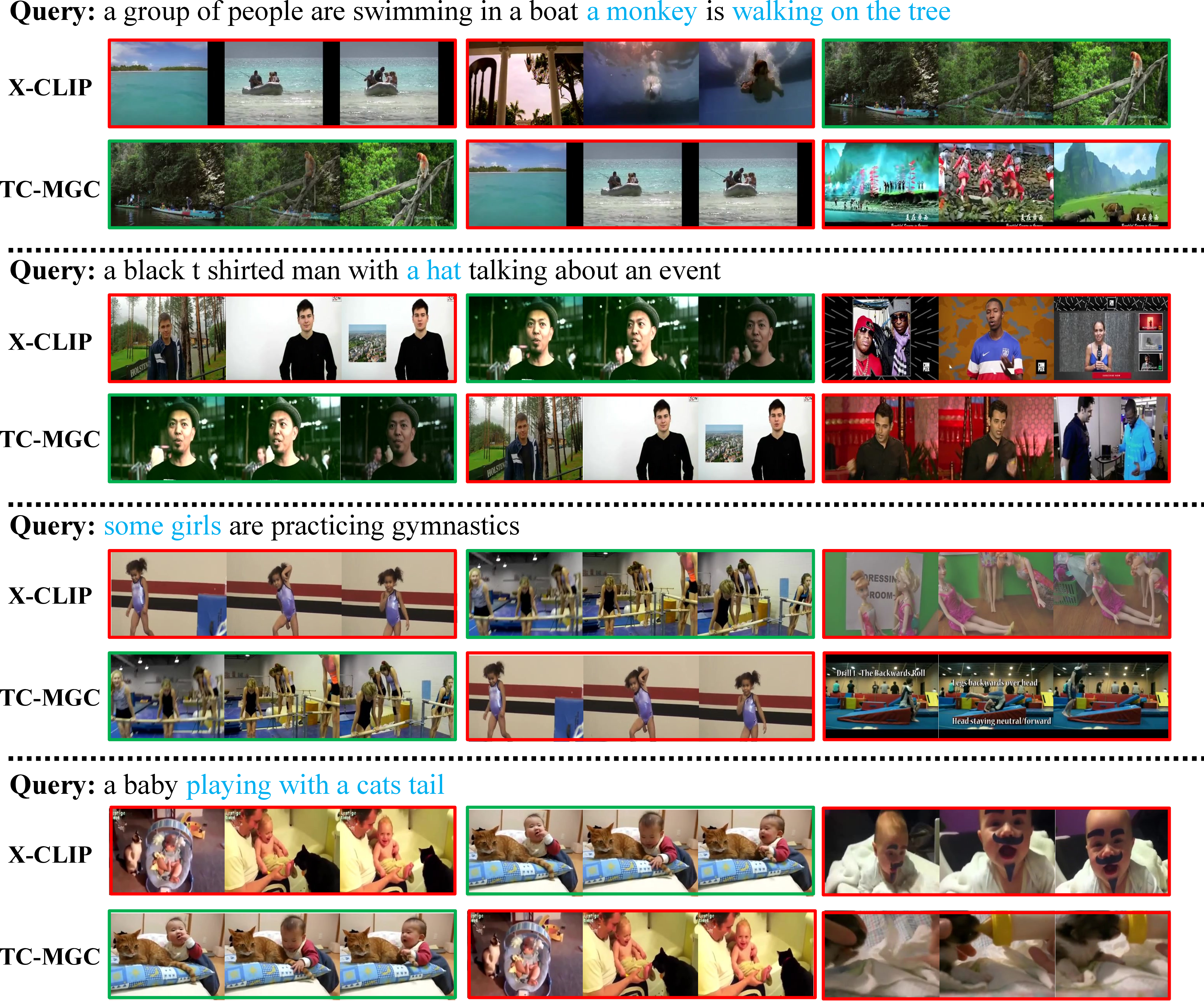}}
    \caption{Visualization of text-to-video retrieval examples on the MSR-VTT dataset. Given the text query, the top-3 retrieved videos are displayed, with ground-truth in green and others in red. Note that the related words are highlighted in blue. }
    \label{fig: retrieval_results_visualization}
\end{figure*}

\begin{figure}[!t]
    \centerline{\includegraphics[width=0.5\textwidth]{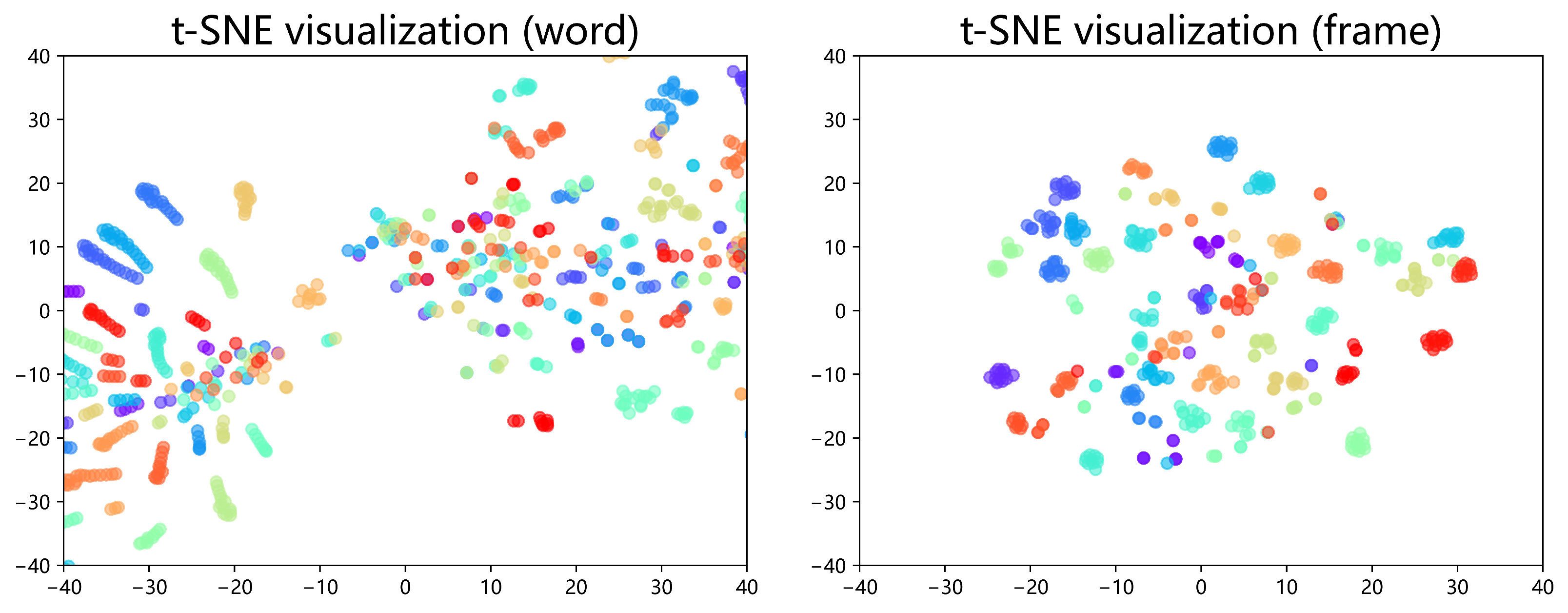}}
    \caption{t-SNE visualization of the word and frame features. Left: word, Right: frame. }
    \label{fig: t_SNE_visualization}
\end{figure}

\begin{figure*}[!t]
    \centerline{\includegraphics[width=\textwidth]{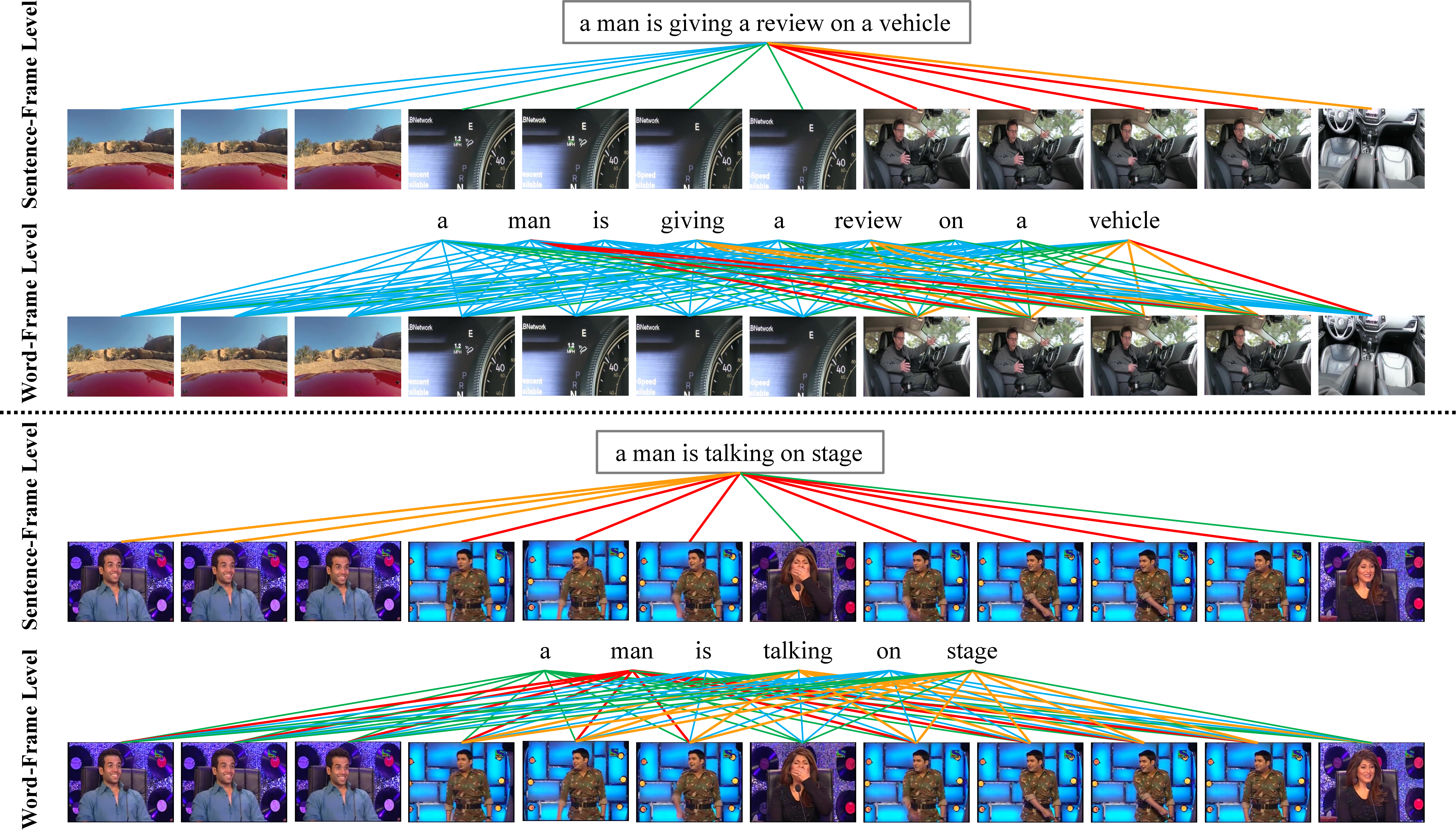}}
    \caption{Visualization of the multi-grained attention. We take video7026 and video8919 in the MSR-VTT dataset as  examples. Based on the attention weights derived from the language-video attention block, a connection is established between the sentence/word and the frame that has different degree of association with it. Note that the degree of relevance from high to low is represented by red, orange, green and blue lines, respectively.}
    \label{fig: lva_visualization}
\end{figure*}

\begin{figure*}[!t]
    \centerline{\includegraphics[width=0.8\textwidth]{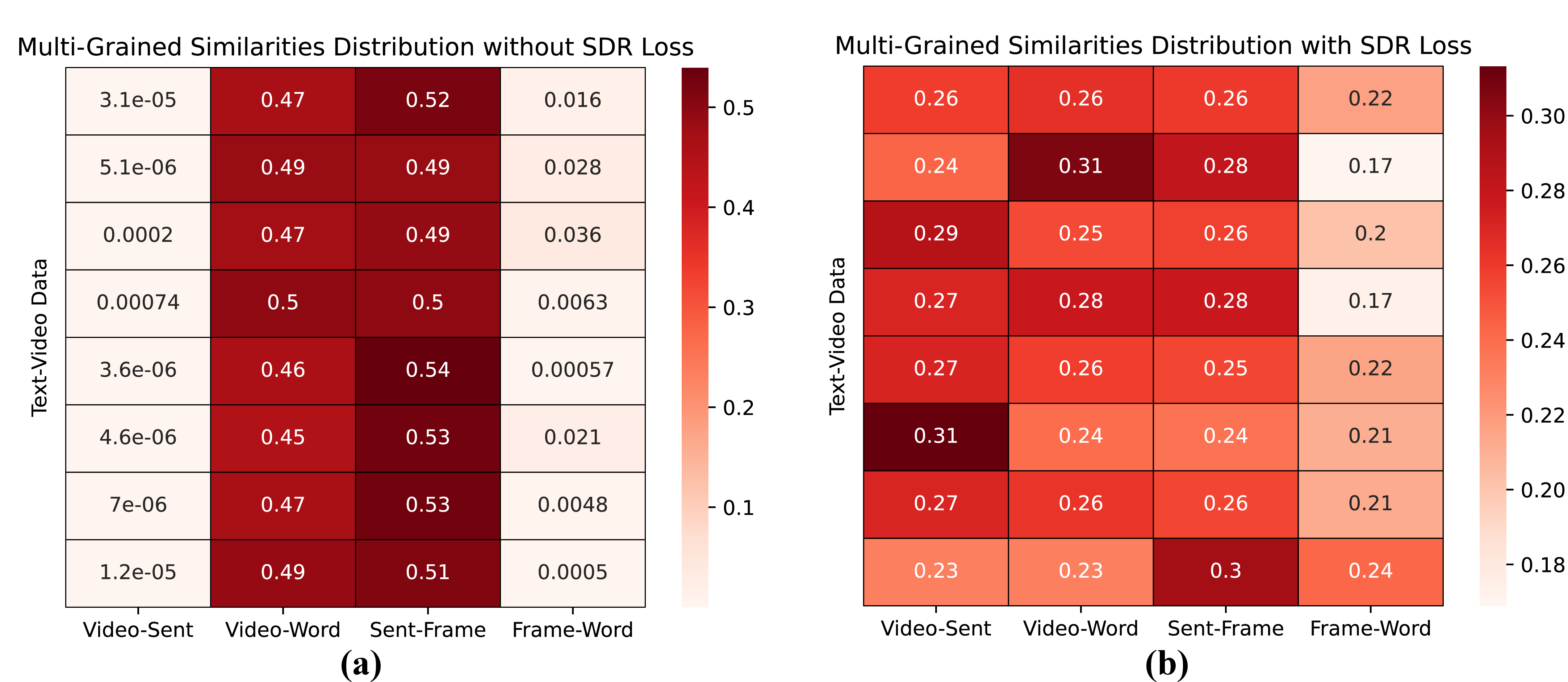}}
    \caption{(a) Qualitative results of the normalized multi-grained similarities distribution w/o SDR loss for matching text-video pairs. (b) Qualitative results of the normalized multi-grained similarities distribution w/ SDR loss for matching text-video pairs.}
    \label{fig: qualitative results of sdr visualization}
\end{figure*}

\begin{figure*}[!t]
    \centerline{\includegraphics[width=0.8\textwidth]{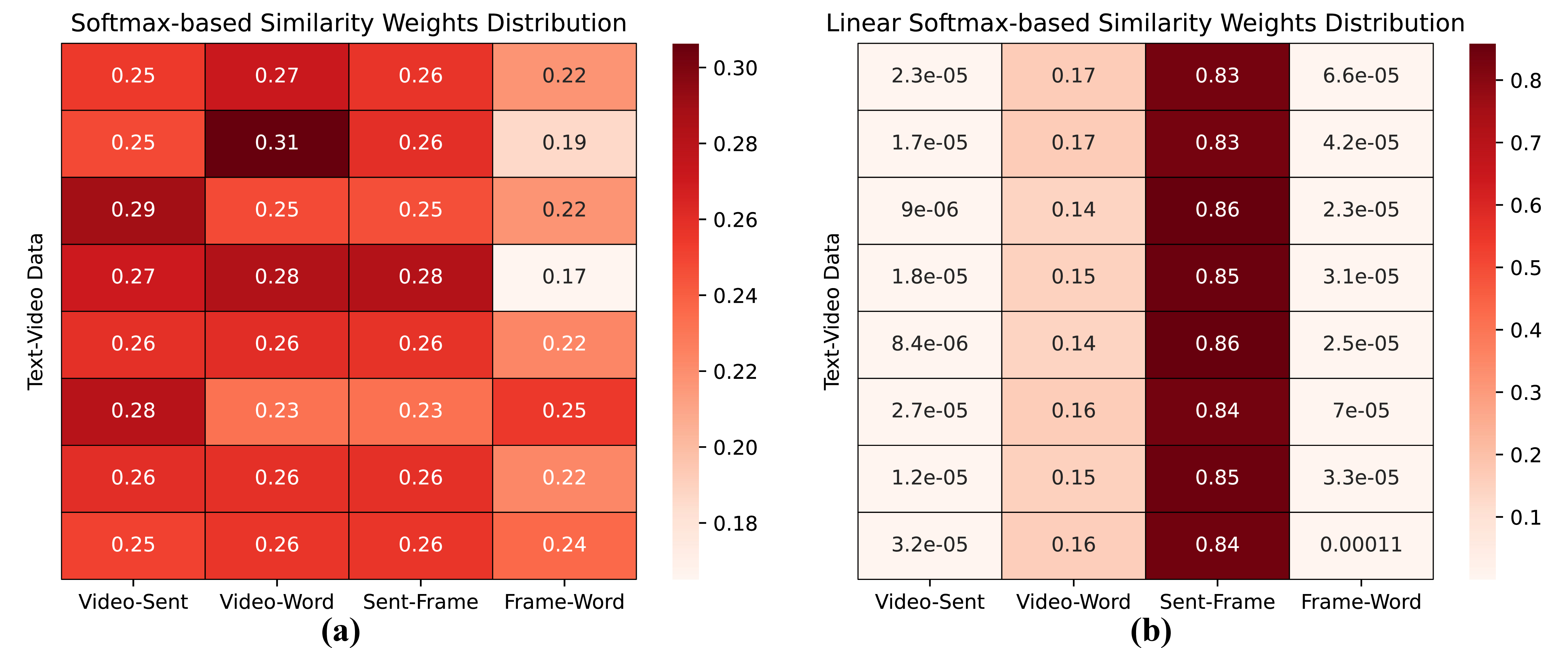}}
    \caption{(a) Qualitative results of softmax-based similarity weights distribution for matching text-video pairs. (b) Qualitative results of linear softmax-based similarity weights distribution for matching text-video pairs.}
    \label{fig: qualitative results of lsa visualization}
\end{figure*}

\begin{figure*}[!t]
    \centerline{\includegraphics[width=0.8\textwidth]{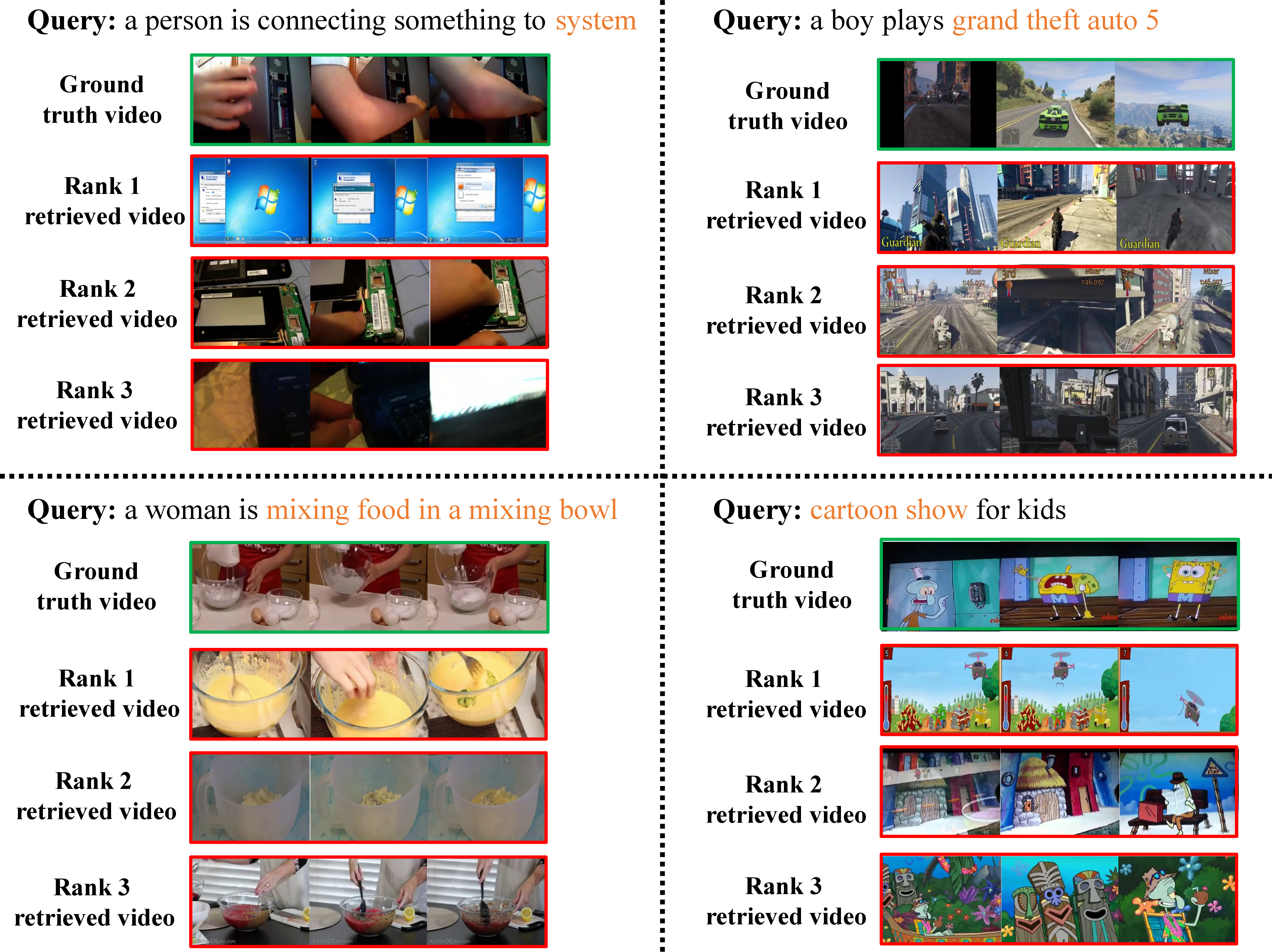}}
    \caption{Visualization of some failure text-to-video retrieval examples on the MSR-VTT dataset. Given the text query, the ground-truth in green and top-3 incorrectly retrieved videos in red are shown, respectively. Note that the ambiguous words are highlighted in orange.}
    \label{fig: fail_case_visualization}
\end{figure*}

\subsubsection{Retrieval Results}
Fig. \ref{fig: retrieval_results_visualization} shows a series of text-to-video retrieval results with two models, including X-CLIP and TC-MGC. In the 1st result, we find that both top-1 results contain ``boat'' but only TC-MGC fully understands the detailed visual semantics, such as ``a monkey'' and ``walking on the tree''. In the 2nd result, we observe that both top-1 results are about an event but only TC-MGC fully fits the word semantics ``a hat''. In the bottom two results, we notice that X-CLIP fails to capture the contents of ``some girls'' and ``playing with a cats tail''. However, TC-MGC successfully retrieves the correct videos from the gallery, showing the powerful fine-grained semantic perception of TC-MGC. These results demonstrate that our method can achieve accurate cross-modal retrieval by semantic alignment between fine-grained textual and visual clues. 

\subsubsection{t-SNE Visualization}
Fig. \ref{fig: t_SNE_visualization} shows the t-SNE \cite{van2008visualizing} visualization of the word and frame features we extracted. We randomly sample 40 text-video pairs. It can be seen that the visualization results present clear clusters, with evenly distributed data points and no obvious overlap. This phenomenon fully reflects the extracted features have great discrimination and representativeness.

\subsubsection{Multi-Grained Attention}
The core of our proposed TC-MGC is to capture partial  relevance between multi-grained textual and visual features for video/frame representations refinement. To better understand this capability, we show the visualization of multi-grained attention weights for both sentences and words on video frames in Fig. \ref{fig: lva_visualization}. As observed, both sentence and word mainly assign weights to semantic-related frames while others are largely ignored. In particular, as shown in the top example, the sentence exhibits the highest relevance with frames 8-11 that effectively represent the ``review-related'' visual scenarios. Furthermore, the word ``man'' is most associated with frames 8-11 that contain analogous semantic content, and only the last frame fully fits the word semantics ``vehicle''. In the bottom example, we find only the frames 4-6 and 8-11 encapsulate the related appearance characteristics of the sentence. Additionally, the word ``man'' is semantic-aligned to all frames except frames 7 and 12, and only frames 4-6 and 8-11 are about stage. These examples prove that the proposed method can achieve precise semantic alignments between sentence/word and frame representations via computed attention weights.

\subsubsection{Similarity Decorrelation Regularization}
In Fig. \ref{fig: qualitative results of sdr visualization}, we perform visualizations of normalized multi-grained similarities distribution under average-based aggregation for matching text-video pairs to qualitatively the effectiveness of SDR loss. From the similarity comparison between plots (a) and (b), it is clear that the SDR loss can greatly mitigate the serious imbalance problem among different similarity scores. Specifically, in the first text-video pair, the model without SDR loss outputs substantial weights to video-word and sentence-frame similarities while the remaining two similarities are approximately not considered, which seriously hinders the utilization of cooperative relationships among multi-grained similarities. However, the auxiliary SDR loss can effectively correct this issue and generate moderate weights for different similarities, thus promoting the cooperation of multi-grained information and achieving better retrieval performance.    
	
\subsubsection{Linear Softmax Aggregation}
Given matching text-video pairs, some qualitative examples of similarity weights output by the softmax and linear-softmax layers are illustrated in Fig. \ref{fig: qualitative results of lsa visualization}. Notably, the SDR loss is used in these experiments. From plot (a), it can be seen that the softmax-based weights are similar, making it hard for the model to identify the importance of similarities with different granularity. However, when our model is equipped with the linear softmax aggregation module, we observe that the output similarity weights in the plot (b) tend to be discriminative. More precisely, the video-word and sentence-frame similarities are assigned to larger weights while the other similarities are virtually neglected in the aggregation process. In this way, the aggregated similarity can be regarded as a comprehensive contrastive score, which enables the model to better capture precise cross-modal semantic correspondence.

\subsubsection{Other Case Study}
We also show some failure cases as well in Fig. \ref{fig: fail_case_visualization}, where the ground truth does not rank first in the retrieval results. As shown in Fig. \ref{fig: fail_case_visualization}, we observe that the top-1 video retrieved by our model seems to be very consistent with the query text although it's not ground truth. This is due to the ambiguous and general annotations in the datasets, such as ``grand left auto 5'' and ``cartoon show'' in the right two examples. The left two examples have similar properties. We find that this problem is common in text-video retrieval, and the polysemous annotations account for 1-2\% of the datasets. Based on the above analysis, our model has a potential to correctly retrieve more videos once more discriminative annotations with enough details are given in the datasets.

\section{Limitations and Discussions}
\label{Limitations and Discussions}
In this paper, our proposed TC-MGC achieves remarkable performance on three popular text-video retrieval benchmarks. However, this approach only conducts cross-modal interactions between word-sentence and semantic-relevant frame-video representations. Generally, texts and videos can be regarded as word/phrase/sentence and frame/clip/video combinations. Without considering phrase-level textual and clip-level visual representations, the model can not fully capture cross-modal correspondence, thus greatly decreasing the retrieval model's recall in practical applications.
A common solution to this issue is to introduce phrase-level textual and clip-level visual representations into the TVR framework. Note that the retrieval performance could be further improved by the usage of phrase-conditioned visual representations in the hierarchical cross-modal interactions.

Moreover, the unified keep rate in the similarity reorganization module exhibits two significant limitations in practical applications. Firstly, the static keep rate may lead to optimization challenges. When the data distribution or task requirement changes, the keep rate must be re-tuned to achieve optimal results. Secondly, the static keep rate may make the model overfit certain datasets, thus limiting the application scope and generalization ability. To address this issue, we consider utilizing the dynamic strategy to control the similarity reorganization at a finer granularity and perform more precise reorganization. We will investigate these in the future study.

\section{Conclusion}
\label{Conclusion}
This work proposes Text-Conditioned Multi-Grained Contrast framework (TC-MGC), a novel method exploring how to effectively investigate text-conditioned multi-grained contrasts for text-video retrieval. As a multi-grained training framework, our TC-MGC performs aggregated video and frame representations generation through a language-video attention block and better aligns text-video semantics. Then, to effectively filter noisy interactions and reduce trainable parameters in the Interactive Similarity Aggregation (ISA) module, we design a Similarity Reorganization (SR) module and the bidirectional variant (Bi-SR) module to preserve attentive similarities in the cross-modal similarity vectors/matrices. Next, we introduce an auxiliary Similarity Decorrelation Regularization (SDR) loss to alleviate over- and under-representation issues among different similarities via variance minimization optimization. Finally, we present a new Linear Softmax Aggregation (LSA) module, which aims to facilitate similarity interactions with different granularity before aggregation. The remarkable performance and thorough ablation studies clearly confirm the effectiveness and superiority of our method. 

In the future, it would be interesting to test whether the text-conditioned video representations can boost the performance of hierarchical cross-modal interaction models as we only integrate them into multi-grained contrastive learning framework. Furthermore, static keep rates in the SR module inevitably limit domain generalization. It would be interesting to investigate the dynamic strategy that conducts more precise similarity reorganization with different keep rates. We also plan to apply our approach to other popular video-language tasks such as video question answering and visual storytelling, as part of future work.

\section*{Acknowledgments}
This research was partially funded by the China National Key  R\&D Research Program (2020YFB1711200) and (2019YFB1705801).

\appendices
\setcounter{table}{0}   
\setcounter{figure}{0}  
\setcounter{equation}{0} 

\renewcommand{\thetable}{A.\arabic{table}}
\renewcommand{\thefigure}{A.\arabic{figure}}
\renewcommand{\theequation}{A.\arabic{equation}}

\section*{Appendix}
In this appendix, \& \ref{sec: sdr loss ablation} contains details of SDR loss ablation: variance minimization data (\& \ref{sec: variance minimization data}) and weighting parameter (\& \ref{sec: sdr loss weight}). \& \ref{sec: keep rate in SR} contains retrieval performance comparison at various keep rates in the SR module. \& \ref{sec: SR_Pseudocode} contains pseudocodes of the SR module.

\section{SDR Loss Ablation} 
\label{sec: sdr loss ablation}
In this subsection, we give a detailed analysis on the SDR loss from two perspectives. On the one hand, we conduct an ablative study to validate the effectiveness of matching text-video data selection for variance minimization in Section \ref{sec: variance minimization data}. On the other hand, we perform experiments to explore the impact of different SDR loss weighting parameter $\alpha$ in Section \ref{sec: sdr loss weight}. Note that all experimental results are obtained with the LSA module for multi-grained similarity scores aggregation.

\subsection{Effect of Variance Minimization Data}
\label{sec: variance minimization data}
\begin{table}[!t]
	\caption{Retrieval results with different variance minimization data in the SDR loss computation. The SDR loss weighting parameter is set to 0.5. VMD is short for variance minimization data. }
	\label{table: ablation study of different variance minimization data}
	\centering
		\resizebox{0.5\textwidth}{!}{
			\Huge
			\renewcommand{\arraystretch}{1.1}
			\begin{tabular}{l|cccc|cccc|c}
				\hline
				\multirow{2}*{VMD} & \multicolumn{4}{c|}{Text$\rightarrow$Video} & \multicolumn{4}{c|}{Video$\rightarrow$Text} & \multirow{2}*{SumR$\uparrow$}\\ 
				\cline{2-5} \cline{6-9}
				& R@1$\uparrow$  & R@5$\uparrow$  & R@10$\uparrow$ & RSum$\uparrow$
				& R@1$\uparrow$  & R@5$\uparrow$  & R@10$\uparrow$ & RSum$\uparrow$\\
				\hline
				Positives (ours) & \textbf{47.4} & 74.8 & \textbf{84.2} & \textbf{206.4} & \textbf{45.9} & \textbf{74.5} & 83.3 & \textbf{203.7} & \textbf{410.1} \\
				Negatives & 45.9 & \textbf{75.0} & 83.1 & 204.0 & 45.3 & 73.0 & 82.1 & 200.4 & 404.4 \\
				All & 46.0 & 74.1 & 83.4 & 203.5 & \textbf{45.9} & 72.8 & \textbf{83.4} & 202.1 & 405.6  \\
				\hline
			\end{tabular}
		}
\end{table}

\begin{figure}[!t]
	\centerline{\includegraphics[width=0.5\textwidth]{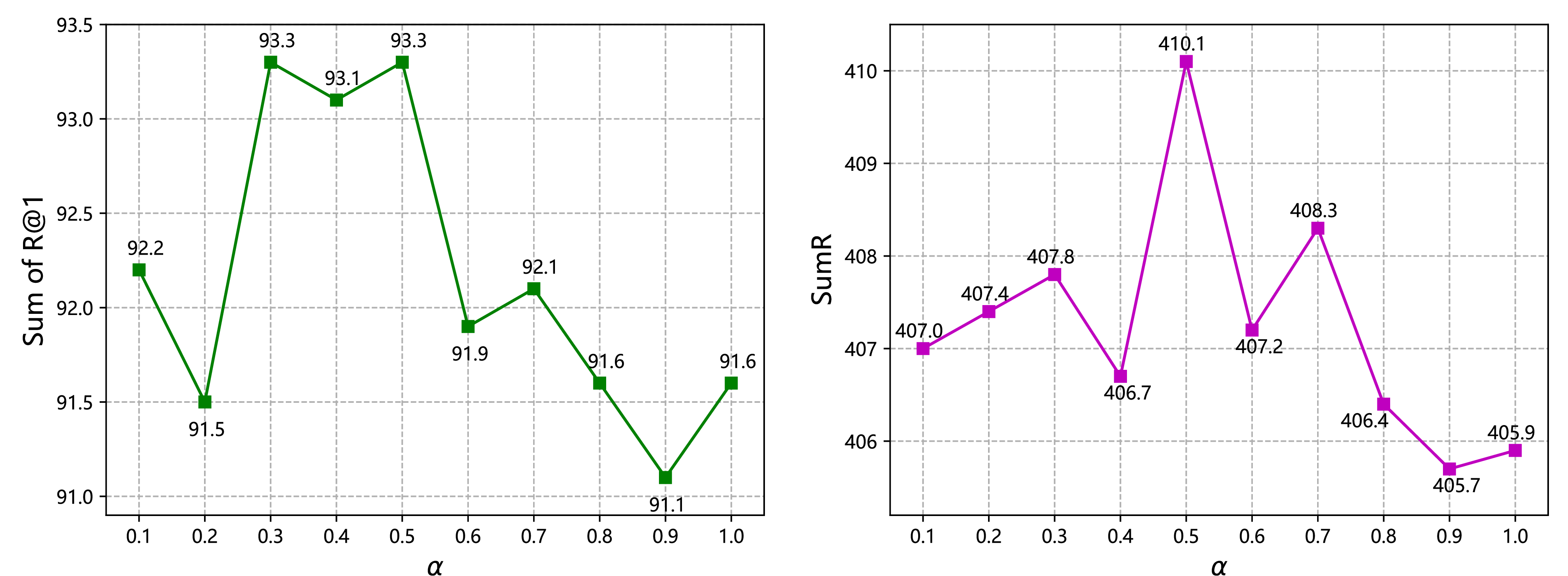}}
	\caption{Comparisons of performance with different $\alpha$ in Eq. \ref{eq: total_loss}.}
	\label{fig: different sdr_loss_weight}
\end{figure}

To justify the superiority of the matching text-video data (positives) in SDR loss calculation, we compare it with other variants (\textit{i.e.,} negatives and all combinations). As shown in Table \ref{table: ablation study of different variance minimization data}, we observe that the adoption of positives performs best (\textit{i.e.,} 410.1 SumR) while the negatives selection performs worst (\textit{i.e.,} 404.4 SumR). This may be because metric consistency on matching text-video pairs can greatly mitigate the serious imbalance problem and enhance the cooperative relationships utilization. However, when the metric consistency is applied to the negatives, it may introduce too many noisy optimization objectives and make it hard for the model to capture precise cross-modal correspondence. Therefore, we leverage matching text-video data in the SDR loss optimization.

\subsection{Effect of Weighting Parameter}
\label{sec: sdr loss weight}
We test the weighting parameter range setting $\alpha \in [0.1, 1]$ to evaluate the effect of different $\alpha$ . In Fig. \ref{fig: different sdr_loss_weight}, we find that $\alpha=0.5$ gives the best retrieval performance. More specifically, with $\alpha$ increasing from 0.1 to 0.5, the sum of R@1 is improved from 92.2 to 93.3 with +1.1\% absolute performance gain. Meanwhile, when $\alpha=0.5$, we can obtain the best SumR of 410.1. Therefore, we adopt $\alpha=0.5$ in our experiments to achieve the best retrieval performance. 

\setcounter{table}{0}   
\setcounter{figure}{0}  
\setcounter{equation}{0} 

\renewcommand{\thetable}{B.\arabic{table}}
\renewcommand{\thefigure}{B.\arabic{figure}}
\renewcommand{\theequation}{B.\arabic{equation}}

\section{Effect of Keep Rate in the SR Module}
\label{sec: keep rate in SR}
\begin{figure}[!t]
    \centerline{\includegraphics[width=0.35\textwidth]{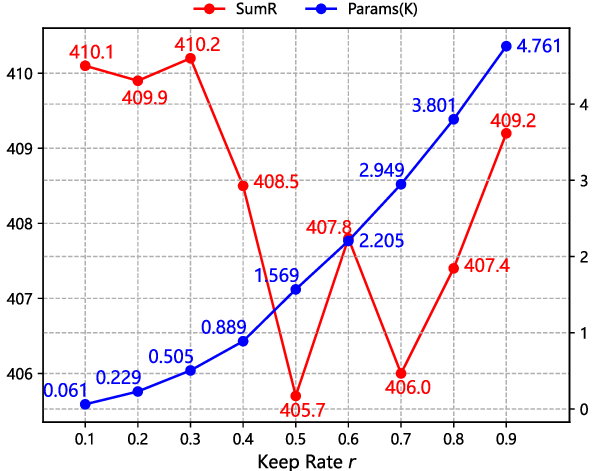}}
    \caption{Comparison of trade-off between performance and trainable parameters with different keep rates. }
    \label{fig: sr_keep_rates_diagram}
\end{figure}

To examine the impact of various keep rates, we evaluate our approach with keep rate range setting $r \in [0.1, 0.9]$. As shown in Fig. \ref{fig: sr_keep_rates_diagram}, we observe that although the best SumR of 410.2 is obtained when $r=0.3$, the parameter statistics in the linear layer of ISA are improved from 0.061k to 0.505k. With the keep rates increasing, the overall SumR suffers a significant decrease. This may be correlated with the excessive noisy interactions in the linear layer of ISA. As a result, considering the trade-off between retrieval performance and trainable parameters, we adopt $r=0.1$ to achieve the best retrieval performance.

\setcounter{table}{0}   
\setcounter{figure}{0}  
\setcounter{equation}{0} 

\renewcommand{\thetable}{C.\arabic{table}}
\renewcommand{\thefigure}{C.\arabic{figure}}
\renewcommand{\theequation}{C.\arabic{equation}}
\renewcommand{\thefigure}{C.\arabic{algorithm}}

\section{Pseudocode for the SR module}
\label{sec: SR_Pseudocode}
Here we present complete pseudocodes for our similarity reorganization (SR) at video-word, sentence-frame and frame-word levels in Algorithm \ref{alg: video_word_sr_algorithm}-\ref{alg: frame_word_sr_algorithm}. We believe the pseudocode would aid researchers to better understand and replicate the proposed similarity reorganization.

\begin{algorithm*}[!htbp]
\caption{PyTorch-style pseudocode for Video-Word Similarity Reorganization.}
\label{alg: video_word_sr_algorithm}
\lstset{style=mystyle}
\begin{lstlisting}
# B_t: batch size of text                          B_v: batch size of video                
# N_t: word size                                   D: dimensionality of the embeddings
# video_feature: video feature (B_t x B_v x D)     word_features: word features (B_t x N_t x D)
# r: keep rate

def video_word_similarity_reorganization(video_feature, word_features):
    # compute video_word_logits
    video_word_logits = torch.matmul(word_features, video_feature.permute(0, 2, 1))  # B_t x N_t x B_v
    N_t = video_word_logits.shape[1]
    
    # compute number of attentive similarities
    attn_sim_num = int(N_t * r)
    
    # obtain index of attentive similarities
    _, attn_sim_index = video_word_logits.topk(attn_sim_num, dim=1, largest=True, sorted=True)
    
    # generate attn_video_word_logits
    attn_video_word_logits = video_word_logits.gather(dim=1, index=attn_sim_index) # B_t x (N_t x r) x B_v
    
    # generate sr_video_word_logits  (B_t x (N_t x r) x B_v)
    sr_video_word_logits = attn_video_word_logits 
    return sr_video_word_logits
\end{lstlisting}
\end{algorithm*}

\begin{algorithm*}[!htbp]
\caption{PyTorch-style pseudocode for Sentence-Frame Similarity Reorganization.}
\label{alg: sent_frame_sr_algorithm}
\lstset{style=mystyle}
\begin{lstlisting}
# B_t: batch size of text                          B_v: batch size of video                
# N_v: frame size                                  D: dimensionality of the embeddings
# sent_feature: sentence feature (B_t x D)         frame_features: frame features (B_t x N_v x B_v x D)
# r: keep rate
        
def sentence_frame_similarity_reorganization(sent_feature, frame_features):
    # compute sent_frame_logits
    sent_frame_logits = torch.einsum('td, twvd -> twv', sent_feature, frame_features) # B_t x N_v x B_v
    N_v = sent_frame_logits.shape[1]
        
    # obtain attn_sent_frame_logits
    attn_sim_num = int(N_v * r)
    _, attn_sim_index = sent_frame_logits.topk(attn_sim_num, dim=1, largest=True, sorted=True)
    attn_sent_frame_logits = sent_frame_logits.gather(dim=1, index=attn_sim_index) # B_t x (N_v x r) x B_v
    
    # obtain fused_sent_frame_score
    fused_sim_num = N_v - attn_sim_num
    _, fused_sim_index = sent_frame_logits.topk(fused_sim_num, dim=1, largest=False, sorted=True)
    fused_sent_frame_logits = sent_frame_logits.gather(dim=1, index=fused_sim_index)
    fused_sent_frame_weights = fused_sent_frame_logits.softmax(dim=1)
    fused_sent_frame_score = (fused_sent_frame_weights * fused_sent_frame_logits).sum(dim=1) # B_t x B_v
    
    # obtain sr_sent_frame_logits (B_t x (N_v x r + 1) x B_v)
    sr_sent_frame_logits = torch.cat([attn_sent_frame_logits, fused_sent_frame_score.unsqueeze(1)], dim=1)
    return sr_sent_frame_logits
    \end{lstlisting}
\end{algorithm*}

\begin{algorithm*}[!htbp]
\caption{PyTorch-style pseudocode for Frame-Word Bidirectional Similarity Reorganization.}
\label{alg: frame_word_sr_algorithm}
\lstset{style=mystyle}
\begin{lstlisting}
# B_t: batch size of text                          B_v: batch size of video                
# N_t: word size                                   N_v: frame size 
# word_features: word features (B_t x N_t x D)     frame_features: frame features (B_t x N_v x B_v x D)
# r: keep rate                                     D: dimensionality of the embeddings
        
def frame_word_bidirectional_similarity_reorganization(frame_features, word_features):
    # compute frame_word_logits (B_t x N_t x B_v x N_v)
    frame_word_logits = torch.einsum('twd, tfvd -> twvf', word_features, frame_features) 
    _, N_t, _, N_v = frame_word_logits.shape
        
    # similarity reorganization on word direction to generate attn_frame_word_logits (B_t x (N_t x r) x B_v x N_v)
    attn_word_num = int(N_t * r)
    _, attn_word_index = frame_word_logits.topk(attn_word_num, dim=1, largest=True, sorted=True)
    attn_frame_word_logits = frame_word_logits.gather(dim=1, index=attn_word_index) 
    
    # similarity reorganization on frame direction to generate sr_frame_word_logits (B_t x (N_t x r) x B_v x (N_v x r))
    attn_frame_num = int(N_v * r)
    _, attn_frame_index = attn_frame_word_logits.topk(attn_frame_num, dim=-1, largest=True, sorted=True)
    sr_frame_word_logits = attn_frame_word_logits.gather(dim=-1, index=attn_frame_index)
    return sr_frame_word_logits
    \end{lstlisting}
\end{algorithm*}



\end{document}